\documentclass[journal,twoside,web]{ieeecolor}
\usepackage{generic}
\usepackage{cite}
\usepackage{amsmath,amssymb,amsfonts}
\usepackage{algorithmic}
\usepackage{graphicx}
\usepackage{textcomp}
\def\BibTeX{{\rm B\kern-.05em{\sc i\kern-.025em b}\kern-.08em
    T\kern-.1667em\lower.7ex\hbox{E}\kern-.125emX}}
\usepackage{float}
\usepackage{epsfig} 
\usepackage{euscript}
\usepackage{color}
\usepackage{ccaption}
\usepackage{listings}
\usepackage{booktabs} 
\usepackage{longtable}
\usepackage{multirow}
\usepackage{epstopdf}
\usepackage{bm}
\usepackage{makecell}
\usepackage{enumerate}
\usepackage{mathrsfs} 
\usepackage{subcaption}

\newtheorem{theorem}{\it {Theorem}}

\newtheorem{proposition}{\it {Proposition}}
\newtheorem{assumption}{\it {Assumption}}
\newtheorem{corollary}{\it {Corollary}}
\newtheorem{lemma}{\it {Lemma}}
\newtheorem{remark}{\it {Remark}}

\newcommand{\Z}{{\mathbb Z}}
\newcommand{\E}{{\mathbb E}}
\newcommand{\COV}{{\mathbb C\mathbb O\mathbb V}}

\newcommand{\R}{{\mathbb R}}

\newcommand{\SURE}{\text{SURE}}

\makeatletter

\newcommand{\Rmnum}[1]{\expandafter \@slowromancap\romannumeral #1@}
\makeatother

\DeclareMathOperator*{\rank}{rank}

\DeclareMathOperator*{\cond}{cond}
\DeclareMathOperator*{\Tr}{Tr}

\DeclareMathOperator*{\MSE}{MSE}
\DeclareMathOperator*{\LS}{LS}
\DeclareMathOperator*{\GCV}{GCV}
\DeclareMathOperator*{\EB}{EB}

\DeclareMathOperator*{\TR}{R}
\DeclareMathOperator*{\tvec}{vec}
\DeclareMathOperator*{\tb}{b}
\DeclareMathOperator*{\AR}{AR}
\DeclareMathOperator*{\ALS}{ALS}

\DeclareMathOperator*{\tH}{H}
\DeclareMathOperator*{\SMSE}{SMSE}

\DeclareMathOperator*\argmin{arg\,min}

\allowdisplaybreaks[4]

\title{Asymptotic Theory for Regularized System Identification Part I: Empirical Bayes Hyper-parameter Estimator
}
\author{Yue Ju$^{1}$, Biqiang Mu$^{2}$, Lennart Ljung$^{3}$ and Tianshi Chen$^{1,*}$
	\thanks{*A preliminary version  of this work \cite{JCML20} was published in the 59th IEEE Conference on Decision and Control (CDC), 2020.
		This work was supported in part by NSFC under contract No. 62273287 and 61773329, by the Shenzhen Science and Technology Innovation Council under contract No. JCYJ20220530143418040 and JCY20170411102101881, the Thousand Youth Talents Plan funded by the central government of China, the Swedish Research Council, contract 2019-04956, the Vinnova’s center LINKSIC and {the Strategic Priority Research Program of Chinese Academy of Sciences under Grant No. XDA27000000}.}
	\thanks{$^{1}$Yue Ju and Tianshi Chen* (corresponding author) are with the School of Data Science and Shenzhen Research Institute of Big Data, The Chinese University of Hong Kong, Shenzhen, 518172, China, {\tt\small yueju@link.cuhk.edu.cn, tschen@cuhk.edu.cn}.}%
	\thanks{$^{2}$Biqiang Mu is with Key Laboratory of Systems and Control, Institute of Systems Science, Academy of Mathematics and System Science, Chinese Academy of Sciences, Beijing 100190, China {\tt\small bqmu@amss.ac.cn}.}
	\thanks{$^{3}$Lennart Ljung is with the Division of Automatic Control, Department of Electrical Engineering, Link$\ddot{\text{o}}$ping University, Link$\ddot{\text{o}}$ping SE-58183, Sweden {\tt\small ljung@isy.liu.se}.}
}

\begin{document}
\maketitle

\begin{abstract}
	Regularized {techniques, also named as kernel-based techniques, are the major advances} in system identification in the last decade. Although many promising results have been achieved, {their theoretical analysis} is far from complete and there are still many key problems to be solved. One of them is the asymptotic theory, which is about convergence properties of the model estimators as the sample size goes to infinity. The existing related results for regularized system identification are about the almost sure convergence of various hyper-parameter estimators. A common problem of those results is that they do not contain information on the factors that affect the convergence properties of those hyper-parameter estimators, e.g., the regression matrix. In this paper, we tackle problems of this kind for the regularized finite impulse response model estimation with the empirical Bayes (EB) hyper-parameter estimator and filtered white noise input. In order to expose and find those factors, we study the convergence in distribution of the EB hyper-parameter estimator, and the asymptotic distribution of its corresponding model estimator. For illustration, we run Monte Carlo simulations to show the efficacy of our obtained theoretical results. 
\end{abstract}



\section{Introduction}

\IEEEPARstart{I}{n} the last decade, there has been a surge of interests to study linear time-invariant (LTI) system identification problems by estimating impulse response models of LTI systems with regularized least squares (RLS) methods, and this research direction is often called \emph{regularized system identification}. Many results have been reported in this direction, {addressing, e.g., the regularization design and analysis \cite{Chen18,MSS16,BP2020mathematical,ZC18,Z21}, and the efficient implementations \cite{CL2013,CA21}; for more references, the interested readers are referred to the survey/tutorial papers} \cite{PDCDL2014,Chiuso16,LCM20} {and the book \cite{PCCDL2022}}. These results make the regularized system identification become not only {a complement to classic system identification based on the maximum likelihood/prediction error methods (ML/PEM)} and its asymptotic theory \cite{Ljung1999}, but also an emerging new system identification paradigm \cite{LCM20}. {The success of RLS} is due to at least the following three factors. First, {the} underlying model structure is determined by a carefully designed regularization term that incorporates the prior knowledge of the system to be identified, such as stability and dominant dynamics, {thus enhancing the estimation performance}. Second, {model complexity is governed by a continuous hyper-parameter used to parametrize the regularization term, which can be tuned more flexibly w.r.t. discrete orders used in classic system identification.} Third, the connections {of RLS with kernel methods} \cite{CS02} and Bayesian methods \cite{RasmussenW:06} {enable the usage of ideas and tools from the latter, which enriches our ideas and tools, and} enhances our capability in dealing with system identification problems.

{Although many promising results have been achieved,
there are still many key problems to be solved.} One of them is the asymptotic theory, which is the theory of convergence properties of the model estimators as the sample size goes to infinity and {that is} widely used to assess the quality of model estimators. {For classic system identification, asymptotic theory has been widely studied \cite{Ljung1999}. However, for regularized system identification, the study of the asymptotic theory just started and only} very few results have been reported so far \cite{PC15,MCL2018,MCL2018gcv,Hnotes,H2020}. 
In particular, the almost sure convergence of the empirical Bayes ($\EB$) hyper-parameter estimator has been studied in \cite{PC15} and \cite{MCL2018} for scalably and generally parameterized regularization term, respectively. In \cite{MCL2018,MCL2018gcv}, we studied the almost sure convergence of the Stein's unbiased risk estimator ($\SURE$) and the generalized cross validation ($\GCV$) hyper-parameter estimators, and showed that they are both asymptotically optimal in the sense of minimizing the mean square error (MSE). {In \cite{Hnotes,H2020}, one asymptotic approximation of MSE is studied.} A common problem of these results is that they do not contain information on the factors that affect the convergence properties of the hyper-parameter estimators. For example, it has been shown in \cite{MCL2018} that the EB hyper-parameter estimator converges to its limit with a rate of $1/\sqrt{N}$, where $N$ is the sample size, but this information is rough. In fact, it is well known from numerical simulations, e.g., \cite{PDCDL2014,MCL2018}, that the more ill-conditioned the regression matrix, the more samples needed to get an RLS estimator with good quality. A conjecture is that the more ill-conditioned the regression matrix, the more slowly the EB hyper-parameter estimator converges to its limit, 
but there have been no theoretical results to support this conjecture so far.

In this paper, we tackle problems of this kind and try to build up the asymptotic theory for the regularized system identification {based on some fundamental results in \cite{JCL2021}\footnote{\cite{JCL2021} is a tutorial and not submitted for publication anywhere, but only uploaded to arXiv for the review of this series of papers. It includes the most fundamental results on the asymptotic properties of the least squares estimator and the regularized least square estimator.}}. In particular, we consider the regularized finite impulse response (FIR) model estimation with the EB hyper-parameter estimator and filtered white noise input. In order to expose and find the factors that affect the convergence properties of the EB hyper-parameter estimator and the corresponding RLS estimator, we first study the convergence in distribution of the EB hyper-parameter estimator and then the asymptotic distribution of the RLS estimator. Moreover, we make the analysis in the following order: first generally parameterized regularization, and then the ridge regression \cite{Hastieetal:01} as an illustration. Finally, we run Monte Carlo simulations to show the efficacy of our theoretical results.

The remaining parts of this paper are organized as follows. In Section \ref{sec:RLS for FIR model}, we first introduce some preliminary materials and then the problem statement. 
We then study in Section \ref{sec:effects on hyper-parameter estimators} the convergence in distribution of the EB hyper-parameter estimator to its limit, and in Section \ref{sec:asmptotic analysis for RLS estimator}, the asymptotic distribution of the corresponding RLS estimator to the true model parameter. In Section \ref{sec:case study}, we consider the ridge regression with filtered white noise inputs as an illustration.  In Section \ref{sec:sim}, we run Monte Carlo simulations to demonstrate our theoretical results. All proofs of theorems and propositions are included in Appendix A.



\section{Preliminary and Problem Statement}\label{sec:RLS for FIR model}

In this section, we first introduce some preliminary {background} materials and then the problem statement of this paper. 

\subsection{FIR Model Estimation}

{We focus on the finite impulse response (FIR) model}
\begin{align}\label{eq:liner regression model at time t}
y(t)=\sum_{i=1}^{n}g_{i}u(t-i)+v(t),\ t=1,\cdots,N,
\end{align}
where $n$ is the order of FIR model, $t\in\mathbb N$ is the time index, $N$ is the sample size and usually assumed to be larger than $n$, $u(t)\in\R$, $y(t)\in\R$, and $v(t)\in\R$ are the input, output and measurement noise at time $t$, respectively, and $g_{1},\cdots,g_{n}\in\R$ are FIR model parameters to be estimated.

{Model} \eqref{eq:liner regression model at time t} can be rewritten in a vector-matrix format:
\begin{align}\label{eq:vector-matrix form linear regression model}
Y=\Phi\theta+V,
\end{align}
where
\begin{subequations}
	\begin{align}
	Y=&\left[\begin{array}{cccc}y(1)& y(2)& \cdots & y(N)\end{array}\right]^{T},\\
	\Phi=&\left[\begin{array}{cccc}\phi(1)& \phi(2) & \cdots & \phi(N)\end{array}\right]^{T},\\
	\theta=&\left[\begin{array}{cccc}g_{1}& g_{2} & \cdots & g_{n}\end{array}\right]^{T},\\
	V=&\left[\begin{array}{cccc}v(1)& v(2)& \cdots & v(N) \end{array}\right]^{T}
	\end{align}
\end{subequations}
with $\phi(t)=\left[\begin{array}{cccc}u(t-1)& u(t-2) & \cdots&u(t-n)\end{array} \right]^{T}$ and $u(t)=0$ for $t< 0$. Here, $\Phi$ is often known as the regression matrix. The FIR model estimation is to estimate the unknown $\theta$ as ``well" as possible based on data $\{y(t),\phi(t)\}_{t=1}^{N}$. 

%

The theoretical analysis of the FIR model estimation is often done in a probabilistic framework. To this goal, we first make assumptions on the input $u(t)$ and the measurement noise $v(t)$.
\begin{assumption}\label{asp:input}
	The input $u(t)$ with $t=1-n,\cdots,N-1$ is {white noise filtered} with the stable filter $H(q)$, i.e.,
	\begin{subequations}\label{eq:asp of ut}
		\begin{align}
		H(q)=&\sum_{k=0}^{\infty}h(k)q^{-k}\ \text{with}\ \sum_{k=0}^{\infty}|h(k)|<\infty,\\
		\label{eq:def of ut}
		u(t)=&H(q)e(t)=\sum_{k=0}^{\infty}h(k)e(t-k),
		\end{align}
	\end{subequations}
	where $q^{-1}$ represents the backward shift operator{, i.e.,} $q^{-1}u(t)=u(t-1)$, and $e(t)$ is {independent} and identically distributed ($i.i.d.$) with zero mean, variance $\sigma_{e}^{2}>0$, bounded moments of order $4+\delta$ for some $\delta>0$, {and $\E[e(t)^4]=c\sigma_{e}^4$ with a constant $c>0$}. Moreover, we let 
	\begin{align}\label{eq:Sigma}
\Sigma=\COV(\left[\begin{array}{cccc}u(0)&u(1)&\cdots&u(n-1)\end{array}\right]^{T}),
	\end{align} where $\COV(\cdot)$ denotes the covariance matrix,
	and assume that $\Sigma$ is positive definite, i.e. $\Sigma\succ 0$. 
\end{assumption}

\begin{assumption}\label{asp:noise}
	The measurement noise $v(t)$ is $i.i.d.$ with zero mean, variance $\sigma^2>0$, and bounded moments of order $4+\delta$ for some $\delta>0$.
\end{assumption}

\begin{assumption}\label{asp:independence between input and noise}
	$\{e(t)\}_{t=-\infty}^{N-1}$ and $\{v(t)\}_{t=1}^{N}$ are mutually independent, which means that for $i=-\infty,\cdots,N-1$ and $j=1,\cdots,N$, $e(i)$ and $v(j)$ are independent. 

\end{assumption}


\begin{remark} \label{rmk:statistical properties of ut}

By Assumptions \ref{asp:input} and \ref{asp:independence between input and noise}, it is easy to verify that $u(t)$ is a stationary stochastic process, independent of $v(t)$, with
\begin{subequations}\label{eq:asp of vt}
	\begin{align}
	\E[u(t)]&=0,\\
	\label{eq:definition of R_u_tau}
	\E[u(t)u(t+\tau)]&\triangleq R_{u}(\tau)=\sigma_{e}^2\sum_{k=0}^{\infty}h(k)h(k+\tau),
	\end{align}
\end{subequations}
where $\E(\cdot)$ denotes the mathematical expectation, $\tau\geq0$, and $R_{u}(\tau)=R_{u}(-\tau)${. Moreover}, the $(i,j)$th element of $\Sigma$ in \eqref{eq:Sigma} is $R_{u}(|i-j|)$, which is determined by the filter $H(q)$ and often has no closed-form expression. 
\end{remark}

\begin{assumption}\label{asp:full column rank of regression matrix}
	The regression matrix $\Phi\in\R^{N\times n}$ with $N>n$ has full column rank, i.e. $\rank(\Phi)=n$.
\end{assumption}

We use the mean square error (MSE) in relation to the impulse response estimation {to assess how ``good" an estimator $\hat{\theta}\in\R^{n}$ of the true parameter $\theta_{0}=\left[\begin{array}{cccc}g_{1}^{0} & g_{2}^{0} & \cdots & g_{n}^{0}\end{array} \right]^{T}\in\R^{n}$ is \cite{COL2012,PC15}.
The MSE} is defined as follows,
	\begin{align}\label{eq:MSEg criterion}
	{\MSE}_{g}(\hat{\theta})=&\E(\|\hat{\theta}-\theta_{0}\|_{2}^{2}),
	\end{align}
where $\|\cdot\|_{2}$ denotes the Euclidean norm. The smaller MSE indicates the better quality of $\hat{\theta}$.

\begin{remark}\label{rmk:selection of model order}
We make the assumption that the dimension $n$ should be large enough to capture the dynamics of the underlying system to be identified. This is made possible, because the model complexity of the regularized FIR model estimator is governed by the hyper-parameter and tuned in a continuous way {(see, e.g., \cite{PCCDL2022})}.

\end{remark}

\subsection{The Least Squares Method}

Under Assumption \ref{asp:full column rank of regression matrix}, the simplest method for FIR model estimation is the Least Squares (LS):
\begin{subequations}\label{eq:LS estimator}
	\begin{align}\label{eq:LS form1}
	\hat{\theta}^{\LS}=&\argmin_{\theta\in\R^{n}}\|Y-\Phi\theta\|_{2}^{2}\\
	=&(\Phi^{T}\Phi)^{-1}\Phi^{T}Y.
	\end{align}
\end{subequations}
Recall the convergence in distribution in statistics\footnote{A sequence of random variables $\xi_{N}\in\R^{d}$ converges in distribution to a random variable $\xi\in\R^{d}$, if $\lim_{N\to\infty}\text{Pr}(\xi_{N}\leq x)=\text{Pr}(\xi\leq x)$ for every $x$ at which the limit distribution function $\text{Pr}(\xi\leq x)$ is continuous, where the map $x\mapsto \text{Pr}(\xi\leq x)$ denotes the distribution function of $\xi$ and $\text{Pr}(\cdot)$ is a probability function. It can be written as $\xi_{N} \overset{d.}\to \xi$.
} and let \begin{align}
			\label{eq:def of V_ALS_1}
		V_{1}^{\ALS}=&\sigma^2\Sigma^{-1}.
\end{align}
Then  it is well known that 
	\begin{align}\label{eq:convergence in dist of ls estimate}
		\sqrt{N}(\hat{\theta}^{\LS}-\theta_{0})\overset{d.}\to& \mathcal{N}(0,V^{\ALS}_{1}),
		\end{align} 
		which indicates that if $\Sigma$ is ill-conditioned, then $\sqrt{N}(\hat{\theta}^{\LS}-\theta_{0})$ may have large limiting variance.

\subsection{The Regularized Least Squares Method}

To handle the ill-conditioned problem, one can introduce a regularization term in \eqref{eq:LS form1} to obtain the regularized least squares (RLS) estimator:
\begin{subequations}\label{eq:RLS estimator}
	\begin{align}
	\hat{\theta}^{\TR}=&\argmin_{\theta\in\R^{n}}\|Y-\Phi\theta\|_{2}^{2}+\sigma^2\theta^{T}P^{-1}\theta\\
	\label{eq:rls estimator form1}
	=&(\Phi^{T}\Phi+\sigma^2 P^{-1})^{-1}\Phi^{T}Y\\
	=&P\Phi^{T}Q^{-1}Y,
	\end{align}
\end{subequations}
where $P\in\R^{n\times n}$ is positive semidefinite, its $(i,j)$th element $[P]_{i,j}$ can be designed through a positive semidefinite \emph{kernel} $\kappa(i,j;\eta):\mathbb N\times \mathbb N\to \R$, {thus, $P$ is often called the kernel matrix with $\eta\in\Omega\subset\R^{p}$ being the hyper-parameter,} and
\begin{align}\label{eq:def of Q}
Q=\Phi P\Phi^{T}+\sigma^2I_{N},
\end{align}
and $I_{N}$ denotes the $N$-dimensional identity matrix.

There are two key issues for the RLS method: 
the kernel design and the hyper-parameter estimation.



\subsubsection{Kernel Design}

The goal of kernel design is to embed the prior knowledge of the system to be identified in the kernel $\kappa(i,j;\eta)$ by parameterization of the kernel with the hyper-parameter $\eta$.

The mostly widely used kernels include
\begin{subequations}
	\begin{align}\label{eq:SS kernel}
	\text{SS}:&\kappa(i,j;\eta)=c\left(\frac{\alpha^{i+j+\max(i,j)}}{2}-\frac{\alpha^{3\max(i,j)}}{6}\right)\nonumber\\
	&\eta=[c,\alpha]\in\Omega=\{c\geq 0,\ \alpha\in[0,1)\},\\
	\label{eq:DC kernel}
	\text{DC}:&\kappa(i,j;\eta)=c\alpha^{(i+j)/2}\rho^{|i-j|},\nonumber\\
	&\eta=[c,\alpha,\rho]\in\Omega=\{c\geq 0,\ \alpha\in[0,1),\ |\rho|\leq 1\},\\
	\label{eq:TC kernel}
	\text{TC}:&\kappa(i,j;\eta)=c\alpha^{\max(i,j)},\nonumber\\
	&\eta=[c,\alpha]\in\Omega=\{c\geq 0,\ \alpha\in[0,1)\},
	\end{align}
\end{subequations}
where the stable spline (SS) kernel \eqref{eq:SS kernel} is introduced in \cite{PD2010}, the diagonal correlated (DC) kernel \eqref{eq:DC kernel} and the tuned-correlated (TC) kernel \eqref{eq:TC kernel} (also named as the first order stable spline kernel) are introduced in \cite{COL2012}.

\subsubsection{Hyper-parameter Estimation}

Given a designed kernel, the next step is to estimate the hyper-parameter $\eta$. There are many methods, {such as the empirical Bayes (EB), Stein's unbiased risk estimation (SURE) and generalized cross validation (GCV) method (see, e.g., \cite{PDCDL2014})}.

In the sequel, we consider the $\EB$ method, which assumes that $\theta$ and $V$ are independent and Gaussian distributed, i.e.,
\begin{align}
&\theta\sim \mathcal{N}({0},P),\ V\sim \mathcal{N}({0},\sigma^2I_{N}),\\
\Rightarrow &Y\sim \mathcal{N}({0},\Phi P\Phi^{T}+\sigma^2I_{N}).
\end{align}
Then, {hyper-parameters are found by maximizing the marginal likelihood function of $Y$ given $\eta$, which} is equivalent to minimizing 
\begin{align}\label{eq:original EB cost function}
\mathscr{F}_{\EB}=&Y^{T}Q^{-1}Y+\log\det(Q),
\end{align}
where $\det(\cdot)$ denotes the determinant of a square matrix {and $Q$ is defined as {in} \eqref{eq:def of Q}}. Moreover, {as the noise variance is unknown in practice,} we use an unbiased estimator of $\sigma^2$,
\begin{align}\label{eq:noisevariance_est}
\widehat{\sigma^2}=&\frac{\|Y-\Phi\hat{\theta}^{\LS}\|_{2}^{2}}{N-n}
=\frac{Y^{T}\left[I_{N}-\Phi(\Phi^{T}\Phi)^{-1}\Phi^{T}\right]Y}{N-n}.
\end{align}

{
\begin{remark}\label{rmk:treat sigma2 as hyper-parameter}
	As mentioned in \cite[Remark 5]{PDCDL2014}, one alternative way to estimate the unknown $\sigma^2$ is to consider it as an additional ``hyper-parameter'' included in $\eta$. In such case, the convergence properties of the corresponding hyper-parameter estimator and RLS estimator will be different from those shown in this paper, which will be shown in another paper of this series of papers.
\end{remark}
}

Replacing $\sigma^2$ with $\widehat{\sigma^2}$, the $\EB$ hyper-parameter estimator can be represented as
\begin{subequations}\label{eq:EB hp and its cost function}
\begin{align}\label{eq:EB hyperparameter estimator}
\text{EB}:\hat{\eta}_{\text{EB}}=&\argmin_{\eta\in\Omega}\widehat{\mathscr{F}_{\EB}}(\eta)\\ \label{eq:EB1}
\widehat{\mathscr{F}_{\EB}}(\eta)=&Y^{T}\hat{Q}(\eta)^{-1}Y+\log\det(\hat{Q}(\eta)),
\end{align}
\end{subequations}
where 
\begin{align}
\hat{Q}(\eta)=\Phi P(\eta)\Phi^{T}+\widehat{\sigma^2} I_{N}.
\end{align}
With \eqref{eq:EB hyperparameter estimator} and $\widehat{\sigma^2}$, the RLS estimator \eqref{eq:RLS estimator} becomes
\begin{align}\label{eq:RLS estimator2}
\hat{\theta}^{\TR}(\hat{\eta}_{\text{EB}})=P(\hat{\eta}_{\text{EB}})\Phi^T\hat Q(\hat{\eta}_{\text{EB}})^{-1}Y.	
\end{align}

\subsection{Problem Statement}\label{subsec:problem statement}

In this paper, we study the convergence properties of the EB hyper-parameter estimator $\hat{\eta}_{\text{EB}}$ in \eqref{eq:EB hyperparameter estimator} and the corresponding RLS estimator $\hat{\theta}^{\TR}(\hat{\eta}_{\text{EB}})$ in \eqref{eq:RLS estimator2} as the sample size $N$ goes to infinity. 
In fact, we have studied in \cite{MCL2018} the almost sure convergence\footnote{A sequence of random variables $\xi_{N}\in\R^{d}$ converges almost surely to a random variable $\xi\in\R^{d}$ if {for all $\epsilon>0$, $\Pr(\underset{N\to\infty}{\lim\sup}\{\|\xi_{N}-\xi\|_{2}>\epsilon\})=0$}, which can be written as $\xi_{N}\overset{a.s.}\to \xi$. {More generally, when $\xi$ is a set, the almost sure convergence of $\xi_{N}$ to $\xi$ \cite[(8.25)]{Ljung1999} is defined as $\inf_{\zeta\in\xi}\|\xi_{N}-\zeta\|_{2}\overset{a.s.}\to 0$ and still written as $\xi_{N}\overset{a.s.}\to \xi$ for simplicity.}} of $\hat{\eta}_{\text{EB}}$, and then we {realized} that it does not contain information on the factors that affect the convergence properties of $\hat{\eta}_{\text{EB}}$. To be more specific, we briefly recall the convergence result of $\hat{\eta}_{\text{EB}}$ in \cite{MCL2018}. To state the result, we make the following assumptions, which are also needed in this paper. 

\begin{assumption}\label{asp:interior points assumption of hat_eta}
	The hyper-parameter estimator $\hat{\eta}_{\text{EB}}$ is an interior point of $\Omega$ and $\Omega$ is a compact set, {where $\Omega$ is irrespective of $N$}.
\end{assumption}

	
	

{
\begin{remark}\label{rmk:estimator on the boundary}
Assumption \ref{asp:interior points assumption of hat_eta} is common in both classical system identification, e.g., \cite{Ljung1999}, and the regularized system identification, e.g., \cite{MCL2018,PC15}. However,  it must be stressed that the proposed analysis of this paper is  thus subject to some limitations. In particular, the proposed analysis cannot be applied to problems where $\hat{\eta}_{\EB}$ lies on the boundary,
 e.g., the network identification problems in \cite{ZC17,Zorzi22} and the sparse Bayesian learning and its application in \cite{WR2004,CALCP14}. 
\end{remark}
}

\begin{assumption}\label{asp:properties of P}
	$P(\eta)$ is positive definite, {continuously differentiable,} and twice continuously differentiable at every $\eta\in\Omega$.
\end{assumption}

\begin{assumption}\label{asp:isolated optimal sets}
	The set $\eta_{\tb}^{*}$, {defined as}
	\begin{subequations}\label{eq:Wb hp and its cost function}
	\begin{align}\label{eq:opt hyperparameter of Wb}
	\eta_{\tb}^{*}=&\argmin_{\eta\in\Omega}W_{\tb}(P,\theta_{0})\\
	\label{eq:Wb}
	W_{\tb}(P,\theta_{0})=&\theta_{0}^{T}P^{-1}\theta_{0}+\log\det(P),
	\end{align}
	\end{subequations}
	contains interior points of $\Omega$ and is made of isolated points.
\end{assumption}

Then by Assumptions \ref{asp:noise} and \ref{asp:full column rank of regression matrix}-\ref{asp:isolated optimal sets}, when we considered the deterministic inputs satisfying $\lim_{N\to\infty}\Phi^{T}\Phi/N=\Sigma$ and used the true noise variance $\sigma^2$, it was shown in \cite[Theorems 1 and 2]{MCL2018}
\begin{enumerate}
	\item the almost sure convergence of $\hat{\eta}_{\EB}$, i.e.,
	\begin{align}\label{eq:almost sure convergence of hat_eta_eb}
	\hat{\eta}_{\EB} \overset{a.s.}\to& \eta_{\tb}^{*};
	\end{align}
	\item how fast the convergence of $\hat{\eta}_{\EB}$ to $\eta_{\tb}^{*}$ only depends on $\|\hat{\theta}^{\LS}-\theta_{0}\|_{2}=O_{p}(1/\sqrt{N})${, as shown in \cite[Theorem 3]{JCL2021},} and is at a rate\footnote{For a sequence of random variables $\xi_{N}\in\R^{d}$ and a nonzero constant sequence $\{a_{N}\}$, we let ${\xi}_{N}=O_{p}(a_{N})$ denote that ${\xi_{N}}/a_{N}$ is bounded in probability, which means that $\forall \epsilon>0$, $\exists L>0$ such that 
		{$\underset{N\to\infty}{\lim\sup}\ \text{Pr}(\|{\xi}_{N}/a_{N}\|_{2}>L)<\epsilon$}.} of $1/\sqrt{N}$, i.e., 
	\begin{align}\label{eq:convergence rate of EB}
	\|\hat{\eta}_{\EB}-\eta_{\tb}^{*}\|_{2}=O_{p}(1/\sqrt{N}). 
	\end{align} 
	
	For convenience, this rate is called the ``convergence rate'' of $\hat{\eta}_{\EB}$ to $\eta_{\tb}^{*}$ in the sequel.
\end{enumerate}


Now it is clear to see that \eqref{eq:almost sure convergence of hat_eta_eb} and \eqref{eq:convergence rate of EB} do not contain any information on the factors that affect the convergence properties of $\hat{\eta}_{\text{EB}}$ to $\eta_{\tb}^{*}$, e.g., the regression matrix $\Phi$ and the kernel matrix $P$. It must be stressed that to know such information has both theoretical and practical significance. For instance, it is well known from numerical simulations {(see, e.g., \cite{PDCDL2014,MCL2018})} that when the filter $H(q)$ in \eqref{eq:asp of ut} is low-pass ($\Phi$ is thus ill-conditioned), it takes more samples to obtain $\hat{\theta}^{\TR}(\hat{\eta}_{\text{EB}})$ with good quality. A conjecture is that the more ill-conditioned $\Phi$, the {more slowly} $\hat{\eta}_{\text{EB}}$ converges to $\eta_{\tb}^{*}$. However, there have been no theoretical results to support this so far. In this paper, we try to tackle problems of this kind and in particular, we study how to expose and find the factors that affect the convergence properties of $\hat{\eta}_{\text{EB}}$ to $\eta_{\tb}^{*}$, and $\hat{\theta}^{\TR}(\hat{\eta}_{\text{EB}})$ to $\theta_0$. {
It is also well known from numerical simulations (see, e.g., \cite{PDCDL2014,MCL2018}) that, $\hat{\theta}^{\TR}(\hat{\eta}_{\text{EB}})$ and $\hat{\theta}^{\LS}$ may behave quite differently, especially when $\Phi$ is ill-conditioned. However, when checking the asymptotic properties of $\hat{\theta}^{\TR}(\hat{\eta}_{\text{EB}})$ and $\hat{\theta}^{\LS}$, and in particular, the frequently used almost sure convergence and convergence in distribution, we find that they have the same convergence properties. This finding gives us the intuition to consider instead the high order asymptotic distribution\footnote{For a sequence of random variables $\xi_N\in\R^{d}$, an $m$th order
		expansion {(e.g., \cite{G94})} of $\xi_N$ is expressed as
		\begin{align*}
			\xi_N = X_{N,1} + \frac{1}{\sqrt{N}} X_{N,2} + \frac{1}{N}X_{N,3} + . . . + \frac{1}{(\sqrt{N})^{m-1}} X_{N,m}
		\end{align*}
		where $(X_{N,1},\cdots,X_{N,m})$ jointly converges in distribution to a nontrivial distribution $(X_{1},\cdots,X_{m})$ (i.e., $X_{1},\cdots,X_{m}$ are all nonzero), $X_{N,i}, X_{i}\in\R^{d}$ for $i=1,\cdots,m$, and moreover, $X_{1} + \frac{1}{\sqrt{N}} X_{2} + \frac{1}{N}X_{3} + . . . + \frac{1}{(\sqrt{N})^{m-1}} X_{m}$ is called the $m$th order asymptotic distribution of $\xi_N$ and denoted by $\xi_N\overset{m\text{th}\ d.}\longrightarrow X_{1} + \frac{1}{\sqrt{N}} X_{2} + \frac{1}{N}X_{3} + . . . + \frac{1}{(\sqrt{N})^{m-1}} X_{m}$. In what follows, the convergence in distribution of $\xi_N$ will be also called the first order asymptotic distribution of $\xi_N$.
	}  to explain the different behaviors between $\hat{\theta}^{\TR}(\hat{\eta}_{\text{EB}})$ and $\hat{\theta}^{\LS}$. }

\subsection{Some Preliminary Results}\label{subsec:preliminary results}

Before proceeding to the discussions of the convergence properties of $\hat{\eta}_{\EB}$ in \eqref{eq:EB hyperparameter estimator} and $\hat{\theta}^{\TR}(\hat{\eta}_{\EB})$ in \eqref{eq:RLS estimator2}, we first state the following lemma, whose proof can be found in \cite{JCL2021}.

\begin{lemma}\label{lem:some preliminary results about inputs and ls estimates}
	For the FIR model \eqref{eq:liner regression model at time t} mentioned in Section \ref{sec:RLS for FIR model}, under Assumptions \ref{asp:input}-\ref{asp:independence between input and noise}, if $P(\eta)$ is differentiable for every $\eta\in\Omega$, we have the following results.
	\begin{enumerate}[1)]
		\item As mentioned in \cite[Theorems 1-2]{JCL2021}, we have
		\begin{gather}
			\label{eq:almost sure convergence of PP_N_inv}
			N(\Phi^{T}\Phi)^{-1}\overset{a.s.}\to \Sigma^{-1},\\
			\label{eq:almost sure convergence of sqrtPVN}
			\frac{\Phi^{T}V}{N}\overset{a.s.}\to  0,\\
			\label{eq:almost sure convergence of noise variance estimator}
			\widehat{\sigma^2}\overset{a.s.}\to \sigma^2,\\
			\left(\sqrt{N}\left(N(\Phi^{T}\Phi)^{-1}-\Sigma^{-1}\right), \sqrt{N}\Phi^{T}V/N,
			\sqrt{N}(\widehat{\sigma^2}-\sigma^2) \right)\nonumber\\
			\label{eq:joint convergence in distribution of ls estimate and other terms}
			\overset{d.}\to \left(-\Sigma^{-1}\Gamma\Sigma^{-1}, \upsilon, \rho \right),
		\end{gather}
		where $\Gamma\in\R^{n\times n}$, $\upsilon\in\R^{n}$ and $\rho\in\R$ are jointly Gaussian distributed with
		\begin{subequations}\label{eq:covariances of Gamma, upsilon and rho}
			\begin{align}
				\label{eq:mean of Gamma, upsilon, rho}
				&\E(\Gamma)=0,\E(\upsilon)=0, \E(\rho)=0,\\
				\label{eq:def of C_Gamma}
				&{C_{\Gamma}=\E(\Gamma\otimes\Gamma),}\nonumber\\
				&{\ \quad =\lim_{N\to\infty}N\E\left[\left(\frac{\Phi^{T}\Phi}{N}-\Sigma\right)\otimes \left(\frac{\Phi^{T}\Phi}{N}-\Sigma\right)\right],}\\
				\label{eq:covariance of Gamma, upsilon, rho}
				&\E(\upsilon\upsilon^{T})=\sigma^2\Sigma,\E(\rho^2)=\E[(v(t))^{4}]-\sigma^4,\\
				\label{eq:correlation of Gamma,upsilon,rho}
				&\E(\upsilon\otimes \Gamma)=0, \E(\rho\upsilon)=0, \E(\rho\Gamma)=0,
			\end{align}
		\end{subequations}
		{and $\otimes$ denotes the Kronecker product.}
		Moreover, for $i,j=1,\cdots,n$, the $(i,j)$th element of $\Sigma$ can be represented as
		$
		\left[\Sigma \right]_{i,j}=R_{u}(|i-j|);
		$
		{for $i,j=1,\cdots,n^2$, the $(i,j)$th element of $C_{\Gamma}$ is
		\begin{align}\label{eq:explicit representation of Expectation of Gamma_o_Gamma}
			&[C_{\Gamma}]_{i,j}
			=\left\{\E[e(t)]^{4}/\sigma_{e}^4-3 \right\} R_{u}(k)R_{u}(l)\\
			&+\sum_{\tau=-\infty}^{\infty}\left[R_{u}(\tau)R_{u}(\tau+k-l)+R_{u}(\tau+k)R_{u}(\tau-l)  \right],\nonumber
		\end{align}
		where $R_{u}(\tau)$ is defined in \eqref{eq:definition of R_u_tau}, 
		\begin{subequations}\label{eq:def of k and l}
			\begin{align}
				k=&\left|\lfloor{(i-1)/n}\rfloor-\lfloor{(j-1)/n}\rfloor\right|,\\
				l=&|i-j-\lfloor{(i-1)/n}\rfloor n+\lfloor{(j-1)/n}\rfloor n|,
			\end{align}
		\end{subequations}
		$|\cdot|$ denotes the absolute value, and $\lfloor{\cdot}\rfloor$ denotes the floor operation, i.e. $\lfloor{x}\rfloor=\max\{\tilde{x}\in\Z|\tilde{x}\leq x\}$.}

		
		
		\item As mentioned in \cite[Theorem 7, (44)-(47)]{JCL2021}, for any given $\eta\in\R^{p}$ {and}
		\begin{align}
		\label{eq:def of hat_S_inv}
		{\hat{S}(\eta)=P({\eta})+\widehat{\sigma^2}(\Phi^{T}\Phi)^{-1},}
		\end{align}
		it holds that
		\begin{gather}
			\label{eq:almost sure convergence of S_inv}
			\hat{S}(\eta)^{-1}\overset{a.s.}\to  P(\eta)^{-1},\\
			\label{eq:almost sure convergence of sqrtN_Sinv_Pinv}
			\sqrt{N}(\hat{S}(\eta)^{-1}-P(\eta)^{-1})\overset{a.s.}\to 0,\\
			\label{eq:almost sure convergence of 1st order derivatives of S_inv and P_inv}
			\frac{\partial \hat{S}(\eta)^{-1}}{\partial \eta_{k}} \overset{a.s.}\to \frac{\partial P(\eta)^{-1}}{\partial \eta_{k}},\\
			\label{eq:almost sure convergence of sqrtN difference of 1st order derivatives of S_inv and P_inv}
			\sqrt{N}\left(\frac{\partial \hat{S}(\eta)^{-1}}{\partial \eta_{k}}-\frac{\partial P(\eta)^{-1}}{\partial \eta_{k}}\right) \overset{a.s.}\to 0,
		\end{gather}
		where $\eta_{k}$ denotes the $k$th element of $\eta$ and $k=1,\cdots,p$.
		
		
		\item As mentioned in \cite[Theorem 7, (49)]{JCL2021}, for any estimator $\hat\eta_N\in\R^{p}$ of $\eta\in\R^{p}$ with $\hat\eta_N\overset{a.s.}\to\eta^*\in\R^{p}$, it holds that
		\begin{align}
			\label{eq:diff of S_inv and P_inv for general P}
			&\hat{S}(\hat{\eta}_{N})^{-1}-P(\eta^{*})^{-1}\nonumber\\
			=&-\hat{S}(\hat{\eta}_{N})^{-1}
			\left[\sum_{k=1}^{p}\left.\frac{\partial P(\eta)}{\partial \eta_{k}}\right|_{\eta=\tilde{\eta}_{N}}e_{k}^{T}(\hat{\eta}_{N}-\eta^{*})\right] P(\eta^{*})^{-1}\nonumber\\
			&-\widehat{\sigma^2}\hat{S}(\hat{\eta}_{N})^{-1}(\Phi^{T}\Phi)^{-1}P(\eta^{*})^{-1},
		\end{align}
		where $e_{k}\in\R^{p}$ denotes a column vector with $k$th element being one and others zero, and $\tilde{\eta}_{N}$ belongs to a neighborhood of $\eta^{*}$ with radius $\|\hat{\eta}_{N}-\eta^{*}\|_{2}$.
		
		\item As mentioned in \cite[Theorem 7, (51)]{JCL2021}, for any given $\eta\in\R^{p}$, it holds that
		\begin{align}\label{eq:difference of S_inv and P_inv}
			\hat{S}(\eta)^{-1}-P(\eta)^{-1}=-\frac{1}{N}\widehat{\sigma^2}\hat{S}(\eta)^{-1}N(\Phi^{T}\Phi)^{-1}P(\eta)^{-1}.
		\end{align}
	\end{enumerate}		
\end{lemma}

\section{Convergence in Distribution of the $\EB$ Hyper-parameter Estimator}\label{sec:effects on hyper-parameter estimators}

To expose the factors that affect the convergence properties of $\hat{\eta}_{\text{EB}}$ to $\eta_{\tb}^{*}$, we study the convergence in distribution  of $\sqrt{N}(\hat{\eta}_{\EB}-\eta_{\tb}^{*})$ under the following additional assumptions.
\begin{assumption}\label{asp:unique minimizing point of eta_b_star}
	$\eta_{\tb}^{*}$ consists of only one point.
\end{assumption}

\begin{remark}\label{remark:global minima of eta}
Assumption \ref{asp:unique minimizing point of eta_b_star} is common in the analysis of convergence in distribution of model estimators, {(see, e.g., \cite[Theorem 9.1]{Ljung1999})}.
\end{remark}

\begin{assumption}\label{asp:nozero 2nd order derivative of P with respect to eta}
	The first-order derivative of $P(\eta)$ with respect to $\eta$ at $\eta=\eta_{\tb}^{*}$ is nonzero, i.e. at least one of $k=1,\cdots,p$, $\left.{\partial P(\eta)}/{\partial \eta_{k}}\right|_{\eta=\eta_{\tb}^{*}}\not=0$.
\end{assumption}

{
	\begin{remark}\label{rmk:meaning of nonzero 1st order derivative}
		If for all $k=1,\cdots,p$, $\left.{\partial P(\eta)}/{\partial \eta_{k}}\right|_{\eta=\eta_{\tb}^{*}}=0$, then $A_{\tb}(\eta^{*}_{\tb})$ in \eqref{eq:element of A eta_b_star} and $B_{\tb}(\eta^{*}_{\tb})$ in \eqref{eq:element of B eta_b_star} will be zero matrices, leading to the meaningless convergence in distribution results in the following theorem. 
\end{remark}}


\begin{theorem}\label{thm:asymptotic normality of eta_eb difference}
	 Under Assumptions \ref{asp:input}-\ref{asp:isolated optimal sets}, \eqref{eq:almost sure convergence of hat_eta_eb} {holds} true.
	 Moreover, under additional Assumptions \ref{asp:unique minimizing point of eta_b_star}-\ref{asp:nozero 2nd order derivative of P with respect to eta}, we have
		\begin{align}\label{eq:convegence in distribution of hat_eta_eb}
		&\sqrt{N}(\hat{\eta}_{\EB}-\eta^{*}_{\tb})\overset{d.}\to\mathcal{N}(0,V_{\tb}^{\tH}(\eta_{\tb}^{*})),\\
		\label{eq:def of Vb_etab_star}
		&V_{\tb}^{\tH}(\eta_{\tb}^{*})=4\sigma^2A_{\tb}(\eta^{*}_{\tb})^{-1}B_{\tb}(\eta^{*}_{\tb})\Sigma^{-1}B_{\tb}(\eta^{*}_{\tb})^{T}A_{\tb}(\eta^{*}_{\tb})^{-1},
		\end{align}
		where for $k,l=1,\cdots,p$, the $(k,l)$th element of $A_{\tb}(\eta^{*}_{\tb})\in\R^{p\times p}$ can be represented as 
		\begin{align}\label{eq:element of A eta_b_star}
		\left[A_{\tb}(\eta^{*}_{\tb})\right]_{k,l}=&\left\{\theta_{0}^{T}\frac{\partial^2 P^{-1}}{\partial\eta_{k}\partial\eta_{l}}\theta_{0}
		+{\Tr}\left(\frac{\partial P^{-1}}{\partial \eta_{l}}\frac{\partial P}{\partial \eta_{k}}\right)\right.\nonumber\\
		&+\left.\left.\Tr\left(P^{-1}\frac{\partial^2 P}{\partial\eta_{k}\partial\eta_{l}} \right)\right\}\right|_{\eta=\eta^{*}_{\tb}},
		\end{align}
		the $k$th row of $B_{\tb}(\eta^{*}_{\tb})\in\R^{p\times n}$ can be represented as
		\begin{align}
		\label{eq:element of B eta_b_star}
		\left[B_{\tb}(\eta^{*}_{\tb})\right]_{k,:}=&\theta_{0}^{T}\left.\frac{\partial P^{-1}}{\partial \eta_{k}}\right|_{\eta=\eta_{\tb}^{*}}.
		\end{align}	
		
\end{theorem}

\begin{remark}\label{rmk: EB hyper-parameter convergence in dist. for deterministic inputs}
	If we consider deterministic inputs and assume that $\lim_{N\to\infty}\Phi^{T}\Phi/N=\Sigma$, Theorem \ref{thm:asymptotic normality of eta_eb difference} still holds.
\end{remark}


Clearly, Theorem \ref{thm:asymptotic normality of eta_eb difference} shows that the limiting covariance matrix $V_{\tb}^{\tH}(\eta_{\tb}^{*})$ contains the factors that affect the convergence properties of $\hat{\eta}_{\EB}$ to $\eta_{\tb}^{*}$, including the limit of $\Phi^T\Phi/N$, i.e., $\Sigma$, the kernel matrix $P$, and the true value of $\theta$, i.e., $\theta_0$. Moreover, the following proposition  shows that as the condition number of $\Sigma$ increases, $\Tr[V_{\tb}^{\tH}(\eta_{\tb}^{*})]$ becomes or tends to become larger, indicating that the more slowly $\hat{\eta}_{\EB}$ converges to $\eta_{\tb}^{*}$.

\begin{proposition}\label{prop:bounds of EB hyper-parameter estimator}
	
Define the {eigenvalue decomposition (EVD)} of $\Sigma$ as follows,
\begin{align}\label{eq:evd of Sigma}
\Sigma=&\sum_{i=1}^{n}\lambda_{i}(\Sigma)e_{\Sigma,i}e_{\Sigma,i}^{T},
\end{align} 
where $\lambda_{1}(\Sigma)\geq \cdots \geq \lambda_{n}(\Sigma)>0$ denote eigenvalues of $\Sigma$ and $e_{\Sigma,i}\in\R^{n}$ denotes the eigenvector of $\Sigma$ associated with $\lambda_{i}(\Sigma)${; moreover,} the condition number of $\Sigma$ is defined as $\cond(\Sigma)=\lambda_{1}(\Sigma)/\lambda_{n}(\Sigma)$. If $\sigma^2$, $\theta_{0}$, $P$, $e_{\Sigma,1},\cdots,e_{\Sigma,n}$ and $\lambda_{1}(\Sigma),\cdots,\lambda_{n-1}(\Sigma)$ are fixed, as $\lambda_{n}(\Sigma)$ decreases ($\cond(\Sigma)$ increases), $\Tr[V_{\tb}^{\tH}(\eta_{\tb}^{*})]$
will increase. More generally, if
\begin{align}\label{eq:condition of bounds of Vb}
e_{\Sigma,n}^{T}B_{\tb}(\eta_{\tb}^{*})^{T}A_{\tb}(\eta_{\tb}^{*})^{-1}\not=0,
\end{align}
there exist $B_{L}^{\tb},B_{U}^{\tb}>0$, irrespective of $\cond(\Sigma)$, such that
\begin{align}\label{eq:bounds of trace of Vb}
\frac{B_{L}^{\tb}}{\lambda_{1}(\Sigma)}\cond(\Sigma)\leq \Tr[V_{\tb}^{\tH}(\eta_{\tb}^{*})] \leq \frac{B_{U}^{\tb}}{\lambda_{1}(\Sigma)}\cond(\Sigma).
\end{align}	
\end{proposition} 

\section{High Order Asymptotic Distributions of RLS Estimator with $\EB$ Hyper-Parameter Estimator}\label{sec:asmptotic analysis for RLS estimator}

By the almost sure convergence of $\hat{\eta}_{\EB}$ as shown in \eqref{eq:almost sure convergence of hat_eta_eb}, we can derive the convergence in distribution of $\sqrt{N}(\hat{\theta}^{\TR}(\hat{\eta}_{\text{EB}})-\theta_0)$, where $\hat{\theta}^{\TR}(\hat{\eta}_{\text{EB}})$ is defined in \eqref{eq:RLS estimator2}. 

\begin{proposition}\label{prop:convergence in distribution of rls estimate using EB}
	 Under Assumptions \ref{asp:input}-\ref{asp:isolated optimal sets}, we have
		\begin{align}\label{eq:convergence in distribution of rls estimate using EB}
		\sqrt{N}(\hat{\theta}^{\TR}(\hat{\eta}_{\text{EB}})-\theta_{0})\overset{d.}\to& \mathcal{N}(0,V^{\ALS}_{1}),	
	   \end{align}	
    {where $V^{\ALS}_{1}$ and $\Sigma$ are defined in \eqref{eq:def of V_ALS_1} and \eqref{eq:Sigma}, respectively.}
\end{proposition} 
 
Proposition \ref{prop:convergence in distribution of rls estimate using EB} shows that  $\sqrt{N}(\hat{\theta}^{\TR}(\hat{\eta}_{\text{EB}})-\theta_0)$ and $\sqrt{N}(\hat{\theta}^{\LS}-\theta_0)$ converge in distribution to the same limiting distribution $\mathcal{N}(0,\sigma^2\Sigma^{-1})$. Clearly, this result is not so interesting and we need a better tool to disclose the difference  between 
$\sqrt{N}(\hat{\theta}^{\TR}(\hat{\eta}_{\text{EB}})-\theta_0)$ and $\sqrt{N}(\hat{\theta}^{\LS}-\theta_0)$ in their convergence properties. To this goal, we study below their high order asymptotic distributions instead of their first order asymptotic distributions, i.e., the convergence in distributions \eqref{eq:convergence in distribution of rls estimate using EB} and \eqref{eq:convergence in dist of ls estimate}. Before proceeding to the details, it is worth to note from{,} e.g., \cite{G94} that, for a sequence of random variables, its high order {expansions and asymptotic distributions} may not be unique. To ensure the uniqueness of high order {expansions and asymptotic distributions} of $\sqrt{N}(\hat{\theta}^{\TR}(\hat{\eta}_{\text{EB}})-\theta_0)$ and $\sqrt{N}(\hat{\theta}^{\LS}-\theta_0)$, we first stress that the information required to differentiate them is {given by} the following three building blocks 
\begin{align}\label{eq:building blocks}
	\sqrt{N}[N(\Phi^{T}\Phi)^{-1}-\Sigma^{-1}], \sqrt{N}\Phi^{T}V/N, \sqrt{N}(\widehat{\sigma^2}-\sigma^2),
\end{align} and their convergences in distribution as shown in Lemma \ref{lem:some preliminary results about inputs and ls estimates}{. Then} we require that in the $m$th order asymptotic {expansions} of $\sqrt{N}(\hat{\theta}^{\TR}(\hat{\eta}_{\text{EB}})-\theta_0)$ and $\sqrt{N}(\hat{\theta}^{\LS}-\theta_0)$, all low order terms up to the $(m-1)$th order have no more than first order {expansion and asymptotic distribution} with respect to \eqref{eq:building blocks} {(see also Remarks \ref{rmk:why we need one rule to ensure uniqueness} and \ref{rmk:uniqueness of high order asymptotics} for more details)}.


\subsection{{The second order asymptotic distribution of $\sqrt{N}(\hat{\theta}^{\LS}-\theta_0)$}}\label{subsec:2nd order asy.dist.}

{We first study the second order asymptotic distribution of $\sqrt{N}(\hat{\theta}^{\LS}-\theta_0)$ and the result is summarized below.}


{
\begin{theorem}\label{thm:joint convergence distribution of decomp. of rls estimate up to OpsqrtN}
Consider $\hat{\theta}^{\LS}$ defined in \eqref{eq:LS estimator}. Suppose Assumptions \ref{asp:input}-\ref{asp:full column rank of regression matrix} hold. Then the second order {expansion} of $\hat{\theta}^{\LS}$  takes the form of
\begin{align}\label{eq:split form of difference of the theta_ls and theta0}
	\sqrt{N}\left( \hat{\theta}^{\LS}-\theta_{0}\right)
	=&\hat{\theta}^{\ALS}_{1}+\frac{1}{\sqrt{N}}\hat{\theta}^{\ALS}_{2},
\end{align}
where
\begin{align}\label{eq:def of theta_ALS_1}
	\hat{\theta}^{\ALS}_{1}=&\Sigma^{-1}\sqrt{N}\frac{\Phi^{T}V}{N},\\
	\label{eq:def of theta_ALS_2}
	\hat{\theta}^{\ALS}_{2}=&\sqrt{N}\left[N(\Phi^{T}\Phi)^{-1}-\Sigma^{-1} \right]\sqrt{N}\frac{\Phi^{T}V}{N}.
\end{align} 
{Moreover, we have}
\begin{align}
	\label{eq:2nd order asymptotic distribution of the theta_ls}
	\sqrt{N}\left( \hat{\theta}^{\LS}-\theta_{0}\right)
	\overset{2\text{nd}\ d.}\longrightarrow {\vartheta}^{\ALS}_{1}+\frac{1}{\sqrt{N}}{\vartheta}^{\ALS}_{2},
\end{align}
where
\begin{align}
	\label{eq:def of vartheta_ALS_1}
	{\vartheta}^{\ALS}_{1}=&\Sigma^{-1}\upsilon,\\
	\label{eq:def of vartheta_ALS_2}
	{\vartheta}^{\ALS}_{2}=&-\Sigma^{-1}\Gamma\Sigma^{-1}\upsilon,\\
	\label{eq:expectation of vartheta_als}
	\E\left(\left[ \begin{array}{c}{\vartheta}^{\ALS}_{1}\\ {\vartheta}^{\ALS}_{2} \end{array}\right] \right)=&\left[\begin{array}{c}0\\ 0 \end{array}\right],\\
	\label{eq:covariance matrix of vartheta_als}
	\COV\left( \left[ \begin{array}{c}{\vartheta}^{\ALS}_{1}\\ {\vartheta}^{\ALS}_{2} \end{array}\right] \right)
	=&\left[\begin{array}{cc}V^{\ALS}_{1} & 0 \\ 0 & V^{\ALS}_{2} \end{array}\right],
\end{align} 
with $\upsilon$, $\Gamma$ defined in \eqref{eq:covariances of Gamma, upsilon and rho}, $V^{\ALS}_{1}$ defined in \eqref{eq:def of V_ALS_1}, and 
\begin{align}
	\label{eq:def of V_ALS_2}
	&V^{\ALS}_{2}=\sigma^2{\tvec}^{-1}\left[(\Sigma^{-1}\otimes \Sigma^{-1})C_{\Gamma}\tvec(\Sigma^{-1}) \right]\succeq 0.
\end{align} 
Here, for a matrix $A\in\R^{n\times n}$, $A\succeq 0$ denotes that $A$ is positive semidefinite, $\tvec(A)$ denotes the vectorization of $A$, which stacks columns of $A$ as an $n^2$-dimensional column vector, ${\tvec}^{-1}(\cdot)$ denotes the inverse operation of the vectorization {into a square matrix}, and $C_{\Gamma}$ is defined in \eqref{eq:def of C_Gamma}.
\end{theorem}}

\begin{remark}\label{rmk:mean and covariance of theta_LS}
For convenience, we define the mean and covariance of the second order asymptotic distribution of $\sqrt{N}(\hat{\theta}^{\LS}-\theta_{0})$ as follows,
\begin{align}
&\E\left(\vartheta^{\ALS}_{1}+\frac{1}{\sqrt{N}}\vartheta^{\ALS}_{2} \right)=0,\\
\label{eq:def of V_ALS}
&\COV\left({\vartheta}^{\ALS}_{1}+\frac{1}{\sqrt{N}}\vartheta^{\ALS}_{2}\right)= V_{1}^{\ALS}+\frac{1}{N}V_{2}^{\ALS}\triangleq V^{\ALS}\succeq 0.
\end{align}

\end{remark}

{Since both $\hat{\theta}^{\ALS}_{1}$ and $\hat{\theta}^{\ALS}_{2}$ have no more than first order {expansions} and distributions with respect to \eqref{eq:building blocks}, we know that $\sqrt{N}(\hat{\theta}^{\LS}-\theta_0)$ has no {expansions} and distributions with order higher than 2 with respect to \eqref{eq:building blocks}.}

\subsection{The {second and third order asymptotic distributions} of $\sqrt{N}(\hat{\theta}^{\TR}(\hat{\eta}_{\text{EB}})-\theta_0)$}\label{subsec:high order asy. dist. of rls using EB}

{

Then we study the second and third order asymptotic distributions of $\sqrt{N}(\hat{\theta}^{\TR}(\hat{\eta}_{\text{EB}})-\theta_0)$.

	The second order asymptotic distribution of $\sqrt{N}(\hat{\theta}^{\TR}(\hat{\eta}_{\EB})-\theta_{0})$ is defined as
	\begin{align}\label{eq:2nd order asymptotic distribution of the theta_rls_EB}
		\sqrt{N}\left( \hat{\theta}^{\TR}(\hat{\eta}_{\EB})-\theta_{0}\right)
		\overset{2\text{nd}\ d.}\longrightarrow {\vartheta}^{\ALS}_{1}+\frac{1}{\sqrt{N}}\left({\vartheta}^{\ALS}_{2}+\vartheta^{\AR}_{\text{b2}}\right),
	\end{align}
	where ${\vartheta}^{\ALS}_{1}$ and ${\vartheta}^{\ALS}_{2}$ are defined in \eqref{eq:def of vartheta_ALS_1} and \eqref{eq:def of vartheta_ALS_2}, respectively, and
	\begin{align}
	\label{eq:def of vartheta_b2_AR}
	{\vartheta}^{\AR}_{\text{b2}}=&
	-\sigma^2\Sigma^{-1}P(\eta_{\tb}^{*})^{-1}\theta_{0}.
	\end{align}
	Compared with the second order asymptotic distribution of $\sqrt{N}(\hat{\theta}^{\LS}-\theta_{0})$ in Theorem \ref{thm:joint convergence distribution of decomp. of rls estimate up to OpsqrtN}, that of $\sqrt{N}(\hat{\theta}^{\TR}(\hat{\eta}_{\EB})-\theta_{0})$ has a different mean ${\vartheta^{\AR}_{\text{b2}}}/{\sqrt{N}}$, which is dependent on $P(\eta_{\tb}^{*})$, but the same covariance matrix $V^{\ALS}$ in \eqref{eq:def of V_ALS}, which is independent of $P(\eta_{\tb}^{*})$. It means that the second order asymptotic distribution of $\sqrt{N}(\hat{\theta}^{\TR}(\hat{\eta}_{\EB})-\theta_{0})$ does not take into account the influence of the regularization on the covariance matrix, which contradicts the observation that the regularization can mitigate the possibly large variance of the LS estimator $\hat{\theta}^{\LS}$. Since the second order asymptotic distribution is not enough to expose the influence of the regularization, the third order asymptotic distribution of $\sqrt{N}(\hat{\theta}^{\TR}(\hat{\eta}_{\EB})-\theta_{0})$  is considered.

\begin{remark}\label{rmk:why we need one rule to ensure uniqueness}
	If we have no additional rule that in the $m$th order expansions, ``all low order terms up to the $(m-1)$th order have no more than first order {expansion and asymptotic distribution} with respect to \eqref{eq:building blocks}'' as mentioned before, then for a fixed order, we may have different expansions and asymptotic distributions. For example, apart from \eqref{eq:2nd order asymptotic distribution of the theta_rls_EB}, the second order expansion and asymptotic distribution of $\sqrt{N}(\hat{\theta}^{\TR}(\hat{\eta}_{\text{EB}})-\theta_0)$ could also be 
	\begin{align}
		\label{eq:alternative second order expansion}
		&\sqrt{N}\left( \hat{\theta}^{\TR}(\hat{\eta}_{\EB})-\theta_{0}\right)=N(\Phi^{T}\Phi)^{-1}\sqrt{N}\frac{\Phi^{T}V}{N}+\frac{1}{\sqrt{N}}\hat{\theta}^{\AR}_{\text{b2}},\\
		\label{eq:alternative 2nd asy. dist.}
		&\sqrt{N}\left( \hat{\theta}^{\TR}(\hat{\eta}_{\EB})-\theta_{0}\right)
		\overset{2\text{nd}\ d.}\longrightarrow {\vartheta}^{\ALS}_{1}+\frac{1}{\sqrt{N}}\vartheta^{\AR}_{\text{b2}}.
	\end{align}
	However, the first order term in expansion \eqref{eq:alternative second order expansion} contains a second order expansion, i.e., $N(\Phi^{T}\Phi)^{-1}\sqrt{N}{\Phi^{T}V}/{N}$ can be decomposed as $\hat{\theta}^{\ALS}_{1}+({1}/{\sqrt{N}})\hat{\theta}^{\ALS}_{2}$. If we apply the additional rule to \eqref{eq:alternative second order expansion}, we will still obtain the second order asymptotic distribution \eqref{eq:2nd order asymptotic distribution of the theta_rls_EB}.

\end{remark}
}


\begin{theorem}\label{thm:joint convergence distribution of decomp. of rls estimate up to OpN}
Consider $\hat{\theta}^{\TR}(\hat{\eta}_{\text{EB}})$ defined in \eqref{eq:RLS estimator2}. Suppose Assumptions \ref{asp:input}-\ref{asp:nozero 2nd order derivative of P with respect to eta} hold. Then the third order {expansion} of $\sqrt{N}(\hat{\theta}^{\TR}(\hat{\eta}_{\text{EB}})-\theta_0)$ takes the form of
\begin{align}
	\label{eq:decomposition of sqrt_diff_theta_rls_EB}
	\sqrt{N}\left(\hat{\theta}^{\TR}(\hat{\eta}_{\EB})-\theta_{0}\right)=&\hat{\theta}^{\ALS}_{1}+\frac{1}{\sqrt{N}}(\hat{\theta}^{\ALS}_{2}+\vartheta^{\AR}_{\text{b2}})
	+\frac{1}{N}\hat{\theta}^{\AR}_{\text{b3}},
\end{align}
where $\hat{\theta}^{\ALS}_{1}$, $\hat{\theta}^{\ALS}_{2}$ and $\vartheta^{\AR}_{\text{b2}}$ are defined in \eqref{eq:def of theta_ALS_1}, \eqref{eq:def of theta_ALS_2} and \eqref{eq:def of vartheta_b2_AR}, respectively, and
\begin{align}
\label{eq:def of theta_hat_b3}
	\hat{\theta}^{\AR}_{\text{b3}}=&
	-\sqrt{N}\big[\widehat{\sigma^2}N(\Phi^{T}\Phi)^{-1}\hat{S}(\hat{\eta}_{\EB})^{-1}\hat{\theta}^{\LS}\nonumber\\
	&-\sigma^2\Sigma^{-1}P(\eta_{b}^{*})^{-1}\theta_{0}\big].
\end{align} 
Note that $\hat{S}(\hat{\eta}_{\EB})$ is defined in \eqref{eq:def of hat_S_inv}.
{Moreover, we have}
\begin{align}\label{eq:3rd order asymptotic distribution of the theta_rls_EB}
	&\sqrt{N}\left(\hat{\theta}^{\TR}(\hat{\eta}_{\EB})-\theta_{0}\right) \overset{3\text{rd}\ d.}\longrightarrow\nonumber\\
	&{\vartheta}^{\ALS}_{1}+\frac{1}{\sqrt{N}}({\vartheta}^{\ALS}_{2}+\vartheta^{\AR}_{\text{b2}})+\frac{1}{N}{\vartheta}^{\AR}_{\text{b3}},
\end{align} 
where $\vartheta^{\ALS}_{1}$ and $\vartheta^{\ALS}_{2}$ are defined in \eqref{eq:def of vartheta_ALS_1} and \eqref{eq:def of vartheta_ALS_2}, respectively, and

	\begin{align}\label{eq:def of vartheta_b3}
		{\vartheta}^{\AR}_{\text{b3}}
		=&-\rho\Sigma^{-1}P(\eta_{\tb}^{*})^{-1}\theta_{0}
		+\sigma^2\Sigma^{-1}\Gamma\Sigma^{-1}P(\eta_{\tb}^{*})^{-1}\theta_{0}\nonumber\\
		&-\sigma^2\Sigma^{-1}C_{\tb}(\eta_{\tb}^{*})\Sigma^{-1}\upsilon,\\
		\label{eq:def of Cb}
		C_{\tb}(\eta_{\tb}^{*})
	    =&-2B_{\tb}(\eta_{\tb}^{*})^{T}A_{\tb}(\eta_{\tb}^{*})^{-1}B_{\tb}(\eta_{\tb}^{*})+P(\eta_{\tb}^{*})^{-1},\\
	\label{eq:expectation of ad. in terms of Op1_N}
	&\E\left(\left[\begin{array}{c}{\vartheta}^{\ALS}_{1}\\ {\vartheta}^{\ALS}_{2}+\vartheta^{\AR}_{\text{b2}}\\ {\vartheta}^{\AR}_{\text{b3}} \end{array}\right] \right)
	=\left[\begin{array}{c}0\\ {\vartheta}^{\AR}_{\text{b2}} \\ 0 \end{array}\right],\\
	\label{eq:covariance matrix of ad. in terms of Op1_N}
	&\COV\left(\left[\begin{array}{c}{\vartheta}^{\ALS}_{1}\\ {\vartheta}^{\ALS}_{2}+\vartheta^{\AR}_{\text{b2}}\\ {\vartheta}^{\AR}_{\text{b3}} \end{array}\right] \right)\nonumber\\
	=&\left[\begin{array}{ccc}V^{\ALS}_{1} & 0 & V^{\AR}_{\text{b3},2}(\eta_{\tb}^{*})\\ 0 & V_{2}^{\ALS} & 0\\ V^{\AR}_{\text{b3},2}(\eta_{\tb}^{*})^{T} & 0 & V^{\AR}_{\text{b3},1}(\eta_{\tb}^{*})  \end{array}\right],
\end{align}
with $\upsilon$, $\Gamma$ and $\rho$ defined in \eqref{eq:joint convergence in distribution of ls estimate and other terms}, {$V_{1}^{\ALS}$ and $V_{2}^{\ALS}$ defined in \eqref{eq:def of V_ALS_1} and \eqref{eq:def of V_ALS_2}}, respectively, and
\begin{align}
	V^{\AR}_{\text{b3},1}(\eta_{\tb}^{*})
	=&V^{\AR}_{\text{b3},1,1}(\eta_{\tb}^{*})+V^{\AR}_{\text{b3},1,2}(\eta_{\tb}^{*})+V^{\AR}_{\text{b3},1,3}(\eta_{\tb}^{*})\succeq 0,\nonumber\\
	\label{eq:def of Vb3_AR_1_1}
	V^{\AR}_{\text{b3},1,1}(\eta_{\tb}^{*})
	=&\sigma^6\Sigma^{-1}C_{\tb}(\eta_{\tb}^{*})\Sigma^{-1}C_{\tb}(\eta_{\tb}^{*})\Sigma^{-1}\succeq 0,\\
	\label{eq:def of Vb3_AR_1_2}
	V^{\AR}_{\text{b3},1,2}(\eta_{\tb}^{*})
	=&\sigma^4{\tvec}^{-1}\left\{(\Sigma^{-1}\otimes\Sigma^{-1})C_{\Gamma}(\Sigma^{-1}\otimes\Sigma^{-1})\right.\nonumber\\
	&\left.\tvec\left[P(\eta_{\tb}^{*})^{-1}\theta_{0}\theta_{0}^{T}P(\eta_{\tb}^{*})^{-1} \right] \right\}\succeq 0,\\
	\label{eq:def of Vb3_AR_1_3}
	V^{\AR}_{\text{b3},1,3}(\eta_{\tb}^{*})
	=&\left\{ \E[v(t)]^4-\sigma^4\right\}\nonumber\\
	&\Sigma^{-1}P(\eta_{\tb}^{*})^{-1}\theta_{0}\theta_{0}^{T}P(\eta_{\tb}^{*})^{-1}\Sigma^{-1}\succeq 0,\\
	\label{eq:def of Vb3_AR_2}
	V_{\text{b3},2}^{\AR}(\eta_{\tb}^{*})
	=&-\sigma^4\Sigma^{-1}C_{\tb}(\eta_{\tb}^{*})\Sigma^{-1}\preceq 0.
\end{align}
\end{theorem}

\begin{remark}\label{rmk:mean and covariance of 3rd asy.dist.}

For convenience, we define the mean and covariance of the third order asymptotic distribution of $\sqrt{N}(\hat{\theta}^{\TR}(\hat{\eta}_{\text{EB}})-\theta_0)$ as follows,
\begin{align}\label{eq:expectation of ad. in terms of OpsqrtN}
	&\E\left[{\vartheta}^{\ALS}_{1}+\frac{{\vartheta}^{\ALS}_{2}+\vartheta^{\AR}_{\text{b2}}}{\sqrt{N}}+\frac{{\vartheta}^{\AR}_{\text{b3}} }{N} \right]=E_{\tb}^{\AR}(\eta_{\tb}^{*})\triangleq \frac{\vartheta^{\AR}_{\text{b2}}}{\sqrt{N}},\\
	&\COV\left[{\vartheta}^{\ALS}_{1}+\frac{1}{\sqrt{N}}({\vartheta}^{\ALS}_{2}+\vartheta^{\AR}_{\text{b2}})+\frac{1}{N}{\vartheta}^{\AR}_{\text{b3}} \right]
	=V^{\AR}_{\tb}(\eta_{\tb}^{*})\nonumber\\
		\label{eq:covariance matrix of ad. in terms of OpN}
	\triangleq&V^{\ALS}+\frac{1}{N^2}V_{\text{b3},1}^{\AR}(\eta_{\tb}^{*})
	+\frac{1}{N}\left[ V_{\text{b3},2}^{\AR}(\eta_{\tb}^{*})+ \left(V_{\text{b3},2}^{\AR}(\eta_{\tb}^{*}) \right)^{T}\right]\succeq 0.
\end{align}
\end{remark}

Theorem \ref{thm:joint convergence distribution of decomp. of rls estimate up to OpN} together with Remark \ref{rmk:mean and covariance of 3rd asy.dist.} indicates that the mean and covariance matrix of the third order asymptotic distribution of $\sqrt{N}(\hat{\theta}^{\TR}(\hat{\eta}_{\EB})-\theta_{0})$ both show the influence of the regularization, which is due to that in \eqref{eq:decomposition of sqrt_diff_theta_rls_EB}, $\hat{\theta}^{\AR}_{\text{b3}}$ and $\vartheta^{\AR}_{\text{b2}}$ are dependent on the $P(\hat{\eta}_{\EB})$ and its limit $P(\eta_{\tb}^{*})$. Together with Proposition \ref{prop:convergence in distribution of rls estimate using EB} and \eqref{eq:2nd order asymptotic distribution of the theta_rls_EB}, we can also say that the third order asymptotic distribution of $\sqrt{N}(\hat{\theta}^{\TR}(\hat{\eta}_{\EB})-\theta_{0})$  is the \emph{lowest} order one that exposes the influence of the regularization on both mean and covariance matrix.

\begin{remark}\label{rmk:uniqueness of high order asymptotics}
	It is easy to check that $\hat{\theta}^{\ALS}_{1}$, $\hat{\theta}^{\ALS}_{2}$ and $\vartheta^{\AR}_{\text{b2}}$ all have no more than first order {expansions} and distributions with respect to \eqref{eq:building blocks}. In contrast, $\hat{\theta}^{\AR}_{\text{b3}}$ still has high order {expansions} and distributions with respect to  \eqref{eq:building blocks}, indicating that $\sqrt{N}(\hat{\theta}^{\TR}(\hat{\eta}_{\text{EB}})-\theta_0)$ has {expansions} and distributions with order higher than 3.	
\end{remark} 

\begin{remark}\label{rmk:explicit distributions of asymptotic distributions}
It is worth to note that the high order asymptotic distributions of $\sqrt{N}(\hat{\theta}^{\LS}-\theta_{0})$ and $\sqrt{N}(\hat{\theta}^{\TR}(\hat{\eta}_{\EB})-\theta_{0})$ are \emph{not} Gaussian:
	\begin{itemize}
		\item $\vartheta_{1}^{\ALS}$ and $\vartheta_{\text{b3}}^{\AR}$ are both Gaussian distributed;
		\item $\vartheta_{\text{b2}}^{\AR}$ is a constant;
		\item the distribution of $\vartheta_{2}^{\ALS}$ is more complicated and in fact $\vartheta_{2}^{\ALS}$ is a linear combination of $\chi^2$ distributions. It is hard to derive the exact distribution of $\vartheta_{2}^{\ALS}$, but since it is easy to calculate the moments of $\vartheta_{2}^{\ALS}$, if necessary, it is possible to construct an approximation of the distribution of $\vartheta_{2}^{\ALS}$ based on its moments {(see, e.g., \cite{MH09})}. 
	\end{itemize}
\end{remark}

Finally, it is possible to gain more insights on the relation between $\hat{\theta}^{\TR}(\hat{\eta}_{\EB})$ and $\hat{\theta}^{\LS}$ as shown in the following result.
\begin{corollary}\label{corollary:convergence in distribution of difference of rls and ls estimates}
     Under Assumptions \ref{asp:input}-\ref{asp:nozero 2nd order derivative of P with respect to eta}, we have
     \begin{align*}
     N\left[\sqrt{N}(\hat{\theta}^{\TR}(\hat{\eta}_{\EB})-\hat{\theta}^{\LS})-E_{\tb}^{\AR}(\eta_{\tb}^{*})\right]
     \overset{d.}\to \mathcal{N}\left(0,V_{\text{b3},1}^{\AR}(\eta_{\tb}^{*})\right),
     \end{align*}
  or equivalently,
     \begin{align*}
     N\left[\hat{\theta}^{\TR}(\hat{\eta}_{\EB})-\hat{\theta}^{\LS}\right]\overset{2\text{nd}\ d.}\longrightarrow& \vartheta^{\AR}_{\text{b2}}+\frac{1}{\sqrt{N}}\vartheta^{\AR}_{\text{b3}}.
     \end{align*}
\end{corollary}

\subsection{Discussions}

We make some discussions below on the accuracy of and the influence of ${\cond}(\Sigma)$ on the asymptotic distributions.

\subsubsection{Accuracy of the Asymptotic Distributions}\label{subsubsec:accuracy of asy.dist.}

In contrast with the first order asymptotic distribution \eqref{eq:convergence in distribution of rls estimate using EB},
the high order asymptotic distributions \eqref{eq:2nd order asymptotic distribution of the theta_rls_EB} and \eqref{eq:3rd order asymptotic distribution of the theta_rls_EB} provide more information, and in particular, show more factors that affect  
the convergence properties of $\sqrt{N}\left(\hat{\theta}^{\TR}(\hat{\eta}_{\text{EB}})-\theta_{0}\right)$, e.g., $\Sigma$, $P$, $\theta_0$ and $C_{\Gamma}$. Then one may expect that the high order asymptotic distributions \eqref{eq:2nd order asymptotic distribution of the theta_rls_EB} and \eqref{eq:3rd order asymptotic distribution of the theta_rls_EB} can also provide more accurate approximation of $\sqrt{N}(\hat{\theta}^{\TR}(\hat{\eta}_{\EB})-\theta_{0})$, which however is a quite complicated problem. 

First, as well known from the theory of high order asymptotics {(see, e.g., \cite{G94})}, higher order asymptotic distributions do not necessarily lead to more accurate approximations. For the case studied here, the following specific discussions follow:
\begin{itemize}
\item  for the first order asymptotic distribution \eqref{eq:convergence in distribution of rls estimate using EB}, the approximation error 
\begin{align*}
\sqrt{N}\left( \hat{\theta}^{\TR}(\hat{\eta}_{\EB})-\theta_{0}\right)
- {\vartheta}^{\ALS}_{1}
\end{align*}
	  depends on the convergence property of $\hat{\theta}^{\ALS}_{1}$ to $\vartheta^{\ALS}_{1}$, which essentially depends on the convergence property of $\sqrt{N}(\Phi^{T}V/N)$ to  $\upsilon$, {with $\upsilon$ defined in \eqref{eq:covariances of Gamma, upsilon and rho}};

	\item for the second order asymptotic distribution \eqref{eq:2nd order asymptotic distribution of the theta_rls_EB},
	the approximation error 
\begin{align*}
\sqrt{N}\left( \hat{\theta}^{\TR}(\hat{\eta}_{\EB})-\theta_{0}\right)
- {\vartheta}^{\ALS}_{1}-\frac{1}{\sqrt{N}}({\vartheta}^{\ALS}_{2}+\vartheta^{\AR}_{\text{b2}})
\end{align*}

	  depends on the convergence properties of $\hat{\theta}^{\ALS}_{1}$, $\hat{\theta}^{\ALS}_{2}$ and $-\widehat{\sigma^2}N(\Phi^{T}\Phi)^{-1}\hat{S}(\hat{\eta}_{\EB})^{-1}\hat{\theta}^{\LS}$ to $\vartheta^{\ALS}_{1}$, $\vartheta^{\ALS}_{2}$ and $\vartheta^{\AR}_{\text{b2}}$, respectively, which essentially depend on the convergence properties of $\sqrt{N}[N(\Phi^{T}\Phi)^{-1}-\Sigma^{-1}]$ and $\sqrt{N}(\Phi^{T}V/N)$ to $-\Sigma^{-1}\Gamma\Sigma^{-1}$ and $\upsilon$, respectively, {with $\Gamma$ defined in \eqref{eq:covariances of Gamma, upsilon and rho}};
	\item  for the third order asymptotic distribution \eqref{eq:3rd order asymptotic distribution of the theta_rls_EB},
	the approximation error 
\begin{align*}
\sqrt{N}\left(\hat{\theta}^{\TR}(\hat{\eta}_{\EB})-\theta_{0}\right)-
	{\vartheta}^{\ALS}_{1}-\frac{({\vartheta}^{\ALS}_{2}+\vartheta^{\AR}_{\text{b2}})}{\sqrt{N}}-\frac{1}{N}{\vartheta}^{\AR}_{\text{b3}}
\end{align*}

	 depends on the convergence properties of $\hat{\theta}^{\ALS}_{1}$, $\hat{\theta}^{\ALS}_{2}$ and $\hat{\theta}^{\AR}_{\text{b3}}$ to $\vartheta^{\ALS}_{1}$, $\vartheta^{\ALS}_{2}$ and $\vartheta^{\AR}_{\text{b3}}$, respectively, which essentially depend on the convergence properties of $\sqrt{N}[N(\Phi^{T}\Phi)^{-1}-\Sigma^{-1}]$, $\sqrt{N}(\Phi^{T}V/N)$ and $\sqrt{N}(\widehat{\sigma^2}-\sigma^2)$ to $-\Sigma^{-1}\Gamma\Sigma^{-1}$, $\upsilon$ and $\rho$, respectively, {with $\rho$ defined in \eqref{eq:covariances of Gamma, upsilon and rho}}.
\end{itemize}

To assess the accuracy of the approximations given by the high order asymptotic distributions, we define the $m$th order asymptotic approximation of $\MSE_{g}(\hat{\theta}^{\TR}(\hat{\eta}_{\EB}))$ based on the $m$th order asymptotic distribution of $\sqrt{N}(\hat{\theta}^{\TR}(\hat{\eta}_{\EB})-\theta_{0})$ and denote it by ${\text{AMSE}_{g}^{\tb,m}(\eta_{\tb}^{*})}$ with $m=1,2,3$:
\begin{subequations}\label{eq:asymptotic MSE with orders 1,2,3}
	\begin{align}
		{\text{AMSE}_{g}^{\tb,1}(\eta_{\tb}^{*})}=&\frac{1}{N}\Tr(V^{\ALS}_{1}),\\
		{\text{AMSE}_{g}^{\tb,2}(\eta_{\tb}^{*})}=&\frac{1}{N}\left[\Tr(V^{\ALS})+\|E^{\AR}_{\tb}(\eta_{\tb}^{*})\|_{2}^2\right],\\
	   \label{eq:asymptotic MSE with order 3}
		{\text{AMSE}_{g}^{\tb,3}(\eta_{\tb}^{*})}=&\frac{1}{N}\left[\Tr(V^{\AR}_{\tb}(\eta_{\tb}^{*}))+\|E^{\AR}_{\tb}(\eta_{\tb}^{*})\|_{2}^2 \right].
	\end{align}
\end{subequations}
It is easy to see that ${\text{AMSE}_{g}^{\tb,1}(\eta_{\tb}^{*})}\leq {\text{AMSE}_{g}^{\tb,2}(\eta_{\tb}^{*})}$ due to the positive semidefiniteness of $V^{\ALS}_{2}$ in \eqref{eq:def of V_ALS_2}. However, the relation between ${\text{AMSE}_{g}^{\tb,3}(\eta_{\tb}^{*})}$ and ${\text{AMSE}_{g}^{\tb,1}(\eta_{\tb}^{*})}$ or ${\text{AMSE}_{g}^{\tb,2}(\eta_{\tb}^{*})}$  is unclear.

{
	\begin{remark}\label{rmk:XMSE}
		One different asymptotic approximation of $\MSE_{g}(\hat{\theta}^{\TR}(\hat{\eta}_{\EB}))$ is proposed in \cite{Hnotes,H2020}, which is denoted as $\text{XMSE}(\eta_{\tb}^{*})$. Note that $\text{XMSE}(\eta_{\tb}^{*})$ is defined for deterministic inputs and in contrast with $\text{AMSE}_{g}^{\tb,3}(\eta_{\tb}^{*})$ in \eqref{eq:asymptotic MSE with order 3}, $\text{XMSE}(\eta_{\tb}^{*})$ does not take into consideration how fast the regression matrix converges to its limit and the third order information.
	\end{remark}
}

Obviously, if it {were} possible to get a closed form expression of $\MSE_{g}(\hat{\theta}^{\TR}(\hat{\eta}_{\EB}))$, then comparing $\MSE_{g}(\hat{\theta}^{\TR}(\hat{\eta}_{\EB}))$ with \eqref{eq:asymptotic MSE with orders 1,2,3} would tell which one of the three high order asymptotic distributions gives the best approximation. Unfortunately, it is impossible, and we are only able to calculate  $\MSE_{g}(\hat{\theta}^{\TR}(\hat{\eta}_{\EB}))$ and thus assess the accuracy of the approximations given by the high order asymptotic distributions \emph{numerically}, as will be illustrated in Section \ref{sec:sim}.

\subsubsection{Influence of ${\cond}(\Sigma)$ on the Asymptotic Distributions}
Similar to Proposition \ref{prop:bounds of EB hyper-parameter estimator}, it is also interesting to investigate the influence of ${\cond}(\Sigma)$ on the asymptotic mean and variances of $\sqrt{N}(\hat{\theta}^{\TR}(\hat{\eta}_{\text{EB}})-\theta_{0})$, i.e., $E_{\tb}^{\AR}(\eta_{\tb}^{*})$ in \eqref{eq:expectation of ad. in terms of OpsqrtN},  $V^{\ALS}$ in \eqref{eq:def of V_ALS} and $V^{\AR}_{\tb}(\eta_{\tb}^{*})$ in \eqref{eq:covariance matrix of ad. in terms of OpN}, which is however much harder. Actually,  we are only able to analyze the influence of $\cond(\Sigma)$ on $E_{\tb}^{\AR}(\eta_{\tb}^{*})$, $V_{1}^{\ALS}$, ${V}_{\text{b3},1,1}^{\AR}(\eta_{\tb}^{*})$ and ${V}_{\text{b3},1,3}^{\AR}(\eta_{\tb}^{*})$, except for some special cases as mentioned briefly in Remark \ref{rmk:bounds of trace of covariance matrix of model estimator}. 
In particular, the following proposition shows that 
$\|E_{\tb}^{\AR}(\eta_{\tb}^{*})\|_{2}^2$,
$\Tr(V^{\ALS}_{1})$, $\Tr[V^{\AR}_{\text{b3},1,1}(\eta_{\tb}^{*})]$ and $\Tr[V^{\AR}_{\text{b3},1,3}(\eta_{\tb}^{*})]$ all tend to become larger as $\cond(\Sigma)$ increases.

\begin{proposition}\label{prop:bounds of trace of covariance matrix of model estimator}
Following Proposition \ref{prop:bounds of EB hyper-parameter estimator}, suppose that $e_{\Sigma,n}^{T}P^{-1}(\eta_{\tb}^{*})\theta_{0}\neq 0$ and $e_{\Sigma,n}^{T}C_{\tb}(\eta_{\tb}^{*})e_{\Sigma,n}\not=0$,
	for fixed $N$, $\sigma^2$, $\theta_{0}$, $P$, $e_{\Sigma,n}$ and $\lambda_{1}(\Sigma)$, there exist positive and increasing functions of $\cond(\Sigma)$ such that $\|E_{\tb}^{\AR}(\eta_{\tb}^{*})\|_{2}^2$, $\Tr(V^{\ALS}_{1})$, $\Tr[V^{\AR}_{\text{b3},1,1}(\eta_{\tb}^{*})]$ and $\Tr[V^{\AR}_{\text{b3},1,3}(\eta_{\tb}^{*})]$ can be  lower bounded and upper bounded by those functions, respectively. For example, there exist increasing functions of $\cond(\Sigma)$, denoted as $ f_{L}(\cond(\Sigma)),f_{U}(\cond(\Sigma)):\R\to\R$, such that
	\begin{align}
	0<f_{L}(\cond(\Sigma))\leq& \|E_{\tb}^{\AR}(\eta_{\tb}^{*})\|_{2}^2 \leq f_{U}(\cond(\Sigma)).
	\end{align}
\end{proposition}

\begin{remark}\label{rmk:bounds of trace of covariance matrix of model estimator}
	As shown in Section \ref{sec:case study}, for ridge regression and some specific filters $H(q)$, e.g., \eqref{eq:tf function},  it is possible to represent both $\Sigma$ and $C_\Gamma$ {(and thus $\|E_{\tb}^{\AR}(\eta_{\tb}^{*})\|_{2}^2$, $\Tr(V^{\ALS})$ and $\Tr[V^{\AR}_{\tb}(\eta_{\tb}^{*})]$)} as functions of the parameters of $H(q)$ in closed-form, based on which we are able to calculate $\|E_{\tb}^{\AR}(\eta_{\tb}^{*})\|_{2}^2$, $\Tr(V^{\ALS})$ and $\Tr[V_{\tb}^{\AR}(\eta_{\tb}^{*})]$ numerically and assess their dependence on  $\cond(\Sigma)$ {through the parameters of $H(q)$}.
\end{remark}

\section{A Special Case: Ridge Regression with Filtered White Noise Input}\label{sec:case study}

To gain some concrete ideas on the convergence properties of  $\hat{\eta}_{\text{EB}}$ and the corresponding RLS estimator $\hat{\theta}^{\TR}(\hat{\eta}_{\text{EB}})$ as found in the previous two sections, we consider a special case below, i.e., the ridge regression with filtered white noise input, i.e.,  $P=\eta I_n$ and $u(t)=H(q)e(t)$ with $H(q)$ in the form of  
\begin{align}\label{eq:tf function}
H(q)=c_{u}\frac{1}{(1-aq^{-1})^2},
\end{align}
where $0\leq a <1$, $c_{u}\in\R$ is the coefficient of $H(q)$. The choice of $c_{u}$ for $0\leq a<1$ will be discussed in Section \ref{sec:sim}.

\begin{remark}\label{rmk:illustration of the choice of the 2nd order filter}
Although the simplest choice of $H(q)$ is
	\begin{align}
	H(q)=c_{u}\frac{1}{1-aq^{-1}},
	\end{align}
	 we did not use it, because its corresponding $\cond(\Sigma)$ does not increase faster enough as $a$ increases from 0 to 1, e.g., $\cond(\Sigma)=5.69\times 10^2$ when $a=0.95$. In contrast for the $H(q)$ in the form of \eqref{eq:tf function}, $\cond(\Sigma)=5.51\times 10^5$ when $a=0.95$.
\end{remark}

First, we show that both $\Sigma$ and $C_\Gamma$ {can be represented as a function of $a$ in closed-form, where $a$ is the parameter of $H(q)$}. 

\begin{lemma}\label{lemma:explicit expressions of Sigma and C_gamma}
	Consider $H(q)$ in the form of \eqref{eq:tf function}. Under Assumption \ref{asp:input}, $\Sigma$ defined in \eqref{eq:def of V_ALS_1} and $C_{\Gamma}$ defined in \eqref{eq:def of C_Gamma} have the following closed-form expressions in terms of $a$:
\begin{itemize}
	\item When $a=0$, it can be shown that 
	\begin{align}\label{eq:Sigma for white noise inputs}
    \Sigma&= c_{u}^2\sigma^2_{e}I_{n},\\
		\label{eq:def of C_H for ridge regression case with white noise input}
		[C_{\Gamma}]_{i,j}
		&=\left\{\begin{array}{ll}c_{u}^4\left[\E[e(t)]^4-\sigma_{e}^4\right],&\text{if}\ i=j, \\ c_{u}^4\sigma_{e}^4,&\text{if}\ k=l\not=0,\\ 0,&\text{otherwise} ,\end{array}\right.
		\end{align}
		where $i,j=1,\cdots,n^2$, and $k$ and $l$ satisfy \eqref{eq:def of k and l}.

\item When $a\neq0$, it can be shown that 
		\begin{align}\label{eq:ij_th element of Sigma for filtered white noise inputs}
		 [\Sigma]_{i,j}
		 =&c_{u}^2\sigma^{2}_{e}a^{|i-j|}\left[\frac{2}{(1-a^2)^3} + \frac{|i-j|-1}{(1-a^2)^2}\right],
		 \end{align}
	     where $i,j=1,\cdots,n$, and
	     \begin{align}
		 \label{eq:Cgamma for 2nd order filter}
		 \left[ C_{\Gamma} \right]_{i,j}
	     =&c_{u}^4\left\{\E[e(t)]^{4}/\sigma_{e}^4-3 \right\}
	     \frac{\sigma_{e}^4a^{k+l}}{(1-a^2)^6}\nonumber\\
	     &\left[k(1-a^2)+1+a^2\right]\left[l(1-a^2)+1+a^2\right]\nonumber\\
	     &+\frac{c_{u}^4\sigma_{e}^4}{(1-a^2)^6}\left[f_{\Gamma}(|k-l|)+f_{\Gamma}(k+l)\right],
	     \end{align}
	     where $i,j=1,\cdots,n^2$, and $k$ and $l$ satisfy \eqref{eq:def of k and l}, and for either $x=|k-l|$ or $x=k+l$,
	     \begin{align}\label{eq:f_Gamma}
	     f_{\Gamma}(x)=
	     &\frac{2a^{x+2}}{1-a^2}
	     \left[(1-x)a^4+(5-x)a^2 +4+2x \right]\nonumber\\
	     &-a^{x}(1-a^2)^2\frac{x(x+1)(2x+1)}{6}\nonumber\\
	     &+a^{x}(1-a^2)^2\frac{x^2(x+1)}{2}\nonumber\\
	     &+a^{x}(x+1)\left[(1-x)a^4+2a^2+1+x\right].
			\end{align}
\end{itemize}
\end{lemma}

Based on the closed-form expressions of $\Sigma$ and $C_\Gamma$ in terms of $a$, we are able to derive a closed-form expression of $V_{\tb}^{\tH}(\eta_{\tb}^{*})$ in \eqref{eq:def of Vb_etab_star} in terms of $a$ under the following assumption.
\begin{assumption}\label{asp:nonzero true theta}
	$\theta_{0}$ is nonzero, i.e., $\|\theta_{0}\|_{2}\neq0$.	
\end{assumption}

\begin{corollary}\label{corollary:ridge regression case}
	Under Assumptions \ref{asp:input}-\ref{asp:nonzero true theta}, when $P=\eta I_{n}$ with $\eta>0$, we have
		\begin{align}
		\label{eq:Vb for ridge regression case}
		V_{\tb}^{\tH}(\eta^{*}_{\tb})
		=&\frac{4\sigma^2}{n^2}\theta_{0}^{T}\Sigma^{-1}\theta_{0}.
		\end{align}
\end{corollary}


Moreover, we are also able to derive closed-form expressions of $E_{\tb}^{\AR}(\eta_{\tb}^{*})$ in \eqref{eq:expectation of ad. in terms of OpsqrtN} and $V_{\tb}^{\AR}(\eta_{\tb}^{*})$ in \eqref{eq:covariance matrix of ad. in terms of OpN} in terms of $a$. Noting \eqref{eq:def of V_ALS_1}, \eqref{eq:def of V_ALS_2} and \eqref{eq:Sigma for white noise inputs}-\eqref{eq:Cgamma for 2nd order filter}, $V^{\ALS}$ in \eqref{eq:def of V_ALS} or \eqref{eq:covariance matrix of ad. in terms of OpN} has a closed-form expression in terms of $a$. Therefore, we only consider the terms ${V}_{\text{b3},1}^{\AR}(\eta_{\tb}^{*})$ and ${V}_{\text{b3},2}^{\AR}(\eta_{\tb}^{*})$ in \eqref{eq:covariance matrix of ad. in terms of OpN} below.

\begin{corollary}\label{corollary:asymptotic mean and variance for rls estimate for ridge regression with filtered white noise inputs}
	Under Assumptions \ref{asp:input}-\ref{asp:nonzero true theta}, when $P=\eta I_{n}$ with $\eta>0$, 
	we have
	\begin{align}
	\label{eq:Eb_rls for the ridge regression case with filtered white noise inputs}
	&E_{\tb}^{\AR}(\eta_{\tb}^{*})=-\frac{1}{\sqrt{N}}\frac{n\sigma^2}{\theta_{0}^{T}\theta_{0}}\Sigma^{-1}\theta_{0},\\
	\label{eq:Vb_2_1_rls for the ridge regression case with filtered white noise inputs}
	&{V}_{\text{b3},1,1}^{\AR}(\eta_{\tb}^{*})=\frac{n^2\sigma^6}{(\theta_{0}^{T}\theta_{0})^2}
	\left[\frac{4}{(\theta_{0}^{T}\theta_{0})^2}\Sigma^{-1}\theta_{0}\theta_{0}^{T}\Sigma^{-1}\theta_{0}\theta_{0}^{T}\Sigma^{-1}\right.\nonumber\\
	&\left.+\Sigma^{-3}-\frac{2}{\theta_{0}^{T}\theta_{0}}\Sigma^{-2}\theta_{0}\theta_{0}^{T}\Sigma^{-1}
	-\frac{2}{\theta_{0}^{T}\theta_{0}}\Sigma^{-1}\theta_{0}\theta_{0}^{T}\Sigma^{-2} \right],\\
	\label{eq:Vb_2_2_rls for the ridge regression case with filtered white noise inputs}
	&{V}_{\text{b3},1,2}^{\AR}(\eta_{\tb}^{*})=\frac{n^2\sigma^4}{(\theta_{0}^{T}\theta_{0})^2}\nonumber\\
	&{\tvec}^{-1}\left[(\Sigma^{-1}\otimes \Sigma^{-1})C_{\Gamma}(\Sigma^{-1}\otimes \Sigma^{-1})\tvec(\theta_{0}\theta_{0}^{T}) \right],\\
	\label{eq:Vb_2_3_rls for the ridge regression case with filtered white noise inputs}
	&{V}_{\text{b3},1,3}^{\AR}(\eta_{\tb}^{*})=\frac{n^2\left\{\E\left[v(t)\right]^4-\sigma^4 \right\}}{(\theta_{0}^{T}\theta_{0})^2}\Sigma^{-1}\theta_{0}\theta_{0}^{T}\Sigma^{-1},\\
	\label{eq:Vb_3_rls for the ridge regression case with filtered white noise inputs}
	&V_{\text{b3},2}^{\AR}(\eta_{\tb}^{*})=\frac{2n\sigma^4}{(\theta_{0}^{T}\theta_{0})^2}\Sigma^{-1}\theta_{0}\theta_{0}^{T}\Sigma^{-1}-\frac{n\sigma^4}{\theta_{0}^{T}\theta_{0}}\Sigma^{-2}.
	\end{align}	
\end{corollary}

As shown in Lemma \ref{lemma:explicit expressions of Sigma and C_gamma} and Corollary \ref{corollary:asymptotic mean and variance for rls estimate for ridge regression with filtered white noise inputs}, $\Sigma$, $E_{\tb}^{\AR}(\eta_{\tb}^{*})$, $V^{\ALS}$ and $V_{\tb}^{\AR}(\eta_{\tb}^{*})$ all have closed-form expressions of $a$. Then it is possible to show the influence of $\cond(\Sigma)$ on {them}, respectively, {which will be given in Section \ref{subsec:influence of cond Sigma}.}

\section{Numerical Simulation}\label{sec:sim}

{
In this section, we run Monte Carlo (MC) simulations to show not only the influence of the condition number of $\Sigma$ but also the efficacy of our obtained theoretical results.

\subsection{Influence of $\cond(\Sigma)$ on $E_{\tb}^{\AR}(\eta_{\tb}^{*})$, $V^{\ALS}$ and $V_{\tb}^{\AR}(\eta_{\tb}^{*})$}\label{subsec:influence of cond Sigma}

\subsubsection{Test Systems}\label{subsubsec:test systems T1}

We generate $100$ test systems (20th order FIR models) in the following way: for each test system, we first generate a $\sigma_{g}^2$, which is uniformly distributed in $[0.5,\ 3]$, and then we generate its FIR coefficients $g_{1}^{0},\cdots,g_{20}^{0}$ as independently and identically Gaussian distributed random variables with mean zero and variance $\sigma^{2}_{g}$. These 100 tests systems will be referred to as T1.

\subsubsection{Simulation Setup}

For each test system of T1, we consider the RLS estimator \eqref{eq:RLS estimator} with $P=\eta I_{n}$. As shown in Section \ref{sec:case study}, when $P=\eta I_{n}$, $\Sigma$, $E_{\tb}^{\AR}(\eta_{\tb}^{*})$, $V^{\ALS}$ and $V_{\tb}^{\AR}(\eta_{\tb}^{*})$ can all be represented in functions of $a$, where $a$ is the parameter of $H(q)$ in \eqref{eq:tf function}. Then we are able to shed light on the influence of $\cond(\Sigma)$ on them by considering $a(i)=10^{-3}i$ with $i=1,\cdots,990$. For convenience, for the fixed $a(i)$, we shall replace $\Sigma$, $E_{\tb}^{\AR}(\eta_{\tb}^{*})$, $V^{\ALS}$ and $V_{\tb}^{\AR}(\eta_{\tb}^{*})$ with $\cond(\Sigma(a(i)))$, $E_{\tb}^{\AR}(a(i))$, $V^{\ALS}(a(i))$ and $V_{\tb}^{\AR}(a(i))$ in this subsection, respectively. For fixed $a(i)$, $\theta_{0}$ and $P$, since $\Sigma(a(i))$, $V^{\ALS}(a(i))$, $E_{\tb}^{\AR}(a(i))$ and $V_{\tb}^{\AR}(a(i))$ also depend on $c_{u}^2$, $\sigma^2$ and $N$, we choose the coefficient $c_{u}(a(i))$ such that $\lambda_{1}(\Sigma(a(i)))=1$, and set the measurement noise variance $\sigma^2=1$ and the sample size $N=10^3,10^5$.

\subsubsection{Simulation Results and Discussions}

To show the tendency more clearly, we define
\begin{align}
	\mathcal{I}_{diff}(f(i))=\left\{\begin{array}{ll}1, & \text{if}\ f(i+1)-f(i)\leq0,\\
		0, & \text{otherwise}, \end{array} \right.
\end{align}
for $i=1,\cdots,989$, where $f(i)$ denotes $\cond(\Sigma(a(i)))$, $\|E_{\tb}^{\AR}(a(i))\|_{2}^2$, $\Tr[V^{\ALS}(a(i))]$ or $\Tr[V_{\tb}^{\AR}(a(i))]$. Fig. \ref{fig:cond_Sigma_a} shows that $\cond(\Sigma(a(i)))$ is a strictly increasing function of $i$. Fig. \ref{fig:Bias2b_a_o1 for N=1e3} shows that for the randomly generated $100$ test systems, $\|E_{\tb}^{\AR}(a(i))\|_{2}^2$ is a strictly increasing function of $i$ for both $N=10^3$ and $10^5$. Fig. \ref{fig:Vb_a_o1 for N=1e3} shows that $V^{\ALS}(a(i))$ is a strictly increasing function of $i$ for both $N=10^3$ and $N=10^5$. Fig. \ref{fig:Vb_a_o2 for N=1e3}-\ref{fig:indicator of Vb_a_o2 for N=1e5} show that for the randomly generated $100$ test systems, $V_{\tb}^{\AR}(a(i))$ is a strictly increasing function of $i$ over the interval $1 \leq i \leq 597$ for $N=10^3$, and $1\leq i \leq 868$
for  $N=10^5$, respectively. Fig. \ref{fig:cond_Sigma_a} together with \ref{fig:Bias2b_a_o1 for N=1e3} and \ref{fig:Vb_a_o1 for N=1e3}-\ref{fig:indicator of Vb_a_o2 for N=1e5} show that, for large $N$,
\begin{itemize}
	\item $\|E_{\tb}^{\AR}(a)\|_{2}^2$ and $\Tr[V^{\ALS}(a)]$ are both increasing functions of $a$ or $\cond(\Sigma(a))$,
	\item $\Tr[V_{\tb}^{\AR}(a)]$ tends to be an increasing function of $a$ or  $\cond(\Sigma(a))$ over an interval $[0,c_N]$ ($0<c_{N}<1$) and moreover, as $N$ increases, $c_N$ also increases.
\end{itemize}

	\begin{figure}[!htp]
	\centering
	\begin{subfigure}[b]{0.24\textwidth}
		\centering
		\includegraphics[width=\textwidth, height=0.5\textwidth]{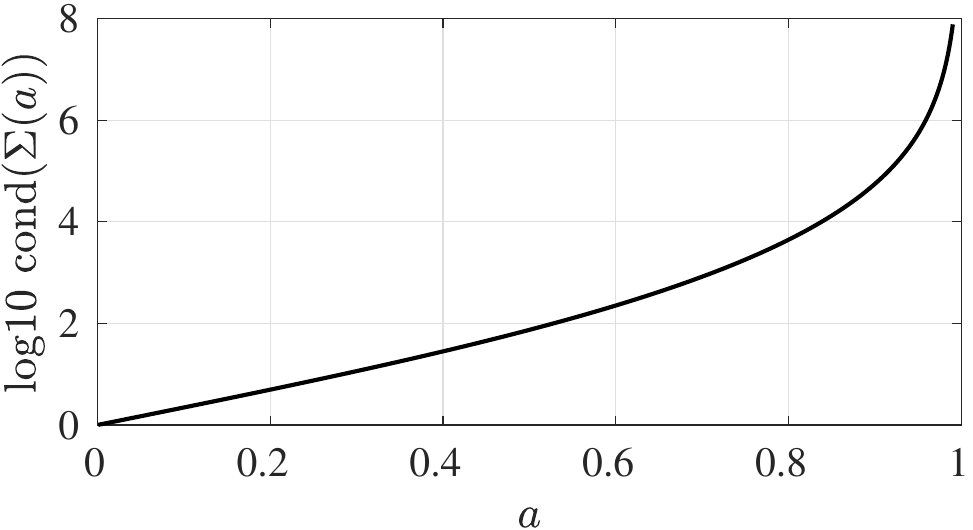}
		\caption{}
		\label{fig:cond_Sigma_a}
	\end{subfigure}
	%
	\begin{subfigure}[b]{0.24\textwidth}
		\centering
		\includegraphics[width=\textwidth, height=0.5\textwidth]{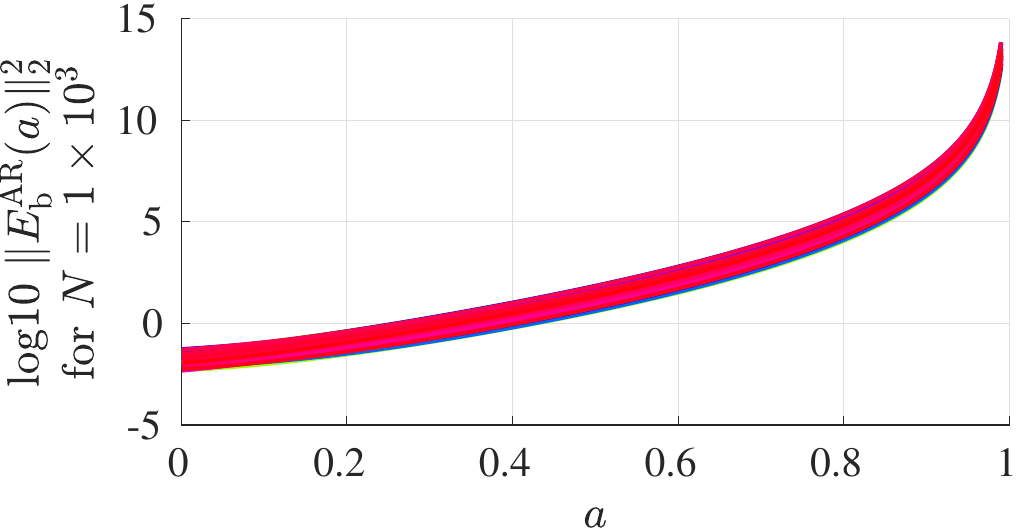}
		\caption{}
		\label{fig:Bias2b_a_o1 for N=1e3}
	\end{subfigure}
	\caption{Profile of $\log_{10}[\cond(\Sigma(a))]$ and $\log_{10}[\|E_{\tb}^{\AR}(a)\|_{2}^2]$ for ridge regression with filtered white noise inputs. Panel (a): $\log_{10}[\cond(\Sigma(a))]$ with $\mathcal{I}_{diff}(\cond(\Sigma(a(i))))=0$ for $i=1,\cdots,990$. Panel (b): $\log_{10}[\|E_{\tb}^{\AR}(a)\|_{2}^2]$ for $100$ test systems and $N=10^3$ with $\mathcal{I}_{diff}(\|E_{\tb}^{\AR}(a(i))\|_{2}^2)=0$ for $i=1,\cdots,990$. {Note that the tendency of $\log_{10}[\|E_{\tb}^{\AR}(a)\|_{2}^2]$ for $N=10^5$ is the same as that for $N=10^3$ and thus omitted.}}
	\label{fig:condSigma_a and indicator function}
\end{figure}

\begin{figure}[!htp]
	\centering
	\begin{subfigure}[b]{0.24\textwidth}
		\centering
		\includegraphics[width=\textwidth, height=0.5\textwidth]{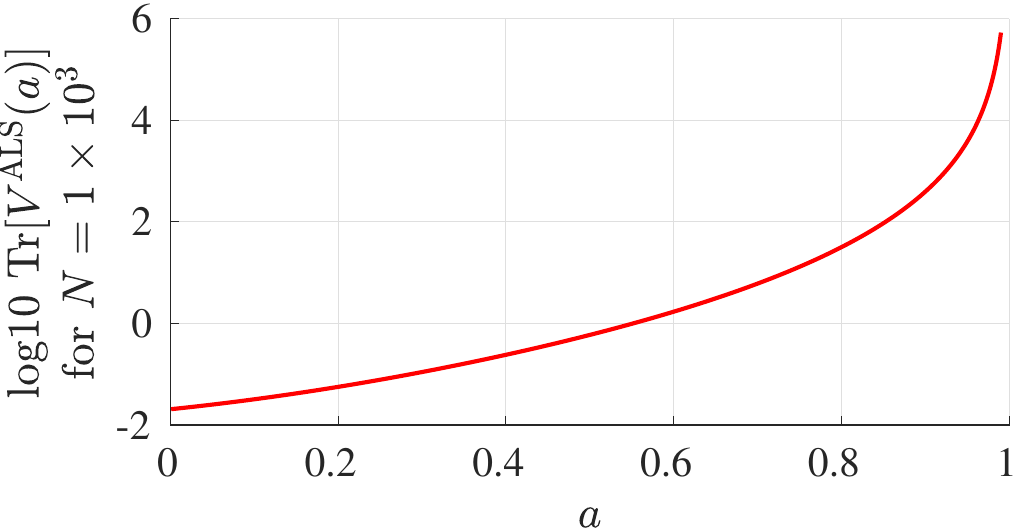}
		\caption{}
		\label{fig:Vb_a_o1 for N=1e3}
	\end{subfigure}
	
	\begin{subfigure}[b]{0.24\textwidth}
		\centering
		\includegraphics[width=\textwidth, height=0.5\textwidth]{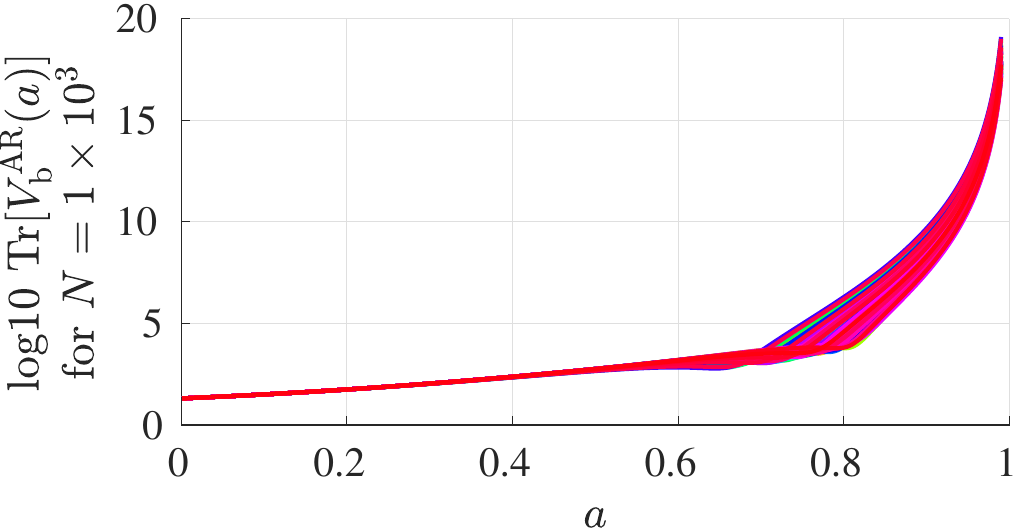}
		\caption{}
		\label{fig:Vb_a_o2 for N=1e3}
	\end{subfigure}
	\begin{subfigure}[b]{0.24\textwidth}
		\centering
		\includegraphics[width=\textwidth, height=0.5\textwidth]{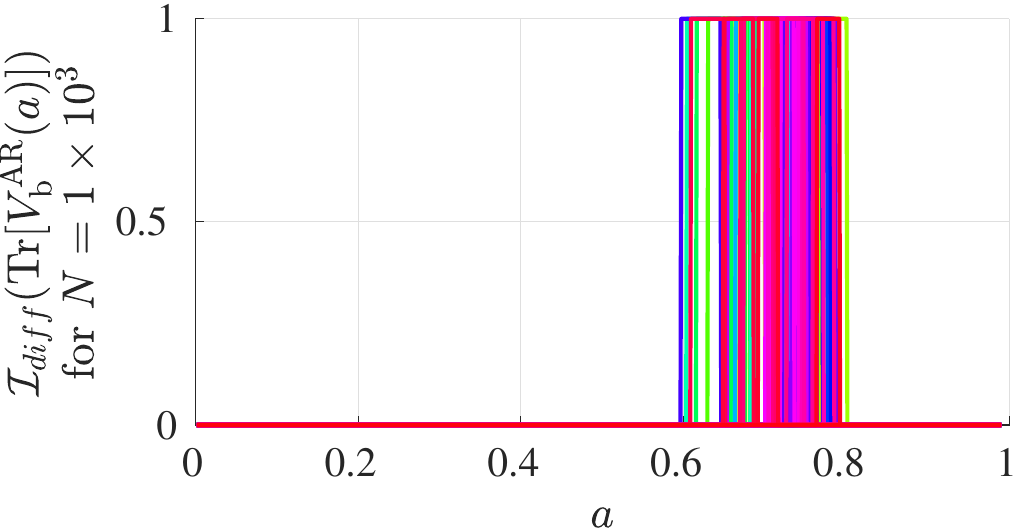}
		\caption{}
		\label{fig:indicator of Vb_a_o2 for N=1e3}
	\end{subfigure}
	\begin{subfigure}[b]{0.24\textwidth}
		\centering
		\includegraphics[width=\textwidth, height=0.5\textwidth]{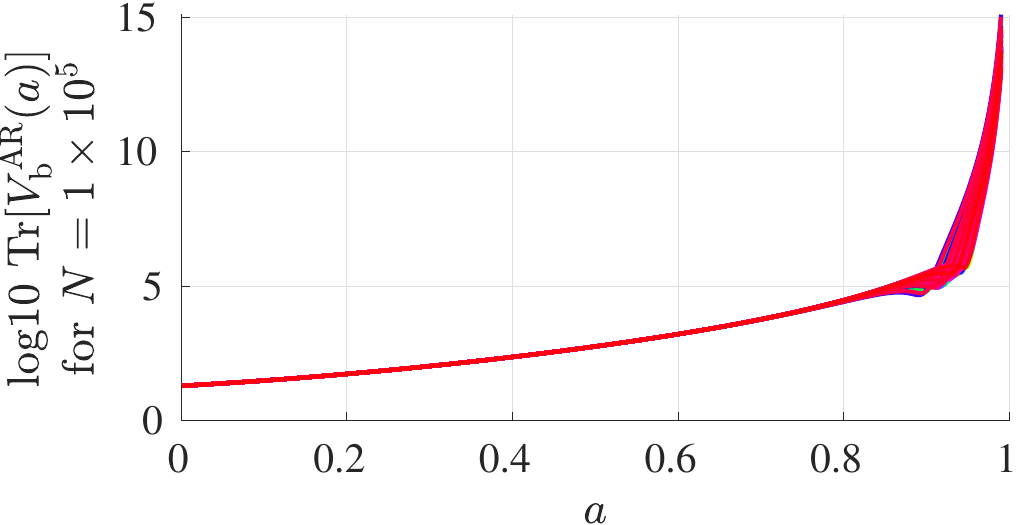}
		\caption{}
		\label{fig:Vb_a_o2 for N=1e5}
	\end{subfigure}
	\begin{subfigure}[b]{0.24\textwidth}
		\centering
		\includegraphics[width=\textwidth, height=0.5\textwidth]{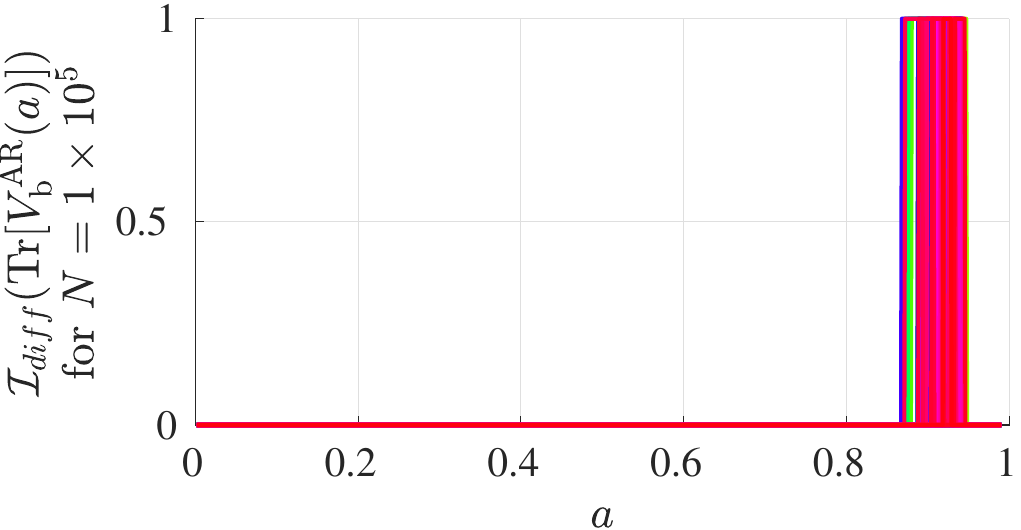}
		\caption{}
		\label{fig:indicator of Vb_a_o2 for N=1e5}
	\end{subfigure}
	\caption{Profile of $\log_{10}\{\Tr[V^{\ALS}(a)]\}$ and $\log_{10}\{\Tr[V_{\tb}^{\AR}(a)]\}$ for ridge regression with filtered white noise inputs and $100$ test systems.  Panel (a): $\log_{10}\{\Tr[V^{\ALS}(a)]\}$ for $N=10^3$ ($\mathcal{I}_{diff}(\Tr[V^{\ALS}(a(i))])=0$ for $i=1,\cdots,990$). {Note that the tendency of $\log_{10}\{\Tr[V^{\ALS}(a)]\}$ for $N=10^5$ is similar to that for $N=10^3$ and is omitted.} Panel (b): $\log_{10}\{\Tr[V_{\tb}^{\AR}(a)]\}$ for $N=10^3$. Panel (c): $\mathcal{I}_{diff}(\Tr[V_{\tb}^{\AR}(a)])$ for $N=10^3$. Panel (d): $\log_{10}\{\Tr[V_{\tb}^{\AR}(a)]\}$ for $N=10^5$. Panel (e): $\mathcal{I}_{diff}(\Tr[V_{\tb}^{\AR}(a)])$ for $N=10^5$. }
	\label{fig:Vb_a and indicator function for N=1e3,1e5}
\end{figure}

}

\subsection{Verification of Theorems \ref{thm:asymptotic normality of eta_eb difference}-\ref{thm:joint convergence distribution of decomp. of rls estimate up to OpN} and Corollaries \ref{corollary:ridge regression case}-\ref{corollary:asymptotic mean and variance for rls estimate for ridge regression with filtered white noise inputs}}\label{subsec:vertification}

\subsubsection{Test Systems}\label{subsubsec:test systems}

We consider two types of test systems, where the first one is T1 as introduced in Section \ref{subsubsec:test systems T1} and the second one is referred to as T2. For T2, we generate each test system as follows: first generate a $30$th order random system using the approach in \cite{COL2012} with its $5$ poles with largest modulus falling in $[0.94,0.96]$ and then truncate its impulse response to a finite one at the order $20$. For each type, we generate $100$ test systems, each of which has an FIR model with order $20$. The impulse response of each test system is then multiplied by a constant such that $\|\theta_{0}\|_{2}=10$ with $\theta_{0}$ defined in \eqref{eq:MSEg criterion}.

\subsubsection{Test Data-bank}\label{subsubsec:test data-bank}

For each test system, we generate the test input signal $u(t)$ as a filtered white noise as described in Assumption \ref{asp:input}, where $e(t)$ is chosen to be \emph{i.i.d.} Gaussian distributed with mean zero and $\sigma_e^2=1$. Noting that the filter $H(q)$ in \eqref{eq:tf function} depends on the choice of $a$ and $c_{u}$, we consider the following values of $a=0.05$, $0.7$ and $0.95$, and
\begin{itemize}
	\item for T1, consider $c_{u}^2=0.02$, $0.1$ and $0.5$;
	\item for T2, consider $c_{u}^2=1$, $10$ and $100$.
\end{itemize}
For each value of $a$ and $c_u^2$, we then simulate each test system with the generated test input signal to get the noise-free output and then corrupt it with an additive measurement noise $v(t)$, which is Gaussian distributed with mean 0 and variance $\sigma^2=1$, leading to the measurement output, and as a result, we collect a data record with $10^3$ pairs of input and measurement output. For each value of $a$ and $c_u^2$ and for each test system, the above procedure is repeated for $6\times10^5$ times, leading to $6\times 10^5$ data records. Therefore, for each test system, there are in total 9 data collections, each with $6\times 10^5$ data records.



\subsubsection{Simulation Setup }

For each test system of type T1 and its associated data collections, we consider the RLS estimator \eqref{eq:RLS estimator} with $P=\eta I_{n}$.  For each test system of type T2 and its associated data collections, we consider the RLS estimator \eqref{eq:RLS estimator} with the TC kernel \eqref{eq:TC kernel}. 

For both cases, the FIR model order $n$ is chosen to be $20$, i.e., $n=20$, the hyper-parameters are estimated using the $\EB$ method \eqref{eq:EB hyperparameter estimator} and the noise variance $\sigma^2$ is estimated by \eqref{eq:noisevariance_est}, and moreover, we have the sample size $N=10^3$.

%
%

\subsubsection{Simulation Results and Discussions}\label{sec:simulation results}

In Table \ref{table: different condition numbers of Sigma for different a}, for $a\in\R$ and $b\in\Z$, we use $a\text{E}b$ to denote $a\times 10^{\tb}$ for convenience.

\begin{itemize}
	
	\item Condition numbers of $\Phi^{T}\Phi$ and $\Sigma$

	\begin{table}[!htp]
		\scriptsize
		\centering
		\caption{Condition number of $\Sigma$ and average condition numbers of $\Phi^{T}\Phi$ for different values of $a$ over the $6\times 10^5$ data records.}
		\label{table: different condition numbers of Sigma for different a}
		\begin{tabular}{cccc}
			\hline
			$a$ &  $0.05$ & $0.7$   & $0.95$\\
			\hline
			$\cond(\Phi^{T}\Phi)$ & 2.00 & 9.10E2 & 5.98E5 \\
			\hline
			$\cond(\Sigma)$ & 1.49 & 8.34E2  & 5.51E5 \\
			\hline
		\end{tabular}
	\end{table}
	
	As shown in Table \ref{table: different condition numbers of Sigma for different a}, 
	as $a$ increases, both $\cond(\Phi^{T}\Phi)$ for fixed $N$ and $\cond(\Sigma)$ increase, i.e. $\Phi^{T}\Phi$ and $\Sigma$ become more ill-conditioned. 

	\item Verification of Theorem \ref{thm:asymptotic normality of eta_eb difference} and Corollary \ref{corollary:ridge regression case}
	
	Fig. \ref{fig:average bias2 and variance of eta for the ridge regression and the TC kernel} shows that for both the ridge regression and the TC kernel,
	\begin{itemize}

\item for fixed $c_{u}^2$, the larger $a$, the larger 
the average\footnote{Hereafter, all ``average'' quantities are referred to as the average of the concerned quantities over the 100 test systems in T1 or T2.} squared norm of mean\footnote{Hereafter, all ``mean'' and ``variance'' quantities are referred to as the sample mean and variance of the concerned quantities over the $6\times 10^5$ data records.} and the average variance of $\hat{\eta}_{\EB}-\eta_{\tb}^{*}$;

\item for fixed $a$, the larger $c_{u}^2$, the smaller the average squared norm of mean and the average variance of $\hat{\eta}_{\EB}-\eta_{\tb}^{*}$;

\item  for fixed $a$ and $c_{u}^2$, the average variance of $\hat{\eta}_{\EB}$ is quite close to $\Tr[V_{\tb}^{\tH}(\eta_{\tb}^{*})]/N$. 
\end{itemize}

	\begin{figure*}[thpb]
		\centering
		\begin{subfigure}[b]{0.4\textwidth}
			\centering
			\includegraphics[width=0.85\textwidth, height=0.58\textwidth]{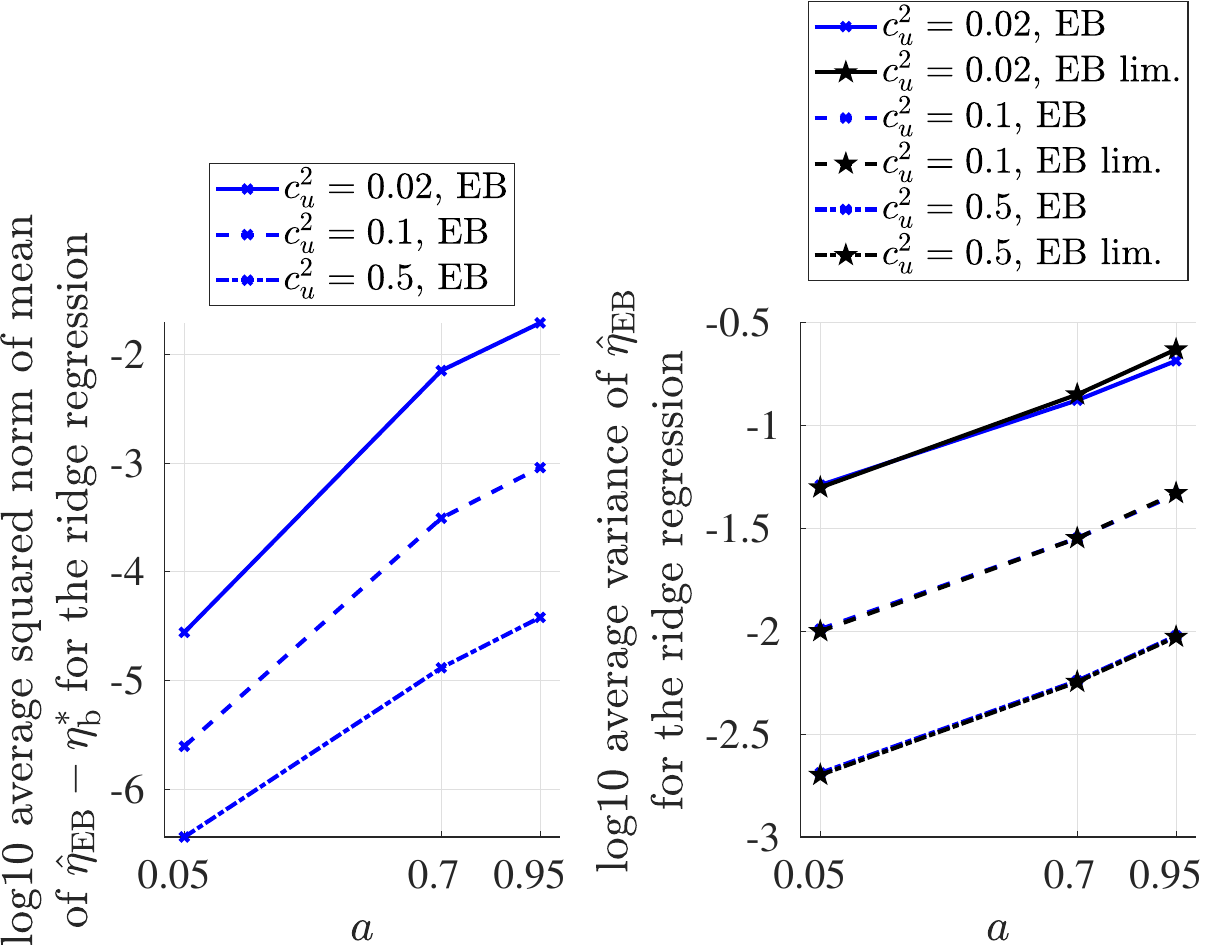}
			\caption{}
			\label{fig:average bias2 and var of etab for the ridge kernel}
		\end{subfigure}
	    \begin{subfigure}[b]{0.4\textwidth}
	    	\centering
	    	\includegraphics[width=0.85\textwidth, height=0.58\textwidth]{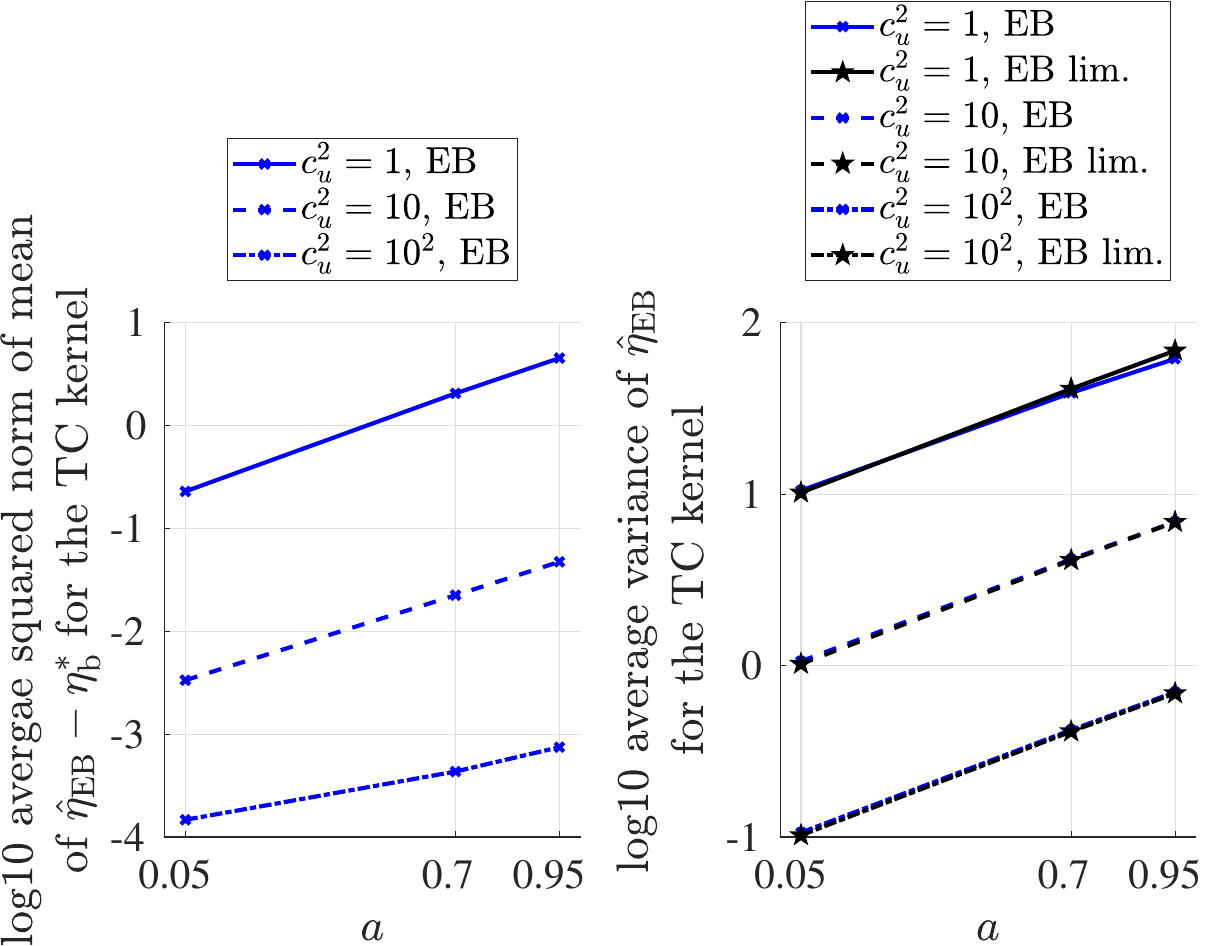}
	    	\caption{}
	    	\label{fig:average bias2 and var of etab for the TC kernel}
	    \end{subfigure}
        \caption{Profile of logarithm base $10$ average squared norm of mean of $\hat{\eta}_{\EB}-\eta^{*}_{\tb}$ and variance of $\hat{\eta}_{\EB}$ for the ridge regression and the TC kernel over $9$ data collections. Panel (a): Logarithm base $10$ of average squared norm of mean of $\hat{\eta}_{\EB}-\eta^{*}_{\tb}$ and average variance of $\hat{\eta}_{\EB}$ for the ridge regression (`` EB lim.'' denotes $V_{\tb}^{\tH}(\eta_{\tb}^{*})/N$ and $V_{\tb}^{\tH}(\eta_{\tb}^{*})$ is defined as \eqref{eq:Vb for ridge regression case}). Panel (b): Logarithm base $10$ of average squared norm of mean of $\hat{\eta}_{\EB}-\eta^{*}_{\tb}$ and average variance of $\hat{\eta}_{\EB}$ for the TC kernel (`` EB lim.'' denotes $\Tr[V_{\tb}^{\tH}(\eta_{\tb}^{*})]/N$ and $V_{\tb}^{\tH}(\eta_{\tb}^{*})$ is defined as \eqref{eq:def of Vb_etab_star}).    }
        \label{fig:average bias2 and variance of eta for the ridge regression and the TC kernel}
	\end{figure*}

    \item Verification of Theorems \ref{thm:joint convergence distribution of decomp. of rls estimate up to OpsqrtN}-\ref{thm:joint convergence distribution of decomp. of rls estimate up to OpN} and Corollary \ref{corollary:asymptotic mean and variance for rls estimate for ridge regression with filtered white noise inputs}
      
Since $\MSE_{g}(\hat{\theta}^{\TR}(\hat{\eta}_{\EB}))$ has no closed form expression, in order to assess the accuracy of the high order asymptotic distributions \eqref{eq:convergence in distribution of rls estimate using EB}, \eqref{eq:2nd order asymptotic distribution of the theta_rls_EB} and \eqref{eq:3rd order asymptotic distribution of the theta_rls_EB},  we calculate for each test system the sample average of ${\MSE}_{g}(\hat{\theta}^{\TR}(\hat{\eta}_{\EB}))$ over its associated $6\times10^5$ Monte Carlo simulations and denote it by ${\SMSE}_{g}(\hat{\theta}^{\TR}(\hat{\eta}_{\EB}))${. Moreover}, we let
  \begin{itemize}
      	\item ${\mathcal{S}_{1}}$ denote the number of systems satisfying
      	\begin{align}
      		 &\left| {\text{AMSE}_{g}^{\tb,2}(\eta_{\tb}^{*})} - {\SMSE}_{g}(\hat{\theta}^{\TR}(\hat{\eta}_{\EB})) \right|\nonumber\\
       \label{eq:def of num_sys1}
      		 <& \left| {\text{AMSE}_{g}^{\tb,1}(\eta_{\tb}^{*})} - {\SMSE}_{g}(\hat{\theta}^{\TR}(\hat{\eta}_{\EB})) \right|;    	
      	 \end{align}

       \item ${\mathcal{S}_{2}}$ denote the number of systems satisfying
      \begin{align}
      	&\left| {\text{AMSE}_{g}^{\tb,3}(\eta_{\tb}^{*})} - {\SMSE}_{g}(\hat{\theta}^{\TR}(\hat{\eta}_{\EB})) \right|\nonumber\\
      	\label{eq:def of num_sys2}
      	<& \left| {\text{AMSE}_{g}^{\tb,1}(\eta_{\tb}^{*})}  - {\SMSE}_{g}(\hat{\theta}^{\TR}(\hat{\eta}_{\EB})) \right|;
      \end{align}
     \item ${\mathcal{S}_{3}}$ denote the number of systems satisfying
     \begin{align}
     	&\left| {\text{AMSE}_{g}^{\tb,3}(\eta_{\tb}^{*})} - {\SMSE}_{g}(\hat{\theta}^{\TR}(\hat{\eta}_{\EB})) \right|\nonumber\\
     	\label{eq:def of num_sys3}
     	<& \left| {\text{AMSE}_{g}^{\tb,2}(\eta_{\tb}^{*})}  - {\SMSE}_{g}(\hat{\theta}^{\TR}(\hat{\eta}_{\EB})) \right|,
     \end{align} 
      \end{itemize}
  among the 100 test systems, {where ${\text{AMSE}_{g}^{\tb,m}(\eta_{\tb}^{*})}$ with $m=1,2,3$ are defined in \eqref{eq:asymptotic MSE with orders 1,2,3}.} Clearly, 
\begin{itemize}
\item  the closer ${\mathcal{S}_{1}}$ or ${\mathcal{S}_{2}}$ to 100, the more accurate the high order asymptotic distributions \eqref{eq:2nd order asymptotic distribution of the theta_rls_EB} or \eqref{eq:3rd order asymptotic distribution of the theta_rls_EB} over {the first order one} \eqref{eq:convergence in distribution of rls estimate using EB}; 

\item  the closer ${\mathcal{S}_{3}}$ to 100, the more accurate {the third order asymptotic distributions \eqref{eq:3rd order asymptotic distribution of the theta_rls_EB} over the second order one \eqref{eq:2nd order asymptotic distribution of the theta_rls_EB}}.

\end{itemize}

\begin{figure*}[thpb]
                \centering
                \begin{subfigure}[b]{0.91\textwidth}
                                \centering
                                \includegraphics[width=\textwidth, height=0.25\textwidth]{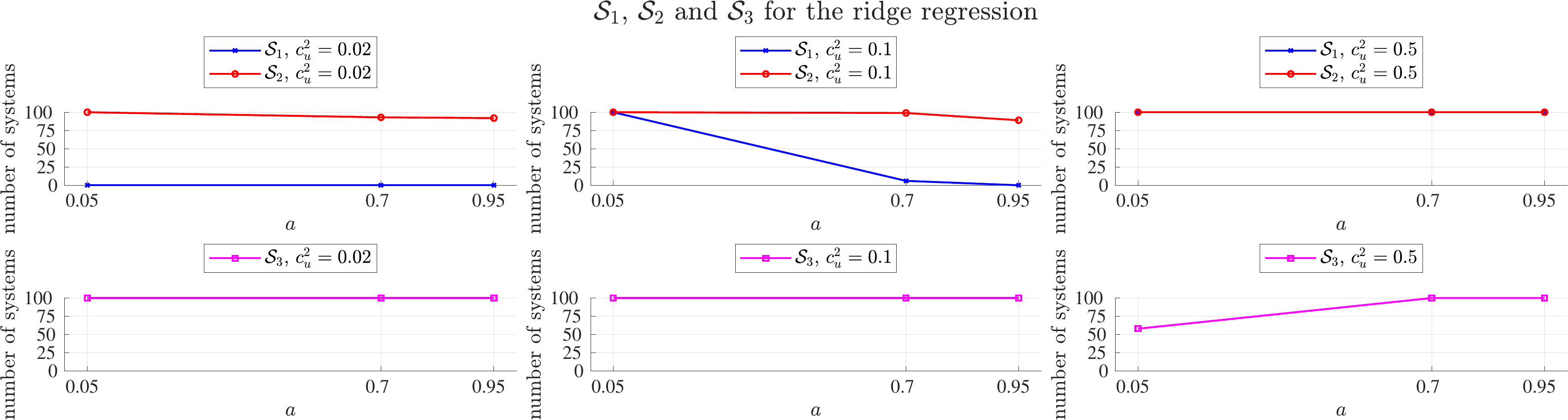}
                                \caption{}
                                \label{fig:num123 of sys for the ridge regression}
                \end{subfigure}
   \begin{subfigure}[b]{0.91\textwidth}
                \centering
                \includegraphics[width=\textwidth, height=0.14\textwidth]{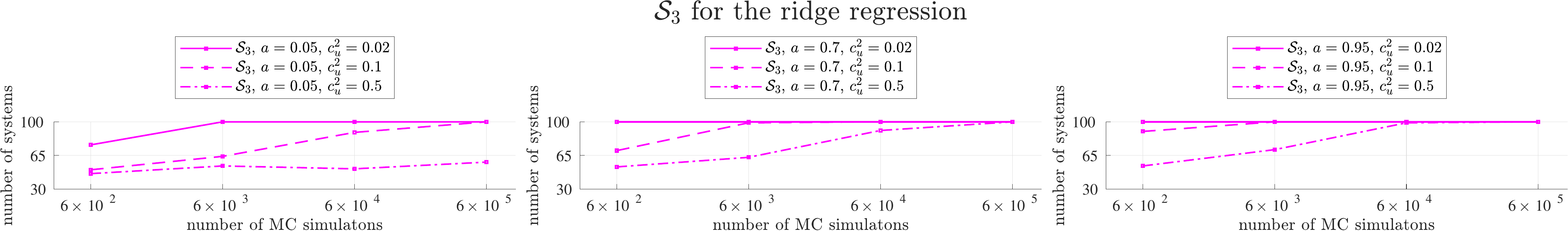}
                \caption{}
                \label{fig:num3 of sys for the ridge regression}
   \end{subfigure}
                                \begin{subfigure}[b]{0.91\textwidth}
                                \centering
                                \includegraphics[width=\textwidth, height=0.25\textwidth]{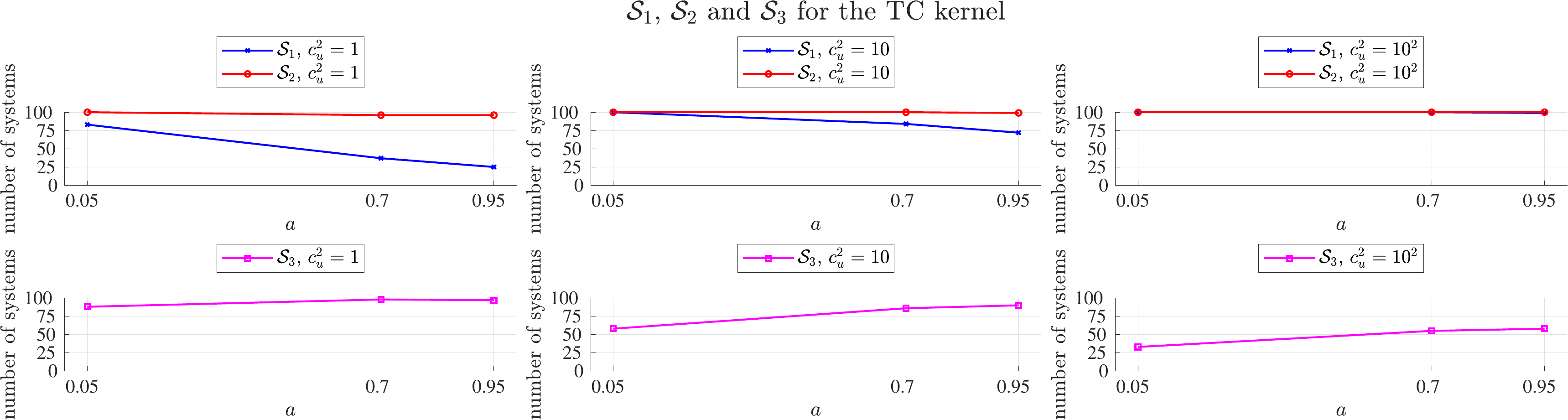}
                                \caption{}
                                \label{fig:num123 of sys for the TC kernel}
                \end{subfigure}
   \begin{subfigure}[b]{0.91\textwidth}
                \centering
                \includegraphics[width=\textwidth, height=0.14\textwidth]{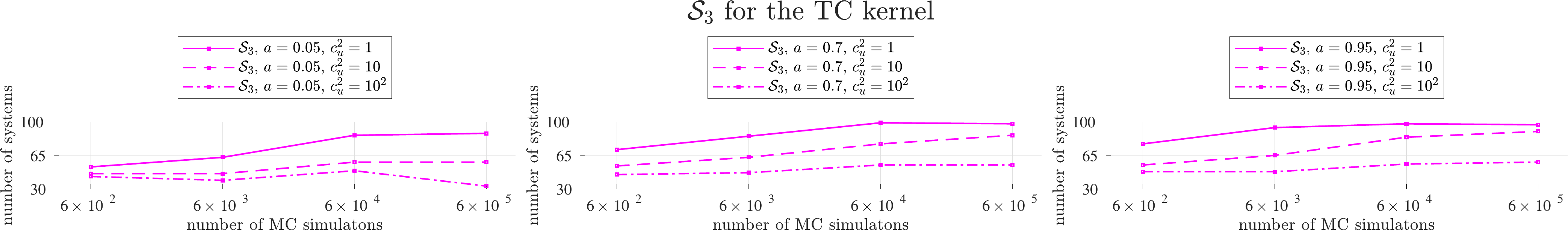}
                \caption{}
                \label{fig:num3 of sys for the TC kernel}
   \end{subfigure}
   \caption{Profile of ${\mathcal{S}_{1}}$ \eqref{eq:def of num_sys1}, ${\mathcal{S}_{2}}$ \eqref{eq:def of num_sys2} and ${\mathcal{S}_{3}}$ \eqref{eq:def of num_sys3} for the ridge regression and the TC kernel. Panel (a): ${\mathcal{S}_{1}}$ and ${\mathcal{S}_{2}}$ (the first row), and ${\mathcal{S}_{3}}$ (the second row) for the ridge regression over $9$ data collections. Panel (b): ${\mathcal{S}_{3}}$ for the ridge regression over different numbers of data records. Panel (c): ${\mathcal{S}_{1}}$ and ${\mathcal{S}_{2}}$ (the first row), and ${\mathcal{S}_{3}}$ (the second row) for the TC kernel over $9$ data collections. Panel (d): ${\mathcal{S}_{3}}$ for TC kernel over different numbers of data records.  }
   \label{fig:num of sys for the ridge regression and the TC kernel}
   \end{figure*}
   Fig. \ref{fig:num of sys for the ridge regression and the TC kernel} shows that for both the ridge regression and the TC kernel:
    \begin{itemize}
    	\item for fixed $c_{u}^2$, as $a$ increases, ${\mathcal{S}_{1}}$ and ${\mathcal{S}_{2}}$ tend to become smaller, and ${\mathcal{S}_{3}}$ tends to become larger (except when $a=0.95$ for the TC kernel);
    	\item for fixed $a$, as $c_{u}^2$ increases, ${\mathcal{S}_{1}}$ and ${\mathcal{S}_{2}}$ tend to become larger, and ${\mathcal{S}_{3}}$ tends to become smaller.
    \end{itemize}

{For the ridge regression and TC kernel, the sums of ${\mathcal{S}_{2}}$ for all values of $a$ and $c_{u}^2$ are $873$ and $891$, respectively. It means that} in contrast with the first order asymptotic distribution \eqref{eq:convergence in distribution of rls estimate using EB}, there are $873$ cases for the ridge regression, and $891$ cases for the TC kernel,
out of the total 900 cases ($9$ data collections) such that the third order one \eqref{eq:3rd order asymptotic distribution of the theta_rls_EB} is more accurate. {Moreover, ${\mathcal{S}_{1}}$ is smaller or equal to ${\mathcal{S}_{2}}$ for both the ridge regression and the TC kernel. It indicates that} the second order one \eqref{eq:2nd order asymptotic distribution of the theta_rls_EB} is less accurate, when $a$ is large or $c_u^2$ is small. This observation is reasonable, because it does not take into account the influence of the regularization, whose role is critical especially when the quality of the data is bad, i.e., when $a$ is large or $c_u^2$ is small. 

{For the ridge regression and TC kernel, the sums of ${\mathcal{S}_{3}}$ for all values of $a$ and $c_{u}^2$ are $858$ and $663$, respectively. It means that} in contrast with the second order asymptotic distribution \eqref{eq:2nd order asymptotic distribution of the theta_rls_EB}, there are $858$ cases for the ridge regression, and $663$ cases for the TC kernel,
out of the total 900 cases  such that the third order one \eqref{eq:3rd order asymptotic distribution of the theta_rls_EB} is more accurate, especially when $a$ is large or $c_u^2$ is small. Moreover, the second order one \eqref{eq:2nd order asymptotic distribution of the theta_rls_EB} \emph{seems} to become more accurate when the quality of the data is getting better, i.e., when $a$ becomes smaller or $c_u^2$ becomes larger. This observation is somewhat against our intuition {that the third order one \eqref{eq:3rd order asymptotic distribution of the theta_rls_EB} should be better, and the reasons might be two-fold}: 
\begin{itemize}
\item First, this may be due to the insufficient number of Monte Carlo simulations. Note that when $a$ becomes smaller or $c_u^2$ becomes larger, the quality of the data becomes better and thus not only $\MSE_{g}(\hat{\theta}^{\TR}(\hat{\eta}_{\EB}))$ but also its high order asymptotic approximations ${\text{AMSE}_{g}^{\tb,m}(\eta_{\tb}^{*})}$, $m=1,2,3$ all become smaller, implying that the differences between them also become smaller. Therefore, we need more Monte Carlo simulations to obtain a more accurate approximation of $\MSE_{g}(\hat{\theta}^{\TR}(\hat{\eta}_{\EB}))$ to differentiate them. This tendency can be seen from most cases in the Panel (b) and Panel (d) in Fig. \ref{fig:num of sys for the ridge regression and the TC kernel},  {where we display the performances} as we increase the number of Monte Carlo simulations to $6\times 10^5$.


\item Second, this may also be due to that the sample size $N=10^3$ might not be large enough such that the approximation error of the {terms involved} might not be negligible
when assessing the difference between ${\SMSE}_{g}(\hat{\theta}^{\TR}(\hat{\eta}_{\EB}))$ and ${\text{AMSE}_{g}^{\tb,m}(\eta_{\tb}^{*})}$, $m=1,2,3$. In addition, it is also worth to note that  the third order asymptotic distribution \eqref{eq:3rd order asymptotic distribution of the theta_rls_EB} has an extra {term} $\sqrt{N}(\widehat{\sigma^2}-\sigma^2)$ in contrast with the second order one \eqref{eq:2nd order asymptotic distribution of the theta_rls_EB}.

\end{itemize}

	\item For reference, we also assess the performance of the RLS estimator $\hat{\theta}^{\TR}(\hat{\eta}_{\EB})$ from the perspective of $\MSE_{g}$ with the ``model fit'' \cite{Ljung1995}:
    \begin{align*}
    	\text{Fit}_{g}=100\times\left(1-\frac{\|\hat{\theta}^{\TR}(\hat{\eta}_{\EB})-\theta_{0}\|_{2}}{\|\theta_{0}-\overline{\theta_{0}}\|_{2}} \right), \overline{\theta_{0}}=\frac{1}{n}\sum_{i=1}^{n}g_{i}^{0}.
    \end{align*}
    In fact, the mean of $\text{Fit}_{g}$ can be seen as a normalized version of $\SMSE_{g}(\hat{\theta}^{\TR}(\hat{\eta}_{\EB}))$, which is equal to {the sum of the squared norm of mean of $\hat{\theta}^{\TR}(\hat{\eta}_{\EB})-\theta_{0}$, and the variance of $\hat{\theta}^{\TR}(\hat{\eta}_{\EB})$}.  Fig. \ref{fig:average Fit mean for the ridge regression and the TC kernel} shows that for both the ridge regression and the TC kernel, 
    \begin{itemize}
    	\item for fixed $c_{u}^2$, the larger $a$, the smaller the {average mean of} $\text{Fit}_{g}$ of $\hat{\theta}^{\TR}(\hat{\eta}_{\EB})$;
    	\item for fixed $a$, the larger $c_{u}^2$, the larger the {average mean of} $\text{Fit}_{g}$ of $\hat{\theta}^{\TR}(\hat{\eta}_{\EB})$.
    \end{itemize}
    
    \begin{figure}[thpb]
    	\centering
    	\begin{subfigure}[b]{0.24\textwidth}
    		\centering
    		\includegraphics[width=\textwidth, height=0.74\textwidth]{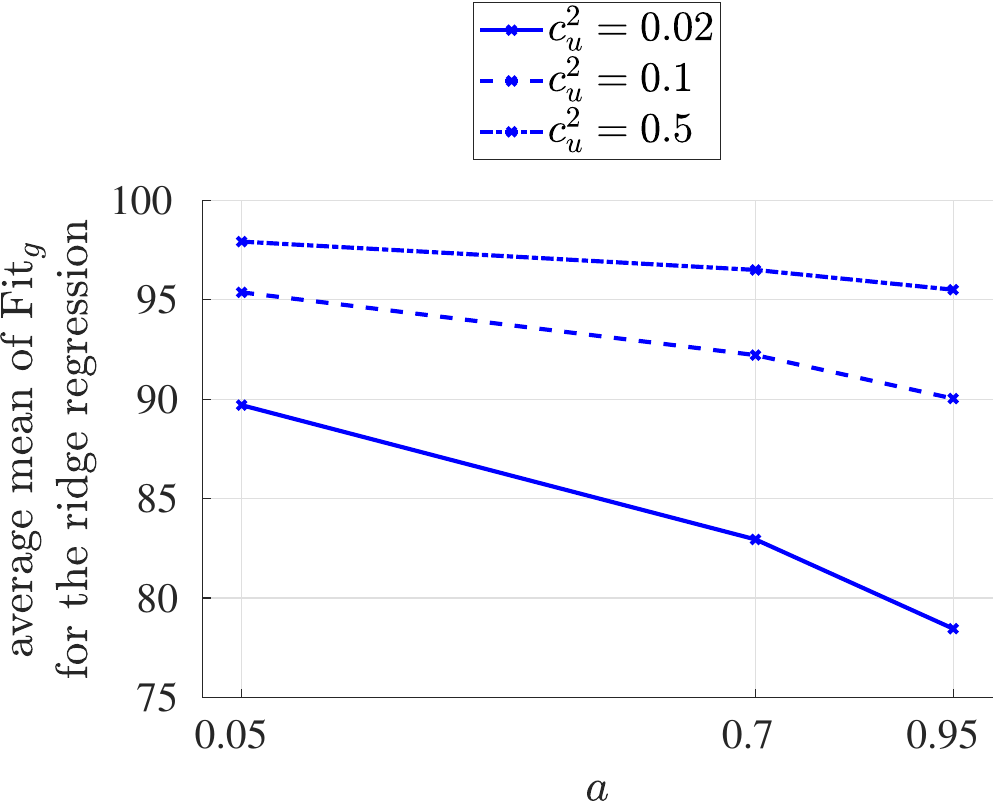}
    		\caption{}
    		\label{fig:average Fit mean for the ridge kernel}
    	\end{subfigure}
    	\begin{subfigure}[b]{0.24\textwidth}
    		\centering
    		\includegraphics[width=\textwidth, height=0.74\textwidth]{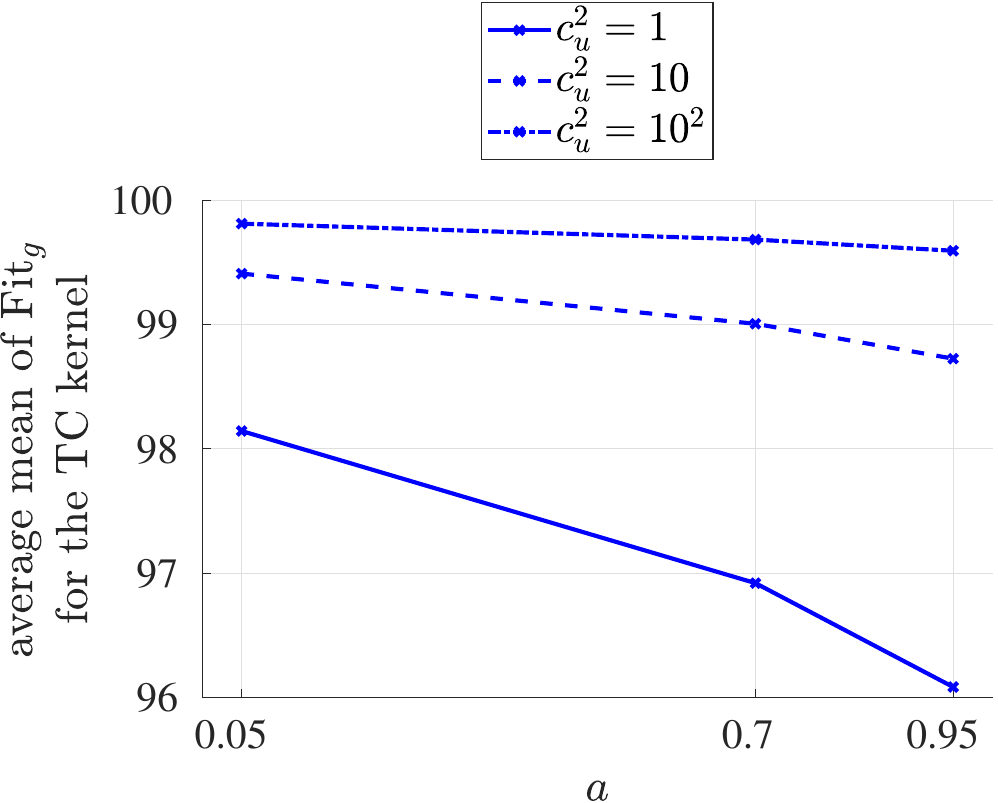}
    		\caption{}
    		\label{fig:average Fit mean for the TC kernel}
    	\end{subfigure}
    	\caption{Profile of average {mean of} $\text{Fit}_{g}$ for the ridge regression and the TC kernel over $9$ data collections. Panel (a): Average {mean of} $\text{Fit}_{g}$ for the ridge regression. Panel (b): Average {mean of} $\text{Fit}_{g}$ for the TC kernel.   }
    	\label{fig:average Fit mean for the ridge regression and the TC kernel}
    \end{figure}

\end{itemize}


\section{Conclusion}\label{sec:conclusion}

Asymptotic theory is a core component for the theory of system identification. In this paper, we studied the asymptotic theory for the regularized system identification{,} and in particular, the regularized finite impulse response (FIR) model estimation with the input signal chosen to be filtered white noise and the hyper-parameter estimator chosen to be the empirical Bayes (EB) method. Our obtained results on the convergence in distribution of the EB hyper-parameter estimator and on the high order asymptotic distributions of the corresponding kernel-based regularized least squares (RLS) estimator, expose the factors (e.g., the regression matrix and the kernel matrix) that affect the convergence properties of the EB hyper-parameter estimator and the corresponding RLS estimator. These  results provide theoretical support to the widely observed numerical simulation results that the more ill-conditioned the regression matrix, the more slowly the EB hyper-parameter estimator and the corresponding RLS estimator converge to their limits, respectively.
These results fill the gaps in the asymptotic theory for the regularized system identification, and have many potential applications, e.g., 
in finding the confidence intervals of the EB hyper-parameter estimator and the corresponding RLS model estimator. 


\def\thesectiondis{\thesection.}                   
\def\thesubsectiondis{\thesection.\arabic{subsection}.}          
\def\thesubsubsectiondis{\thesubsection.\arabic{subsubsection}.}

\setcounter{subsection}{0}

\renewcommand{\thesection}{A}
\setcounter{theorem}{0}
\renewcommand{\thelemma}{A.\arabic{lemma}}

\renewcommand{\theequation}{A.\arabic{equation}}
\setcounter{equation}{0}

\renewcommand{\thesubsection}{\thesection.\arabic{subsection}}

\section*{Appendix A}\label{sec:Appendix A}

Proofs of theorems, propositions and corollaries are included in Appendix A, among which proofs of Corollaries \ref{corollary:convergence in distribution of difference of rls and ls estimates}-\ref{corollary:asymptotic mean and variance for rls estimate for ridge regression with filtered white noise inputs} are omitted because of the limitation of space. 

\subsection{Proof of Theorem \ref{thm:asymptotic normality of eta_eb difference}}\label{subsec:proof of asymptotic normality of eta_eb}

First, let
\begin{align} 
\overline{\mathscr{F}_{\EB}}(\eta)\label{eq:F_eb form1}
=&\widehat{\mathscr{F}_{\EB}}(\eta)-(N-n)-(N-n)\log\widehat{\sigma^2}\nonumber\\
&-\log\det(\Phi^{T}\Phi)\\
\label{eq:F_eb form2}
=&(\hat{\theta}^{\LS})^{T}\hat{S}(\eta)^{-1}\hat{\theta}^{\LS}+\log\det(\hat{S}(\eta)),
\end{align}
where $\hat{S}(\eta)$ is defined in \eqref{eq:def of hat_S_inv}. Since the difference between $\widehat{\mathscr{F}_{\EB}}(\eta)$ in \eqref{eq:F_eb form1} and $\overline{\mathscr{F}_{\EB}}(\eta)$ in \eqref{eq:EB1} is irrespective of $\eta$, we have $\hat{\eta}_{\EB}=\argmin_{\eta\in\Omega}\overline{\mathscr{F}_{\EB}}(\eta)$. Using the analogous idea in the proof of \cite[Theorem 1]{MCL2018}, we can apply \eqref{eq:almost sure convergence of PP_N_inv}-\eqref{eq:almost sure convergence of noise variance estimator}, \eqref{eq:difference of S_inv and P_inv} and \cite[Lemma B3]{MCL2018} to derive \eqref{eq:almost sure convergence of hat_eta_eb}.

Then we will derive the convergence in distribution of $\sqrt{N}(\hat{\eta}_{\EB}-\eta_{\tb}^{*})$ based on the first-order Taylor expansion of ${\partial \overline{\mathscr{F}_{\EB}}}/{\partial\eta}|_{\eta=\hat{\eta}_{\EB}}$ around $\eta_{\tb}^{*}$ as follows,
\begin{align*}
0=\left.\frac{\partial \overline{\mathscr{F}_{\EB}}}{\partial \eta}\right|_{\eta=\hat{\eta}_{\EB}}
=\left.\frac{\partial \overline{\mathscr{F}_{\EB}}}{\partial \eta}\right|_{\eta=\eta_{\tb}^{*}}
+\left.\frac{\partial^2 \overline{\mathscr{F}_{\EB}}}{\partial \eta\partial\eta^{T}}\right|_{\eta=\overline{\eta}_{N}}(\hat{\eta}_{\EB}-\eta_{\tb}^{*}),
\end{align*}
where the remainder term is represented in the Lagrange's form and $\overline{\eta}_{N}$ belongs to a neighborhood of $\eta_{\tb}^{*}$ with radius $\|\hat{\eta}_{\EB}-\eta_{\tb}^{*}\|_{2}$.
It follows that
\begin{align}\label{eq:Taylor expansion form for sqrtN_eta_eb difference}
\sqrt{N}(\hat{\eta}_{\EB}-\eta_{\tb}^{*})=
-\left(\left.\frac{\partial^2 \overline{\mathscr{F}_{\EB}}}{\partial \eta\partial\eta^{T}}\right|_{\eta=\overline{\eta}_{N}}\right)^{-1}\left(\sqrt{N}\left.\frac{\partial \overline{\mathscr{F}_{\EB}}}{\partial \eta}\right|_{\eta=\eta_{\tb}^{*}}\right),
\end{align}
where if $\left.{\partial^2 \overline{\mathscr{F}_{\EB}}}/{\partial \eta\partial\eta^{T}}\right|_{\eta=\overline{\eta}_{N}}$ is not positive definite for small $N$, the pseudo inverse could be used instead.

Now, in what follows, we consider the almost sure convergence of $\left.{\partial^2 \overline{\mathscr{F}_{\EB}}}/{\partial \eta\partial\eta^{T}}\right|_{\eta=\overline{\eta}_{N}}$ and the convergence in distribution of $\sqrt{N}\left.{\partial \overline{\mathscr{F}_{\EB}}}/{\partial \eta}\right|_{\eta=\eta_{\tb}^{*}}$.

\begin{enumerate}
	\item Firstly, we show in two steps that
	\begin{align}\label{eq:almost sure convergence of Hessian matrix}
	\left.\frac{\partial^2 \overline{\mathscr{F}_{\EB}}}{\partial \eta\partial\eta^{T}}\right|_{\eta=\overline{\eta}_{N}}
	\overset{a.s.}\to \left.\frac{\partial^2 W_{\tb}}{\partial \eta\partial\eta^{T}}\right|_{\eta=\eta^{*}_{\tb}}=A_{\tb}(\eta_{\tb}^{*})\succ 0,
	\end{align} 
	where $A_{\tb}(\eta_{\tb}^{*})$ is defined in \eqref{eq:element of A eta_b_star}.
	The first step is to prove
	\begin{align}\label{eq:as convergence of overline_etaN}
	\overline{\eta}_{N}\overset{a.s.}\to \eta_{\tb}^{*},
	\end{align}
	which is true because  $\|\overline{\eta}_{N}-\eta_{\tb}^{*}\|_{2}\leq \|\hat{\eta}_{\EB}-\eta_{\tb}^{*}\|_{2}\overset{a.s.}\to 0$.
	The second step is to prove that ${\partial^2 \overline{\mathscr{F}_{\EB}}}/{\partial \eta\partial\eta^{T}}$ converges to ${\partial^2 W_{\tb}}/{\partial \eta\partial\eta^{T}}$ almost surely and uniformly. Their $(k,l)$th elements are
	\begin{align}
	\frac{\partial^2 \overline{\mathscr{F}_{\EB}}}{\partial\eta_{k}\partial\eta_{l}}
	=&(\hat{\theta}^{\LS})^{T}\frac{\partial^2 \hat{S}^{-1}}{\partial\eta_{k}\partial\eta_{l}}\hat{\theta}^{\LS}
	+\Tr\left(\frac{\partial \hat{S}^{-1}}{\partial\eta_{l}}\frac{\partial P}{\partial\eta_{k}}\right)\nonumber\\
	&+\Tr\left(\hat{S}^{-1}\frac{\partial^2 P}{\partial\eta_{k}\partial\eta_{l}}\right),\\
	\label{eq:2nd derivatives of Wb}
	\frac{\partial^2 W_{\tb}}{\partial\eta_{k}\partial\eta_{l}}
	=&\theta_{0}^{T}\frac{\partial^2 P^{-1}}{\partial\eta_{k}\partial\eta_{l}}\theta_{0}+\Tr\left(\frac{\partial P^{-1}}{\partial\eta_{l}}\frac{\partial P}{\partial\eta_{k}}\right)\nonumber\\
	&+\Tr\left(P^{-1}\frac{\partial^2 P}{\partial\eta_{k}\partial\eta_{l}}\right).
	\end{align}
	Then their difference can be represented as
	\begin{align*}
	{\partial^2 \overline{\mathscr{F}_{\EB}}}/{\partial\eta_{k}\partial\eta_{l}}-{\partial^2 W_{\tb}}/{\partial\eta_{k}\partial\eta_{l}}
	=&\Psi_{1,\tb}+\Tr(\Psi_{2,\tb}),
	\end{align*}
	where 
	\begin{align*}
	\Psi_{1,\tb}
	=&(\hat{\theta}^{\LS}-\theta_{0})^{T}\frac{\partial^2 \hat{S}^{-1}}{\partial\eta_{k}\partial\eta_{l}}\hat{\theta}^{\LS}
	+\theta_{0}^{T}\frac{\partial^2 P^{-1}}{\partial\eta_{k}\partial\eta_{l}}(\hat{\theta}^{\LS}-\theta_{0})\nonumber\\
	&+\theta_{0}^{T}\left(\frac{\partial^2 \hat{S}^{-1}}{\partial\eta_{k}\partial\eta_{l}}-\frac{\partial^2 P^{-1}}{\partial\eta_{k}\partial\eta_{l}}\right)\hat{\theta}^{\LS},\\
	\Psi_{2,\tb}=&\left(\frac{\partial \hat{S}^{-1}}{\partial\eta_{l}}-\frac{\partial P^{-1}}{\partial\eta_{l}}\right)\frac{\partial P}{\partial\eta_{k}}
	+(\hat{S}^{-1}-P^{-1})\frac{\partial^2 P}{\partial\eta_{k}\partial\eta_{l}}.
	\end{align*}
	
	Under Assumption \ref{asp:interior points assumption of hat_eta}, there exists a compact subset $\widetilde{\Omega}_{1}$ of $\Omega$ such that $\eta_{\tb}^{*}\in	\widetilde{\Omega}_{1}\subset\Omega$
	and moreover, for any $k,l=1,\cdots,p$,
	\begin{subequations}\label{eq:boundedness of P,S and their derivatives}
	\begin{align}
	\label{eq:boundedness of P, Sinv, Pinv}
	&\|P\|_{F},\ \|\hat{S}^{-1}\|_{F}<\|P^{-1}\|_{F}\ \text{are}\ \text{bounded},\\
	\label{eq:boundedness of P 1st derivative, P 2nd derivative}
	&\left\|{\partial P}/{\partial\eta_{l}}\right\|_{F},\ \left\|{\partial^2 P}/{\partial \eta_{k}\partial\eta_{l}}\right\|_{F}\ \text{are}\ \text{bounded}.
	\end{align}
	\end{subequations}
	According to \cite[(59) p. 9]{PP2012}, we can see that both ${\partial^2 P^{-1}}/{\partial\eta_{k}\partial\eta_{l}}$ and ${\partial^2 \hat{S}^{-1}}/{\partial\eta_{k}\partial\eta_{l}}$ are made of $P^{-1}$, $\hat{S}^{-1}$, ${\partial P}/{\partial \eta_{k}}$, ${\partial P}/{\partial \eta_{l}}$ and ${\partial^2 P}/{\partial\eta_{k}\partial\eta_{l}}$. 	Hence, using \eqref{eq:boundedness of P,S and their derivatives}, \eqref{eq:difference of S_inv and P_inv},  \cite[(59) p. 9]{PP2012} and matrix norm inequalities in \cite[p. 61-62]{PP2012}, there exists a constant $M_{1}>0$, irrespective of $N$, such that,
	\begin{align}
	\sup_{\eta\in\widetilde{\Omega}_{1}}|\Psi_{1,\tb}|
	\leq&\|\hat{\theta}^{\LS}-\theta_{0}\|_{2}
	\sup_{\eta\in\widetilde{\Omega}_{1}}\left\|\frac{\partial^2 \hat{S}^{-1}}{\partial\eta_{k}\partial\eta_{l}}\right\|_{F}\|\hat{\theta}^{\LS}\|_{2}\nonumber\\
	&+\|\theta_{0}\|_{2}\sup_{\eta\in\widetilde{\Omega}_{1}}\left\|\frac{\partial^2 \hat{S}^{-1}}{\partial\eta_{k}\partial\eta_{l}}-\frac{\partial^2 P^{-1}}{\partial\eta_{k}\partial\eta_{l}}\right\|_{F}\|\hat{\theta}^{\LS}\|_{2}\nonumber\\
	&+\|\theta_{0}\|_{2}\sup_{\eta\in\widetilde{\Omega}_{1}}\left\|\frac{\partial^2 P^{-1}}{\partial\eta_{k}\partial\eta_{l}}\right\|_{F}
	\|\hat{\theta}^{\LS}-\theta_{0}\|_{2}\nonumber\\
	\leq&M_{1}\|\hat{\theta}^{\LS}-\theta_{0}\|_{2}\|\hat{\theta}^{\LS}\|_{2}\nonumber\\
	&+M_{1}\frac{1}{N}\|\theta_{0}\|_{2}\widehat{\sigma^2}\|N(\Phi^{T}\Phi)^{-1}\|_{F}\|\hat{\theta}^{\LS}\|_{2}\nonumber\\
	&+M_{1}\|\theta_{0}\|_{2}\|\hat{\theta}^{\LS}-\theta_{0}\|_{2}\overset{a.s.}\to 0,
	\end{align}
	where the almost sure convergence can be proved using \eqref{eq:almost sure convergence of PP_N_inv}-\eqref{eq:almost sure convergence of noise variance estimator}, the continuous mapping theorem \cite[Theorem 2.3]{Vaart1998} and {Slutsky's theorem} \cite[Theorem 2.8]{Vaart1998}.	Similarly, 
	it can be shown that there exists a constant $M_{2}>0$, irrespective of $N$, such that
	\begin{align*}
	\sup_{\eta\in\widetilde{\Omega}_{1}}|\Tr(\Psi_{2,\tb})|\leq M_{2}\widehat{\sigma^2}\|N(\Phi^{T}\Phi)^{-1}\|_{F}/N\overset{a.s.}\to 0.
	\end{align*}
	
	Since both $\Psi_{1,\tb}$ and $\Tr(\Psi_{2,\tb})$ converge to zero almost surely and uniformly in $\widetilde{\Omega}_{1}$, we have 
\begin{align*}
\sup_{\eta\in\widetilde{\Omega}_{1}}\left|{\partial^2 \overline{\mathscr{F}_{\EB}}}/{\partial\eta_{k}\partial\eta_{l}}-{\partial^2 W_{\tb}}/{\partial\eta_{k}\partial\eta_{l}}\right| 
\overset{a.s.}\to 0
	\end{align*}
by the continuous mapping theorem \cite[Theorem 2.3]{Vaart1998}. Finally, note that $\overline{\eta}_{N}\overset{a.s.}\to \eta_{\tb}^{*}$ and then by {\cite[Lemma B.17]{JCL2021}}, we have \eqref{eq:almost sure convergence of Hessian matrix}, where the positive definiteness of $A_{\tb}(\eta_{\tb}^{*})$ is  due to Assumption \ref{asp:isolated optimal sets}.
	
	\item Secondly, we show that
	\begin{align}\label{eq:convergence in distribution of gradient}
	\sqrt{N}\left.{\partial \overline{\mathscr{F}_{\EB}}}/{\partial \eta}\right|_{\eta=\eta_{\tb}^{*}} \overset{d.}\to \mathcal{N}(0,\sigma^2B_{\tb}(\eta_{\tb}^{*})\Sigma^{-1}B_{\tb}(\eta_{\tb}^{*})^{T}),
	\end{align}
	where  $B_{\tb}(\eta_{\tb}^{*})$ is defined in \eqref{eq:element of B eta_b_star}.
	
	The $k$th elements of ${\partial \overline{\mathscr{F}_{\EB}}}/{\partial \eta}$ and ${\partial W_{\tb}}/{\partial \eta}$ are
	\begin{align}	
	\frac{\partial \overline{\mathscr{F}_{\EB}}}{\partial \eta_{k}}=&(\hat{\theta}^{\LS})^{T}\frac{\partial \hat{S}^{-1}}{\partial \eta_{k}}\hat{\theta}^{\LS}+\Tr\left(\hat{S}^{-1}\frac{\partial P}{\partial \eta_{k}}\right),\\
	\label{eq:1st order derivative of Wb}
	\frac{\partial W_{\tb}}{\partial \eta_{k}}=&\theta_{0}^{T}\frac{\partial P^{-1}}{\partial \eta_{k}}\theta_{0}+\Tr\left(P^{-1}\frac{\partial P}{\partial \eta_{k}}\right).
	\end{align}
	From Assumption \ref{asp:isolated optimal sets} and \eqref{eq:opt hyperparameter of Wb}, we can see that $\eta_{\tb}^{*}$ should satisfy the first-order optimality condition, i.e. for $k=1,\cdots,p$, $\left.{\partial W_{\tb}}/{\partial \eta_{k}}\right|_{\eta=\eta_{\tb}^{*}}=0$.
	It leads to 
	\begin{align}\label{eq:difference of 1st order derivatives of EB}
	\sqrt{N}\left.\frac{\partial \overline{\mathscr{F}_{\EB}}}{\partial \eta_{k}}\right|_{\eta=\eta_{\tb}^{*}}
	=&\sqrt{N}\left.\left(\frac{\partial \overline{\mathscr{F}_{\EB}}}{\partial \eta_{k}}-\frac{\partial W_{\tb}}{\partial \eta_{k}}\right)\right|_{\eta=\eta_{\tb}^{*}}\nonumber\\
	=&[\Upsilon_{\tb,1}]_{k}+[\Upsilon_{\tb,2}]_{k},
	\end{align}
	where for $k=1,\cdots,p$, the $k$th elements of $\Upsilon_{\tb,1}\in\R^{p}$ and $\Upsilon_{\tb,2}\in\R^{p}$ are
	\begin{align}
	\label{eq:Upsilon_1b}
	&{\left[\Upsilon_{\tb,1}\right]_{k}
	=}\\
	&{\left.\left[(\hat{\theta}^{\LS})^{T}\frac{\partial \hat{S}^{-1}}{\partial \eta_{k}}+\theta_{0}^{T}\frac{\partial P^{-1}}{\partial \eta_{k}}\right]\right|_{\eta=\eta_{\tb}^{*}}
	N(\Phi^{T}\Phi)^{-1}\sqrt{N}\frac{\Phi^{T}V}{N},\nonumber}\\
	\label{eq:Upsilon_2b}
	&\left[\Upsilon_{\tb,2}\right]_{k}
	=\left.\theta_{0}^{T}\sqrt{N}\left(\frac{\partial \hat{S}^{-1}}{\partial \eta_{k}}-\frac{\partial P^{-1}}{\partial \eta_{k}}\right)\right|_{\eta=\eta_{\tb}^{*}}\hat{\theta}^{\LS}\nonumber\\
	&\qquad\qquad+\left.\Tr\left[\sqrt{N}(\hat{S}^{-1}-P^{-1})\frac{\partial P}{\partial \eta_{k}}\right]\right|_{\eta=\eta_{\tb}^{*}}.
	\end{align}
	\begin{enumerate}
		\item For $[\Upsilon_{\tb,1}]_{k}$, using \eqref{eq:almost sure convergence of sqrtPVN}, \eqref{eq:joint convergence in distribution of ls estimate and other terms}, \eqref{eq:almost sure convergence of 1st order derivatives of S_inv and P_inv}, the continuous mapping theorem \cite[Theorem 2.3]{Vaart1998}, {Slutsky's theorem} \cite[Theorem 2.8]{Vaart1998} and \cite[Theorem 2.7]{Vaart1998}, we have
		\begin{align}\label{eq:convergence in dist. of Upsilon_b_1}
		[\Upsilon_{\tb,1}]_{k} \overset{d.}\to 2\left[B_{\tb}(\eta^{*}_{\tb})\right]_{k,:}\Sigma^{-1}\upsilon,
		\end{align}
		where $\left[B_{\tb}(\eta^{*}_{\tb})\right]_{k,:}$ denotes the $k$th row of $B_{\tb}(\eta^{*}_{\tb})$.
		
		\item For $[\Upsilon_{\tb,2}]_{k}$, using \eqref{eq:almost sure convergence of sqrtPVN}, \eqref{eq:almost sure convergence of sqrtN_Sinv_Pinv}, \eqref{eq:almost sure convergence of sqrtN difference of 1st order derivatives of S_inv and P_inv}, the continuous mapping theorem \cite[Theorem 2.3]{Vaart1998}, {Slutsky's theorem} \cite[Theorem 2.8]{Vaart1998} and \cite[Theorem 2.7]{Vaart1998}, it can be seen that
		\begin{align}\label{eq:convergence in dist. of Upsilon_b_2}
		[\Upsilon_{\tb,2}]_{k}\overset{d.}\to 0.
		\end{align}

	\end{enumerate}	
	It follows
	$
	\sqrt{N}\left.{\partial \overline{\mathscr{F}_{\EB}}}/{\partial \eta_{k}}\right|_{\eta=\eta_{\tb}^{*}}\overset{d.}\to
	2\left[B_{\tb}(\eta^{*}_{\tb})\right]_{k,:}\Sigma^{-1}\upsilon$.
	Therefore,
	\begin{align}\label{eq:convergence in distribution of 1st order derivative of Feb with respec to eta at eta_star_b}
	\sqrt{N}\left.{\partial \overline{\mathscr{F}_{\EB}}}/{\partial \eta}\right|_{\eta=\eta_{\tb}^{*}}
	\overset{d.}\to 2B_{\tb}(\eta_{\tb}^{*})\Sigma^{-1}\upsilon.
	\end{align}
	Then noting $\E(\upsilon\upsilon^{T})=\sigma^2\Sigma^{-1}$ in \eqref{eq:covariance of Gamma, upsilon, rho}, the covariance matrix of its limiting distribution is nothing but $4\sigma^2B_{\tb}(\eta_{\tb}^{*})\Sigma^{-1}B_{\tb}(\eta_{\tb}^{*})^{T}$.
	
	\item Lastly, we insert \eqref{eq:almost sure convergence of Hessian matrix} and \eqref{eq:convergence in distribution of gradient} into \eqref{eq:Taylor expansion form for sqrtN_eta_eb difference}. Using {Slutsky's theorem}, we complete the proof of \eqref{eq:convegence in distribution of hat_eta_eb}.
\end{enumerate}

\subsection{Proof of Proposition \ref{prop:bounds of EB hyper-parameter estimator}}

We apply \eqref{eq:evd of Sigma} to $V_{\tb}^{\tH}(\eta_{\tb}^{*})$ in \eqref{eq:def of Vb_etab_star} to obtain
\begin{align*}
\Tr[V_{\tb}^{\tH}(\eta_{\tb}^{*})]=\sum_{i=1}^{n}\frac{4\sigma^2}{\lambda_{i}(\Sigma)}e_{\Sigma,i}^{T}B_{\tb}(\eta_{\tb}^{*})^{T}A_{\tb}(\eta_{\tb}^{*})^{-2}B_{\tb}(\eta_{\tb}^{*})e_{\Sigma,i}.
\end{align*}
Since $\eta_{\tb}^{*}$, as defined in \eqref{eq:Wb hp and its cost function}, is irrespective of $\Sigma$ and only depends on $\theta_0$ and $P$, we can obtain {that $\Tr[V_{\tb}^{\tH}(\eta_{\tb}^{*})]$ increases if we fix $\lambda_{i}(\Sigma)$ for all $i=1,\cdots,n-1$ and decrease $\lambda_{n}(\Sigma)$, i.e., if we increase $\cond(\Sigma)$}. For bounds of $\Tr[V_{\tb}^{\tH}(\eta_{\tb}^{*})]$, we use {\cite[(B.66) in Lemma B.23]{JCL2021}}.

\subsection{Proof of Proposition \ref{prop:convergence in distribution of rls estimate using EB}}

We rewrite \eqref{eq:rls estimator form1} as
\begin{align}
\hat{\theta}^{\TR}(\hat{\eta}_{\EB})
=&\left[\Phi^{T}\Phi+\widehat{\sigma^2} P(\hat{\eta}_{\EB})^{-1}\right]^{-1}\Phi^{T}Y\nonumber\\
=&P(\hat{\eta}_{\EB})\hat{S}(\hat{\eta}_{\EB})^{-1}(\Phi^{T}\Phi)^{-1}\Phi^{T}Y\nonumber\\
=&\left[\hat{S}(\hat{\eta}_{\EB})-\widehat{\sigma^2}(\Phi^{T}\Phi)^{-1}\right]\hat{S}(\hat{\eta}_{\EB})^{-1}(\Phi^{T}\Phi)^{-1}\Phi^{T}Y\nonumber\\
\label{eq:split form 1 of rls estimate}
=&\theta_{0}+(\Phi^{T}\Phi)^{-1}\Phi^{T}V-\frac{1}{N}\widehat{\sigma^2}N(\Phi^{T}\Phi)^{-1}\hat{S}(\hat{\eta}_{\EB})^{-1}\hat{\theta}^{\LS}.
\end{align}
Moreover, under Assumption \ref{asp:properties of P}, using \eqref{eq:boundedness of P, Sinv, Pinv}, \eqref{eq:almost sure convergence of hat_eta_eb}, \eqref{eq:almost sure convergence of S_inv} and {\cite[Lemma B.17]{JCL2021}}, we have
\begin{align}\label{eq:middle result of S_inv and P_inv at eta estimate}
\hat{S}(\hat{\eta}_{\EB})^{-1} \overset{a.s.}\to P(\eta^{*}_{\tb})^{-1}.
\end{align}	 
Then applying \eqref{eq:almost sure convergence of PP_N_inv}, \eqref{eq:almost sure convergence of sqrtPVN}, \eqref{eq:almost sure convergence of noise variance estimator}, \eqref{eq:joint convergence in distribution of ls estimate and other terms}, \eqref{eq:almost sure convergence of hat_eta_eb}, \eqref{eq:split form 1 of rls estimate}, \eqref{eq:middle result of S_inv and P_inv at eta estimate}, the continuous mapping theorem \cite[Theorem 2.3]{Vaart1998}, {Slutsky's theorem} \cite[Theorem 2.8]{Vaart1998} and \cite[Theorem 2.7]{Vaart1998}, {we can conclude that $\sqrt{N}(\hat{\theta}^{\TR}(\hat{\eta}_{\EB})-\theta_{0})\overset{d.}\to\mathcal{N}(0,V^{\ALS}_{1})$}.

\subsection{Proof of Theorem \ref{thm:joint convergence distribution of decomp. of rls estimate up to OpsqrtN}}

For $\sqrt{N}(\hat{\theta}^{\LS}-\theta_{0})$, we can rewrite it as
\begin{align}\label{eq:split form of sqrtN diff_LS}
	&\sqrt{N}(\hat{\theta}^{\LS}-\theta_{0})=\sqrt{N}(\Phi^{T}\Phi)^{-1}\Phi^{T}V\nonumber\\
	=&\Sigma^{-1}\sqrt{N}\frac{\Phi^{T}V}{N}+\left[N(\Phi^{T}\Phi)^{-1}-\Sigma^{-1} \right]\sqrt{N}\frac{\Phi^{T}V}{N},
\end{align}
which is nothing but {\eqref{eq:split form of difference of the theta_ls and theta0}-\eqref{eq:def of theta_ALS_2}}. 
Since \eqref{eq:split form of difference of the theta_ls and theta0} contains two building blocks: $\sqrt{N}[N(\Phi^{T}\Phi)^{-1}-\Sigma^{-1}]$ and $\sqrt{N}\Phi^{T}V/N$, we can apply \eqref{eq:joint convergence in distribution of ls estimate and other terms} together with \eqref{eq:almost sure convergence of PP_N_inv}, \eqref{eq:almost sure convergence of sqrtPVN} and the continuous mapping theorem \cite[Theorem 2.3]{Vaart1998} to derive \eqref{eq:2nd order asymptotic distribution of the theta_ls}-\eqref{eq:def of vartheta_ALS_2}. Moreover, according to \eqref{eq:covariances of Gamma, upsilon and rho} and \cite[(511), (520) p. 60]{PP2012}, we can obtain \eqref{eq:expectation of vartheta_als}-\eqref{eq:def of V_ALS_2}.


\subsection{Proof of Theorem \ref{thm:joint convergence distribution of decomp. of rls estimate up to OpN}}

{ For $\sqrt{N}(\hat{\theta}^{\TR}(\hat{\eta}_{\EB})$ $-\theta_{0})$, we first decompose it using \eqref{eq:split form 1 of rls estimate},
	\begin{align}
	&\sqrt{N}\left( \hat{\theta}^{\TR}(\hat{\eta}_{\EB})-\theta_{0}\right)\nonumber\\
			=&\hat{\theta}^{\ALS}_{1}+\frac{1}{\sqrt{N}}\left(\hat{\theta}^{\ALS}_{2}-\frac{1}{\sqrt{N}}\widehat{\sigma^2}N(\Phi^{T}\Phi)^{-1}\hat{S}(\hat{\eta}_{\EB})^{-1}\hat{\theta}^{\LS}\right)\nonumber,
	\end{align}
	where $\hat{\theta}^{\ALS}_{1}$ and $\hat{\theta}^{\ALS}_{2}$ have no more than first order {expansions}, and}
\begin{align}\label{eq: rewritten form of diff of theta_rls_EB and theta0}
	&-\frac{1}{\sqrt{N}}\widehat{\sigma^2}N(\Phi^{T}\Phi)^{-1}\hat{S}(\hat{\eta}_{\EB})^{-1}\hat{\theta}^{\LS}\nonumber\\
	=&-\frac{1}{\sqrt{N}}\sigma^2\Sigma^{-1}P(\eta_{\tb}^{*})^{-1}\theta_{0}\\
	&-\frac{1}{N}\sqrt{N}\Big[\widehat{\sigma^2}N(\Phi^{T}\Phi)^{-1}\hat{S}(\hat{\eta}_{\EB})^{-1}\hat{\theta}^{\LS}
	-\sigma^2\Sigma^{-1}P(\eta_{\tb}^{*})^{-1}\theta_{0}\Big].\nonumber
\end{align}
It leads to the third order {expansion} of $\sqrt{N}(\hat{\theta}^{\TR}(\hat{\eta}_{\EB})$ $-\theta_{0})$ as shown in \eqref{eq:decomposition of sqrt_diff_theta_rls_EB}.

To derive the third order asymptotic distribution of $\sqrt{N}(\hat{\theta}^{\TR}(\hat{\eta}_{\EB})-\theta_{0})$, we decompose $\hat{\theta}^{\AR}_{\text{b3}}$ in \eqref{eq:def of theta_hat_b3} {as follows:}
\begin{align}
	&-\sqrt{N}\Big[\widehat{\sigma^2}N(\Phi^{T}\Phi)^{-1}\hat{S}(\hat{\eta}_{\EB})^{-1}\hat{\theta}^{\LS}
	-\sigma^2\Sigma^{-1}P(\eta_{\tb}^{*})^{-1}\theta_{0}\Big]\nonumber\\
	=&-\left(\Xi_{\tb,1}^{\AR}+\Xi_{\tb,2}^{\AR}+\Xi_{\tb,3}^{\AR}+\Xi_{\tb,4}^{\AR}\right),
\end{align}
where
\begin{align*}
	\Xi_{\tb,1}^{\AR}=&\sqrt{N}(\widehat{\sigma^2}-\sigma^2)N(\Phi^{T}\Phi)^{-1}\hat{S}(\hat{\eta}_{\EB})^{-1}\hat{\theta}^{\LS},\\
	\Xi_{\tb,2}^{\AR}=&\sigma^2{\sqrt{N}\left[N(\Phi^{T}\Phi)^{-1}-\Sigma^{-1}\right]}\hat{S}(\hat{\eta}_{\EB})^{-1}\hat{\theta}^{\LS},\\
	\Xi_{\tb,3}^{\AR}=&-\sigma^2\Sigma^{-1}\hat{S}(\hat{\eta}_{\EB})^{-1}\nonumber\\
	&\left[\sum_{k=1}^{p}\left.\frac{\partial P(\eta)}{\partial \eta_{k}} \right|_{\eta=\tilde{\eta}_{N}} e_{k}^{T}\sqrt{N}(\hat{\eta}_{\EB}-\eta_{\tb}^{*}) \right]P(\eta_{\tb}^{*})^{-1}\hat{\theta}^{\LS}\nonumber\\
	&+\sigma^2\Sigma^{-1}P(\eta_{\tb}^{*})^{-1}N(\Phi^{T}\Phi)^{-1}\sqrt{N}\Phi^{T}V/N,\\
	\Xi_{\tb,4}^{\AR}=&-\frac{1}{\sqrt{N}}\widehat{\sigma^2}\sigma^2\Sigma^{-1}\hat{S}(\hat{\eta}_{\EB})^{-1}{N}(\Phi^{T}\Phi)^{-1}P(\eta_{\tb}^{*})^{-1}\hat{\theta}^{\LS},
\end{align*}
and for the derivation we use \eqref{eq:diff of S_inv and P_inv for general P} and $\tilde{\eta}_{N}$ belongs to a neighborhood of $\eta^{*}_{\tb}$ with radius $\|\hat{\eta}_{\EB}-\eta^{*}_{\tb}\|_{2}$. 
\begin{itemize}
	\item For $\Xi_{\tb,1}^{\TR}$ and $\Xi_{\tb,2}^{\AR}$, it is clear that they contain building blocks $\sqrt{N}(\widehat{\sigma^2}-\sigma^2)$ and $\sqrt{N}[N(\Phi^{T}\Phi)^{-1}-\Sigma^{-1}]$, respectively. 
	\item For $\Xi_{\tb,3}^{\AR}$, inserting \eqref{eq:Taylor expansion form for sqrtN_eta_eb difference}, \eqref{eq:almost sure convergence of Hessian matrix} and \eqref{eq:convergence in distribution of 1st order derivative of Feb with respec to eta at eta_star_b} into $\sqrt{N}(\hat{\eta}_{\EB}-\eta_{\tb}^{*})$, we can see that $\Xi_{\tb,3}^{\AR}$ contains the building block $\sqrt{N}\Phi^{T}V/N$.
	\item For $\Xi_{\tb,4}^{\AR}$, using \eqref{eq:almost sure convergence of PP_N_inv}, \eqref{eq:almost sure convergence of sqrtPVN}, \eqref{eq:almost sure convergence of noise variance estimator}, \eqref{eq:middle result of S_inv and P_inv at eta estimate} and {Slutsky's theorem} \cite[Theorem 2.8]{Vaart1998}, we have $\Xi_{\tb,4}^{\AR}\overset{a.s.}\to 0$. 
\end{itemize}
Hence, we can apply \eqref{eq:joint convergence in distribution of ls estimate and other terms} together with \eqref{eq:almost sure convergence of PP_N_inv}, \eqref{eq:almost sure convergence of sqrtPVN}, \eqref{eq:almost sure convergence of noise variance estimator}, \eqref{eq:middle result of S_inv and P_inv at eta estimate}, {\cite[Lemma B.17]{JCL2021}}, the continuous mapping theorem \cite[Theorem 2.3]{Vaart1998}, {Slutsky's theorem} \cite[Theorem 2.8]{Vaart1998} and \cite[Theorem 2.7]{Vaart1998} to obtain \eqref{eq:3rd order asymptotic distribution of the theta_rls_EB}-\eqref{eq:def of vartheta_b3}, and
\begin{align}
	\label{eq:original form of Cb}
	C_{\tb}(\eta_{\tb}^{*})
	=&2\left[\sum_{k=1}^{p}P(\eta_{\tb}^{*})^{-1}
	\left.\frac{\partial P(\eta)}{\partial \eta_{k}} \right|_{\eta=\eta_{\tb}^{*}}P(\eta_{\tb}^{*})^{-1}\theta_{0}
	e_{k}^{T}\right]\nonumber\\
	&A_{\tb}(\eta_{\tb}^{*})^{-1}B_{\tb}(\eta_{\tb}^{*})+P(\eta_{\tb}^{*})^{-1}
\end{align}
can be rewritten as \eqref{eq:def of Cb} using \eqref{eq:element of B eta_b_star}, \cite[(59) p. 9]{PP2012}, and the fact that for $e_{k}\in\R^{p}$, its $k$th element is one and others zero.

Moreover, according to \eqref{eq:covariances of Gamma, upsilon and rho} and \cite[(511), (520) p. 60]{PP2012}, we can obtain \eqref{eq:expectation of ad. in terms of Op1_N}-\eqref{eq:def of Vb3_AR_2}.

\subsection{Proof of Proposition \ref{prop:bounds of trace of covariance matrix of model estimator}}

{It can be derived by using the EVD of $\Sigma$ in \eqref{eq:evd of Sigma} and {\cite[Lemma B.23]{JCL2021}}.}

{
\subsection{Proof of Lemma \ref{lemma:explicit expressions of Sigma and C_gamma}}\label{subsec:proof of CGamma}

To derive Lemma \ref{lemma:explicit expressions of Sigma and C_gamma}, we use Newton's generalized binomial formula and formulas of mathematical series.

First, if we consider $H(q)$ in the form of \eqref{eq:tf function}, we have 
\begin{align*}
H(q)=c_{u}\frac{1}{(1-aq^{-1})^2}=c_{u}\sum_{k=0}^{\infty}(k+1)a^{k}q^{-k},
\end{align*}
which implies that the impulse response of $H(q)$ is $h(k)=c_{u}(k+1)a^{k}$ for $k\geq 0$ and $h(k)=0$ for $k<0$.
%
Recall Newton's generalized binomial formula, for any $|x|<1$, the following equality holds,
\begin{align}\label{eq:newton generalized binomial formula}
\frac{1}{(1-x)^{\alpha}}=\sum_{k=0}^{\infty}\left(\begin{array}{c}\alpha+k-1\\ k\end{array}\right)x^{k}.
\end{align}
Then it follows that for $\tau\in\Z$, we insert \eqref{eq:definition of R_u_tau} to obtain
\begin{align}\label{eq:middle result of Sigma element}
R_{u}(\tau)
=&2\sigma^{2}_{e}c_{u}^2a^{|\tau|}\sum_{k=0}^{\infty}\left[\frac{(k+1)(k+2)}{2}(a^2)^{k}\right]\nonumber\\
&+(|\tau|-1)\sigma^{2}_{e}c_{u}^2a^{|\tau|}\sum_{k=0}^{\infty}\left[(k+1)(a^2)^{k}\right]\nonumber\\
=&c_{u}^2\sigma_{e}^2a^{|\tau|}\left[\frac{2}{(1-a^2)^3} + \frac{|\tau|-1}{(1-a^2)^2}\right],
\end{align}
which is derived from \eqref{eq:newton generalized binomial formula} and $|a|<1$. 

Then, by using formulas of mathematical series, we can insert \eqref{eq:middle result of Sigma element} into each element of $\Sigma$ and \eqref{eq:explicit representation of Expectation of Gamma_o_Gamma} to obtain \eqref{eq:ij_th element of Sigma for filtered white noise inputs} and \eqref{eq:Cgamma for 2nd order filter}, respectively. Moreover, when $a=0$, we can obtain \eqref{eq:Sigma for white noise inputs} and \eqref{eq:def of C_H for ridge regression case with white noise input}.}

\bibliographystyle{abbrv}
\bibliography{database}

\begin{thebibliography}{10}

\bibitem{BP2020mathematical}
M.~Bisiacco and G.~Pillonetto.
\newblock On the mathematical foundations of stable {RKHS}s.
\newblock {\em Automatica}, 118:109038, 2020.

\bibitem{Chen18}
T.~Chen.
\newblock On kernel design for regularized {LTI} system identification.
\newblock {\em Automatica}, 90:109--122, 2018.

\bibitem{CA21}
T.~Chen and M.~S. Andersen.
\newblock On semiseparable kernels and efficient implementation for regularized
  system identification and function estimation.
\newblock {\em Automatica}, 132:109682, 2021.

\bibitem{CALCP14}
T.~Chen, M.~S. Andersen, L.~Ljung, A.~Chiuso, and G.~Pillonetto.
\newblock System identification via sparse multiple kernel-based regularization
  using sequential convex optimization techniques.
\newblock {\em IEEE Transactions on Automatic Control}, (11):2933--2945, 2014.

\bibitem{CL2013}
T.~Chen and L.~Ljung.
\newblock Implementation of algorithms for tuning parameters in regularized
  least squares problems in system identification.
\newblock {\em Automatica}, 49(7):2213 -- 2220, 2013.

\bibitem{COL2012}
T.~Chen, H.~Ohlsson, and L.~Ljung.
\newblock On the estimation of transfer functions, regularizations and
  {G}aussian processes -- revisited.
\newblock {\em Automatica}, 48(8):1525--1535, 2012.

\bibitem{Chiuso16}
A.~Chiuso.
\newblock Regularization and {B}ayesian learning in dynamical systems: {P}ast,
  present and future.
\newblock {\em Annual Reviews in Control}, 41:24 -- 38, 2016.

\bibitem{CS02}
F.~Cucker and S.~Smale.
\newblock On the mathematical foundations of learning.
\newblock {\em American Mathematical Society}, 39(1):1--49, 2002.

\bibitem{G94}
J.~K. Ghosh.
\newblock Higher order asymptotics.
\newblock {\em NSF-CBMS Regional Conference Series in Probability and
  Statistics}, 4:i--111, 1994.

\bibitem{Hastieetal:01}
T.~Hastie, R.~Tibshirani, and J.~Friedman.
\newblock {\em The Elements of Statistical Learning}.
\newblock Springer, 2001.

\bibitem{Hnotes}
H.~Hjalmarsson.
\newblock Dynamic model learning: A geometric perspective.
\newblock {\em Lecture Notes in FEL3201/FEL3202}.

\bibitem{H2020}
H.~Hjalmarsson.
\newblock Estimation accuracy of kernel-based estimators.
\newblock {\em IFAC WC Workshop on Bayesian and Kernel-Based Methods in
  Learning Dynamical Systems}, 2020.

\bibitem{JCML20}
Y.~Ju, T.~Chen, B.~Mu, and L.~Ljung.
\newblock On the influence of ill-conditioned regression matrix on
  hyper-parameter estimators for kernel-based regularization methods.
\newblock In {\em 2020 59th IEEE Conference on Decision and Control (CDC)},
  pages 300--305, 2020.

\bibitem{JCL2021}
Y.~Ju, T.~Chen, B.~Mu, and L.~Ljung.
\newblock A tutorial on asymptotic properties of regularized least squares
  estimator for finite impulse response model.
\newblock {\em arXiv e-prints: 2112.10319}, 2021.

\bibitem{Ljung1995}
L.~Ljung.
\newblock {\em System Identification Toolbox for Use with MATLAB}.
\newblock The Math Works, 1995.

\bibitem{Ljung1999}
L.~Ljung.
\newblock {\em System Identification: Theory for the User}.
\newblock Upper Saddle River, NJ: Prentice Hall, 12 1999.

\bibitem{LCM20}
L.~Ljung, T.~Chen, and B.~Mu.
\newblock A shift in paradigm for system identification.
\newblock {\em International Journal of Control}, 93(2):173--180, 2020.

\bibitem{MSS16}
A.~Marconato, M.~Schoukens, and J.~Schoukens.
\newblock Filter-based regularisation for impulse response modelling.
\newblock {\em IET Control Theory \& Applications}, 11:194--204, 2016.

\bibitem{MH09}
R.~M. Mnatsakanov and A.~S. Hakobyan.
\newblock Recovery of distributions via moments.
\newblock In {\em Optimality}, pages 252--265. Institute of Mathematical
  Statistics, 2009.

\bibitem{MCL2018gcv}
B.~Mu, T.~Chen, and L.~Ljung.
\newblock Asymptotic properties of generalized cross validation estimators for
  regularized system identification.
\newblock {\em IFAC-PapersOnLine}, 51(15):203--208, 2018.

\bibitem{MCL2018}
B.~Mu, T.~Chen, and L.~Ljung.
\newblock On asymptotic properties of hyperparameter estimators for
  kernel-based regularization methods.
\newblock {\em Automatica}, 94:381--395, 2018.

\bibitem{PP2012}
K.~B. Petersen and M.~S. Pedersen.
\newblock The matrix cookbook, 2012.

\bibitem{PCCDL2022}
G.~Pillonetto, T.~Chen, A.~Chiuso, G.~De~Nicolao, and L.~Ljung.
\newblock {\em Regularized System Identification: Learning Dynamic Models from
  Data}.
\newblock Springer Nature, 2022.

\bibitem{PC15}
G.~Pillonetto and A.~Chiuso.
\newblock Tuning complexity in regularized kernel-based regression and linear
  system identification: The robustness of the marginal likelihood estimator.
\newblock {\em Automatica}, 58:106--117, 2015.

\bibitem{PD2010}
G.~Pillonetto and G.~De~Nicolao.
\newblock A new kernel-based approach for linear system identification.
\newblock {\em Automatica}, 46(1):81--93, 2010.

\bibitem{PDCDL2014}
G.~Pillonetto, F.~Dinuzzo, T.~Chen, G.~De~Nicolao, and L.~Ljung.
\newblock Kernel methods in system identification, machine learning and
  function estimation: A survey.
\newblock {\em Automatica}, 50(3):657--682, 2014.

\bibitem{RasmussenW:06}
C.~E. Rasmussen and C.~K.~I. Williams.
\newblock {\em Gaussian Processes for Machine Learning}.
\newblock MIT Press, Cambridge, MA, 2006.

\bibitem{Vaart1998}
A.~W. v.~d. Vaart.
\newblock {\em Asymptotic Statistics}.
\newblock Cambridge Series in Statistical and Probabilistic Mathematics.
  Cambridge University Press, 1998.

\bibitem{WR2004}
D.~P. Wipf and B.~D. Rao.
\newblock Sparse bayesian learning for basis selection.
\newblock {\em IEEE Transactions on Signal processing}, 52(8):2153--2164, 2004.

\bibitem{Z21}
M.~Zorzi.
\newblock A second-order generalization of {TC} and {DC} kernels.
\newblock {\em arXiv preprint arXiv:2109.09562}, 2021.

\bibitem{Zorzi22}
M.~Zorzi.
\newblock Nonparametric identification of kronecker networks.
\newblock {\em Automatica}, 145:110518, 2022.

\bibitem{ZC17}
M.~Zorzi and A.~Chiuso.
\newblock Sparse plus low rank network identification: A nonparametric
  approach.
\newblock {\em Automatica}, 76:355--366, 2017.

\bibitem{ZC18}
M.~Zorzi and A.~Chiuso.
\newblock The harmonic analysis of kernel functions.
\newblock {\em Automatica}, 94:125--137, 2018.

\end{thebibliography}

\begin{IEEEbiography}[{\includegraphics[width=1in,height=1.25in,clip,keepaspectratio]{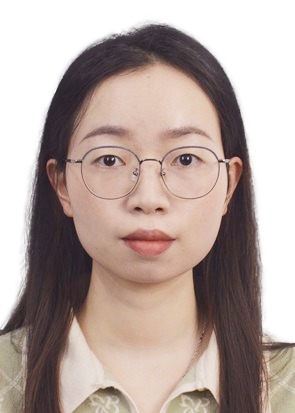}}]{Yue Ju} received her Bachelor degree from Nanjing University of Science $\&$ Technology in 2017 and her Ph.D. from the Chinese University of Hong Kong, Shenzhen, in 2022. She is now a postdoc at the Chinese University of Hong Kong, Shenzhen and Shenzhen Research Institute of Big Data. She has been mainly working in the area of system identification.
\end{IEEEbiography}
\vskip -2\baselineskip plus -1fil
\begin{IEEEbiography}[{\includegraphics[width=1in,height=1.25in,clip,keepaspectratio]{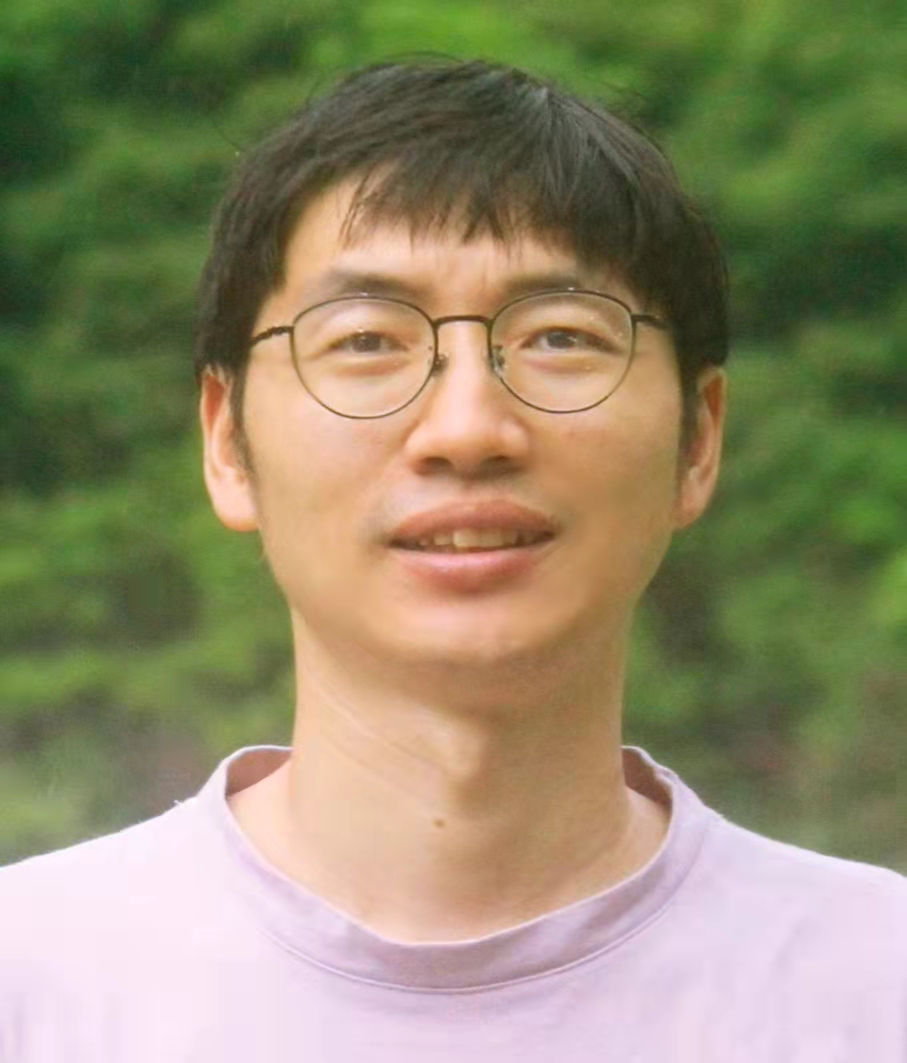}}]{Biqiang Mu} received the Bachelor of Engineering degree from Sichuan University and the Ph.D. degree in Operations Research and Cybernetics from the Academy of Mathematics and Systems Science, Chinese Academy of Sciences. He was a postdoc at the Wayne State University, the Western Sydney University, and the Link\"{o}ping University, respectively. He is currently an associate professor at the Academy of Mathematics and Systems Science, Chinese Academy of Sciences. His research interests include system identification, machine learning, and their applications.
\end{IEEEbiography}
\vskip -2\baselineskip plus -1fil
\begin{IEEEbiography}[{\includegraphics[width=1in,height=1.25in,clip,keepaspectratio]{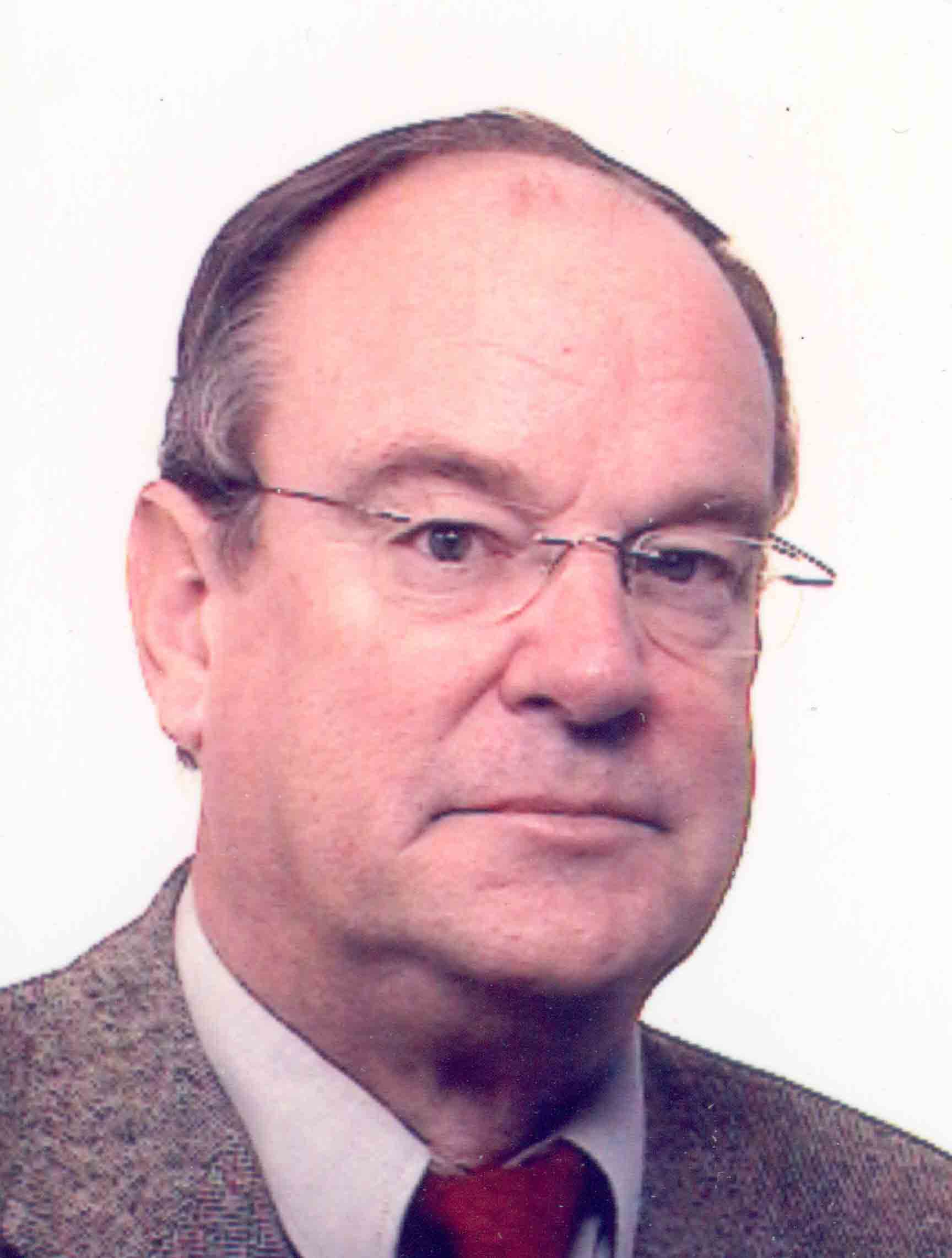}}]
{Lennart Ljung} received his PhD in Automatic Control from Lund Institute of Technology in 1974. Since 1976 he is Professor of the chair of Automatic Control In Link\"{o}ping, Sweden. He has held visiting positions at Stanford and MIT and has written several books on System Identification
and Estimation. He is an IEEE Fellow, an IFAC Fellow and an IFAC Advisor. He is as a member of the Royal Swedish Academy of
Sciences (KVA), a member of the Royal Swedish Academy of Engineering
Sciences (IVA),  an Honorary Member of the Hungarian Academy of Engineering, an Honorary Professor of the Chinese Academy of Mathematics and Systems Science, and a Foreign Member
of the US National Academy of Engineering (NAE). He has received honorary doctorates from the Baltic
State Technical University in St Petersburg, from  Uppsala University, Sweden, from the Technical University of Troyes, France,  from the Catholic
University of Leuven, Belgium and from Helsinki University of Technology, Finland.
He has received both the Quazza Medal (2002) and the Nichols Medal (2017) from IFAC. In 2003 he  received the Hendrik W. Bode Lecture Prize from the IEEE Control Systems Society, and he was the 2007 recipient of the IEEE Control Systems Award.
In 2018 he received the Great Gold Medal from the Royal Swedish Academy of Engineering.

\end{IEEEbiography}
\vskip -2\baselineskip plus -1fil
\begin{IEEEbiography}[{\includegraphics[width=1in,height=1.25in,clip,keepaspectratio]{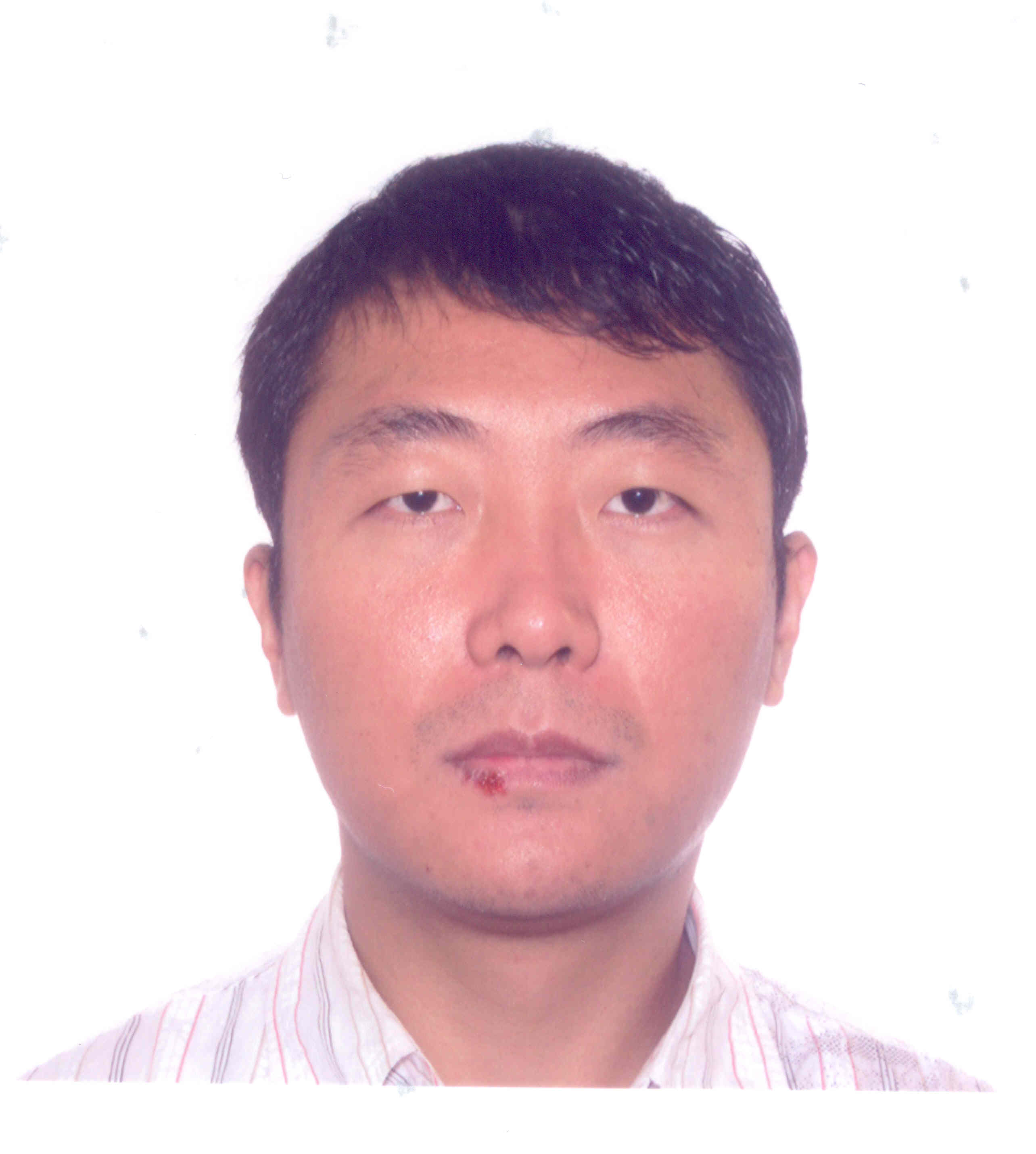}}]{Tianshi Chen} received his Ph.D. in Automation and Computer-Aided Engineering from The Chinese University of Hong Kong in December 2008. From April 2009 to December 2015, he was working in the Division of Automatic Control, Department of Electrical Engineering, Link\"{o}ping University, Link\"{o}ping, Sweden, first as a Postdoc and then (from April 2011) as an Assistant Professor. In May 2015, he received the Oversea High-Level Youth Talents Award of China, and in December 2015, he joined the Chinese University of Hong Kong, Shenzhen (CUHK-SZ), as an Associate Professor.   
His research interests include system identification, state estimation, automatic control, and their applications. He is/was an associate editor for Automatica (2017-present), System \& Control Letters (2017-2020), and IEEE CSS Conference Editorial Board (2016-2019). He received several teaching awards, including the Presidential Examplary Teaching Award of CUHK-SZ in 2021 and the Outstanding Teacher Award of Shenzhen in 2022.  He was a plenary speaker at the 19th IFAC Symposium on System Identification, Padova, Italy, 2021, and he is a coauthor of the book ``Regularized System Identification - Learning Dynamic Models from Data''. 
\end{IEEEbiography}

\end{document}